\documentclass[structabstract]{aa710} 
\usepackage{graphicx}
\usepackage{txfonts}
\usepackage{natbib}
\bibpunct{(}{)}{;}{a}{}{,} 

\defcitealias{Kampert:2011}{Kampert 2011}
\defcitealias{Matthews:2013}{Matthews 2013}

\usepackage[font=small]{caption}

\begin{document}

\title{Anisotropy expectations for ultra-high-energy cosmic rays with future high statistics experiments}

\author{B. Rouill\'e d'Orfeuil \inst{1,2} \and D. Allard \inst{3} \and C. Lachaud \inst{3} \and E. Parizot \inst{3} \and C. Blaksley \inst{3} \and S.~Nagataki \inst{4}}

\institute{Kavli Institute for Cosmological Physics and Enrico Fermi Institute, The University of Chicago, Chicago, IL, USA \and
Laboratoire de l'Acc\'el\'erateur Lin\'eaire (LAL), Universit\'e Paris 11, CNRS-IN2P3, France \\
\email{rouille@lal.in2p3.fr} \and
Laboratoire AstroParticule et Cosmologie (APC), Universit\'e Paris 7, CNRS-IN2P3, Paris, France \and
Astrophysical Big Bang Laboratory, RIKEN, Saitama 351-0198, Japan}

\date{}
 
\abstract
{Ultra-high-energy cosmic rays (UHECRs) have attracted a lot of attention in astroparticle physics and high-energy astrophysics, due to their challengingly high energies, and to their potential value to constrain the physical processes and astrophysical parameters in the most energetic sources of the universe. Current detectors, despite their very large acceptance, have failed to detect significant anisotropies in their arrival directions, which had been expected to lead to the long-sought identification of their sources. Some indications about the composition of the UHECRs, which may become heavier at the highest energies, has even put into question the possibility that such a goal could be achieved in the foreseeable future.}
{We investigate the potential value of a new-generation detector, with an exposure increased by one order of magnitude, to overcome the current situation and make significant progress in the detection of anisotropies and thus in the study of UHECRs. We take as an example the expected performances of the JEM-EUSO detector, assuming a uniform full-sky coverage with a total exposure of 300,000~km$^{2}$~sr~yr.}
{We simulate realistic UHECR sky maps for a wide range of possible astrophysical scenarios allowed by the current constraints, taking into account the energy losses and photo-dissociation of the UHE protons and nuclei, as well as their deflections by intervening magnetic fields. These sky maps, built for the expected statistics of JEM-EUSO as well as for the current Auger statistics, as a reference, are analyzed from the point of view of their intrinsic anisotropies, using the two-point correlation function. A statistical study of the resulting anisotropies is performed for each astrophysical scenario, varying the UHECR source composition and spectrum as well as the source density, and exploring a set of five hundred independent realizations for each choice of the set of parameters.}
{We find that significant anisotropies are expected to be detected by a next-generation UHECR detector, for essentially all the astrophysical scenarios studied, and give precise, quantitative meaning to this statement.}
{Our results show that a gain of one order of magnitude in the total exposure of UHECR detectors would make a significant difference compared to the existing, and allow considerable progress in the study of these mysterious particles and their sources.}


\maketitle


\section{Introduction}
\label{sec:introduction}

Discovering the origin of the ultra-high-energy cosmic rays (UHECRs), with energies of the order of 100~EeV is widely recognized as an important challenge in the field of astroparticle physics, both from the theoretical and observational points of view. Among the main scientific goals are a deeper understanding of particle acceleration in the universe and the search for complementary information and constraints on potential sources, in a multi-messenger approach involving photons, energetic nuclei, neutrinos, and, perhaps, gravitational waves \citep[see][for a recent review]{Kotera:2011}.

This quest has been pursued for decades with larger and larger detectors \citep[see][for reviews]{Nagano:2000,Letessier-Selvon:2011}, and is currently led in the Southern hemisphere by the Pierre Auger Collaboration \citep{Auger:2004a}, and in the Northern hemisphere by the Telescope Array Collaboration \citep{TA:2012a,TA:2012b}. Important milestones have been reached, notably through the confirmation of the so-called GZK-effect \citep{Greisen:1966,Zatsepin:1966}, which causes a rapid decrease of the UHECR flux around 60~EeV \citep{HiRes:2008a,Auger:2010a,TA:2013a} due to the interaction of the UHE protons and/or nuclei with the extragalactic background radiation. Hints of anisotropies in the arrival directions of the UHECRs above $\sim 60$~EeV have also been reported \citep{Auger:2007a,Auger:2008b}, indicating that the intervening magnetic fields do not completely randomize the distribution of UHECRs on the sky, as it does at lower energies, preventing a direct identification of the sources.

However, the initial hope to reveal the sources by observing an accumulation of UHECR events in well-defined directions has been frustrated up to now. This is either because the number of contributing sources is too large and very few events have been observed from any given source, or because the particles observed from a given source are deflected and spread over too large areas of the sky, or both. 

To overcome these difficulties, a natural strategy is to focus on the highest energy cosmic rays, because their magnetic rigidity is expected to be larger (unless they also have a larger charge), and because the number of sources visible from Earth is significantly smaller, due to the GZK-effect. General statistical studies taking into account the relevant propagation effects show that only a handful of sources are responsible for most of the detectable UHECRs around 100~EeV~\citep{Blaksley:2013}. Depending on the source density and the exact contingent distribution of the sources around us, the dominant source in the sky typically account for between 15\% and 60\% of the total flux at 100~EeV \citep{Blaksley:2013}. In such conditions, it seems likely that the very few contributing sources could be distinguished as separate clusters of events on the sky, even if most particles suffer relatively large deflections, provided a sufficient number of UHECRs are observed.

This has not yet occurred with the very limited statistics available. No significant anisotropy has been detected at 100~EeV, and even though the Pierre Auger Observatory has reported a departure from isotropy at lower energy, $E \sim 60$~EeV, at the 99\% confidence level \citep{Auger:2007a,Auger:2010b}, its interpretation is still unclear and it could not be used to gather any decisive information about the sources.

In total, an exposure of the order of $\sim 30,000\,\mathrm{km}^{2}\,\mathrm{sr}\,\mathrm{yr}$ has been accumulated so far, and it appears that decisive progress towards the identification of the sources will not be possible unless a significant increase in exposure is achieved. In this paper, we investigate whether a next-generation detector gathering an exposure of $300,000\,\,\mathrm{km}^{2}\,\mathrm{sr}\,\mathrm{yr}$ at the highest energies can indeed detect a significant anisotropy in the UHECR arrival directions, even though no clear signal emerges with the currently available statistics.

To this end, we investigate several astrophysical scenarios, varying the UHECR source density as well as the injection composition and energy spectrum, and build simulated sky maps taking into account the propagation of the UHECRs in the cosmological photon backgrounds and in the extragalactic and Galactic magnetic fields. These sky maps, built for various total numbers of events, are then analyzed from the point of view of their intrinsic anisotropy, through the analysis of the two-point correlation function. Among the different scenarios, only those which lead to typical sky-maps that are not ruled out by the current data (i.e. when examined with the Auger statistics) are considered as reasonable scenarios, and explored with larger statistics. Many realizations of each astrophysical scenario are simulated, to investigate the range of possible corresponding sky-maps, referred to as the ``cosmic variance''.

To be definite, we use the JEM-EUSO mission as a prospective example of such a next-generation detector, i.e. we assume a (nearly) uniform full-sky coverage and the actual detection efficiency of JEM-EUSO as a function of energy, as described in \cite{JEM-EUSO:2013a}. We also assume a conservative energy resolution of 30\% to demonstrate that the resulting spill-over of lower-energy events (whose sources can be more distant and numerous) due to a mis-reconstructed energy does not compromise the main result of this study: significant anisotropy is indeed expected to be detected at the highest energies for essentially all the models studied, even in the extreme case where the UHECRs are completely dominated by heavy nuclei at the highest energies. Therefore, a full-sky coverage detector achieving an exposure of $300,000\,\,\mathrm{km}^{2}\,\mathrm{sr}\,\mathrm{yr}$ should make a valuable difference in the current state of knowledge regarding UHECRs, and start a new phase of their phenomenological and theoretical study, through the identification of significant anisotropies and the study of individual sources.

In Sect.~\ref{sec:GZK}, we emphasize the key feature of the GZK horizon effect, which makes such a study possible at the highest end of the cosmic ray energy spectrum, even in the presence of relatively large deflections (e.g. with an Fe-dominated composition). In Sect.~\ref{sec:models}, we describe the various astrophysical scenarios explored and their main ingredients. In Sect.~\ref{sec:skymaps}, we describe the magnetic field models implemented in the simulation and the general procedure used to produce representative UHECR sky-maps. In Sect.~\ref{sec:results}, we present and discuss the main results of our investigation, through the quantitative study of the anisotropies that can be expected for the various scenarios. A general summary is given in Sect.~\ref{sec:summary}.


\section{The GZK cutoff and UHECR anisotropies}
\label{sec:GZK}

As recalled in the introduction, current UHECR observatories did not succeed in detecting significant and unambiguous anisotropies, and in reaching the stage of individual source astronomy. Two important elements of the UHECR phenomenology may concur to such a situation: the typical source density, and the angular deflection of the particles. Since the total UHECR flux is known, the effective source density is directly related to the intrinsic power of the sources, which determines how many events can be expected from a given source (depending on its distance). As \cite{Blaksley:2013} have shown, the most intense sources should already have contributed several events to the current dataset. Yet, no obvious multiplets, defined as UHECR events coming from the same source, have been identified so far. If multiplets are present, but not recognized as such, it must be that the typical deflections exceed or approach the angular separation between sources. At energies between 10 and 50~EeV, say, this may be due to the large number of sources, which overlap more or less uniformly over the sky. At higher energies, however, the GZK effect can be extremely valuable because of its most basic consequence, which is to reduce the horizon distance, i.e. the distance beyond which the contribution of UHECR sources to the observed flux is negligible. As the horizon becomes significantly closer with increasing energy (from $\sim$180 Mpc to $\sim$ 75 Mpc between 60~EeV and 100~EeV for protons), the number of visible sources is considerably reduced (by more than a factor of 10 between these two energies) and the background of more distant sources -- which are also more isotropically distributed -- is cut.

For this reason, it is crucial to focus on the highest energy particles. Above 60~EeV or so, the visible sources are located in a region of the universe where matter is not distributed uniformly. Even in the case of large source densities, the source distribution itself would leave a visible imprint on the UHECR sky map if the deflections were small \citep{Decerprit:2012a}. Thus, the absence of clear large-scale anisotropy patterns already puts some constraints on the typical deflection of UHECRs, with deflections probably larger than $\sim 10$--15~degrees, whether due to a large magnetic field or a large electric charge. This is in agreement with the indication that a transition towards heavier nuclei occurs above 10$^{19}$~eV, as shown by the analysis of the Auger data \citep{Auger:2010c}. A small fraction of low-Z nuclei may nevertheless be present among the UHECRs, which would eventually lead to the observation of multiplets on a small angular scale as the statistics increases, especially at the highest energies where the deflections of the particles are smallest. 

To overcome the general problem caused by the large deflections of UHECRs, a natural strategy is thus to extend the exposure of the experiments at energies two or three times higher than currently available with reasonable statistics, not only to reduce the deflections by the same factor (if the composition does not become heavier above 60~EeV), but mostly to reduce drastically the number of sources, and make their separation on the sky easier. At 100~EeV, most of the UHECRs are expected to come from a handful of sources, even in the case of relatively large source densities \citep{Blaksley:2013}. One may thus expect to be able to isolate them on the sky or at least detect a significant anisotropy from the associated multiplets, even if the deflections are relatively large, e.g. in the case of Fe nuclei in a typical Galactic and extragalactic magnetic field.

To verify this basic idea quantitatively, and estimate the required integrated exposure, we perform simulations under various astrophysical assumptions, as described in the next section.

The reason why many different scenarios are possible is that the all-sky UHECR spectrum does not contain much information by itself. The significant, drastic reduction of the flux observed above $\sim 60$~EeV is generically attributed to ``the'' GZK-cutoff. However, there is not a \emph{single} GZK-cutoff, and the current measurements are compatible with a wide range of models, because what gives its shape to the all-sky propagated spectrum is the horizon scale structure (its variation with energy), much more than the individual source spectrum, maximum energies, source density, and even source composition. The coincidental similarity, in first approximation, of the horizon scale structure for protons and Fe nuclei \citep{Allard:2008} is such that the current spectrum can be accounted for by a model assuming pure proton sources, pure Fe sources, or a mixed composition of cosmic rays. Even unrealistic scenarios where the sources accelerate, say, only C nuclei, or only Si nuclei, or essentially any individual nuclear species, can give a good fit of the data (above the ankle, interpreted as the transition from Galactic to extragalactic nuclei in these scenarios), provided that one suitably adjust the source spectrum, which is unknown anyway. This is because Fe nuclei have the same horizon structure as the protons, and the lighter nuclei are rapidly destroyed by photo-disintegration in the intergalactic photon fields, the resulting secondary protons then giving rise to the standard GZK-cutoff, as if the sources were effectively proton sources, i.e. as if they were essentially emitting directly the secondary particles.

This property of UHECR phenomenology is the reason why we do not expect significant progress in this field of research from a refined study of the all-sky spectrum (even though the discovery of a recovery in the spectrum, see \citet{Berezinsky:2006}, or any unexpected feature would be significant), but rather from anisotropy studies at the highest energies, and/or the comparison of the spectra in different regions of the sky.

In order to determine whether significant anisotropies can be observed with future detectors, and to evaluate the confidence level with which they can be expected to be measured, we simulate UHECR sky maps for a wide range of models, as described below.


\section{The source models and their parameters}
\label{sec:models}

\subsection{General astrophysical assumptions}

Standard propagation studies that take into account the energy losses and angular deflections of the UHECRs allow to reproduce the whole sky spectrum with a limited number of parameters, under the simplifying assumption that all UHECR sources are essentially identical, i.e. i) they have the same intrinsic power, ii) they inject UHECRs in the extragalactic medium continuously, iii) with the same power-law energy spectrum, and iv) with the same composition. The remaining free parameters are the source density, the logarithmic slope of the source spectrum, the maximum energy of the UHECRs at the source, and the relative abundances of the various nuclei. These parameters are not independent, and must be chosen so as to reproduce the observed spectrum (see below).

It is likely that, in reality, individual sources are all different. However, the above assumptions allow to explore a large set of models which should be representative of the range of patterns one may expect from the point of view of anisotropies. Relaxing them would introduce more free parameters on which there are no constraints at the moment, neither observationally nor theoretically, without significantly enlarging the range of possibilities explored. The only assumption which we relax in the present study is that of an identical intrinsic power for all sources -- namely, we adopt the same intrinsic luminosity distribution as that of the galaxies in the catalog that we use to represent a realistic source distribution in the local universe (see below). This can be argued to provide a more natural benchmark luminosity distribution than the ``standard candle'' assumption, but it is a non-essential feature of our study, so we stick to this initial assumption throughout and do not explore an extra dimension of the parameter space by varying the luminosity distribution.

A distribution of maximum energies can also be expected in principle, but its net effect would be a difference between the actual spectrum of the UHECRs \emph{at the source} and the \emph{effective source spectrum} convolving the former with the maximum energy distribution, in a way that does not modify the overall phenomenology of UHECRs significantly (e.g. \citealt{Kachelriess:2006}, \citealt{Blaksley:2012}). \citet{Blaksley:2012} have also shown that a distribution of maximum energies results in a modification of the effective source composition, but this effect is already covered by the range of compositions that we explore (see Sect.~\ref{sec:sourceCompo}). In principle, one may also consider a cosmological evolution of the number density and/or power of the sources, e.g. following the star formation rate as a function of redshift or another law characterizing the UHECR sources. However, this is known to modify the constraints imposed by the data on the astrophysical models only (or mostly) by requiring a different intrinsic spectral index at the source \citep{Kotera:2010}, with no significant effect on the observed cosmic rays in the GZK energy range. This is due to the horizon effect, which results in the fact that the UHECR flux in this energy range is completely dominated by nearby, and thus almost contemporary sources (compared to the average source evolution timescale). Therefore, the anisotropy patterns should not be expected to depend significantly on the source evolution.

A more important effect on anisotropies should however be expected if one would relax the assumption of continuous sources. A new set of models with impulsive sources (e.g. in scenarios where GRBs are the sources of UHECRs) could then be simulated. This is not explored in the present paper. We simply note that, leaving the other parameters unchanged, an impulsive source model should in principle give rise to stronger anisotropies than the corresponding continuous source model does, because of the resulting larger effective source power (for the sources contributing to the observed flux at a given time) and the smaller range of energies observed at a given time from a given source (due to the energy-dependent diffusion effects), and thus the smaller range of angular deflections.


\subsection{Source composition and energy spectrum}
\label{sec:sourceCompo}

\begin{table*}[t!]
\caption{Physical parameters used in the different source models}
\begin{tabular}{|c|c|c|c|}
\hline
& & & \\
{\bf Model} & {\bf x} & {\bf $E_{\max}$} & {\bf Comment} \\
\hline
\hline
& & & \\
MC-high & 2.3 &$Z\times 10^{20.5}$~eV & mixed composition, but dominated by protons up to the highest energies \\
\hline
& & & \\
MC-4EeV & 1.4 &$Z\times 4\,10^{18}$~eV & Fe-dominated at high energy $\rightarrow$ good agreement with the Auger composition trend \\
\hline
& & & \\
MC-15EeV & 1.4 &$Z\times 1.5\,10^{19}$~eV & no protons, but some CNO and intermediate nuclei injected above $5 \times10^{19}$ eV \\
\hline
& & & \\
pure-p & 2.6 &$ 10^{20.5}$~eV & only protons at all energies \\
\hline
& & & \\
pure-Fe & 2.3 &$Z\times 10^{20.5}$~eV & only Fe nuclei at the source $\rightarrow$ subdominant but sizable fraction of protons at the highest energies \\
\hline
\end{tabular}
\label{table:compo}
\end{table*}

The relative abundance of the various nuclei accelerated at the UHECR sources -- simply referred to as the \emph{source composition} -- has a strong influence on the level of anisotropies that one can expect to measure. Obviously, scenarios in which UHECRs are dominated by protons are much more favorable than scenarios in which Fe nuclei are dominant: \emph{ceteris paribus}, 
the smaller deflections of protons result in much tighter multiplets observable in the sky map, and facilitate the study of individual sources and their identification. In the absence of prior knowledge about the source composition, we explore a range of possibilities, choosing models with the main requirement that they reproduce the measured cosmic-ray spectrum above the ankle energy, based on either the Pierre Auger Observatory or the HiRes and Telescope Array data, which at first order appear to differ only by a global shift in the energy scale \citep{Dawson:2013,Fukushima:2013}.

The energy spectrum at the source is assumed to be a power-law with a logarithmic index $x$, which is the same for all nuclear species. For each nucleus, $i$, we set a maximum energy, $E_{\max,i}$, proportional to its charge, $Z_{i}$, so that $E_{\max,i} = Z_{i} \times E_{\max,\mathrm{p}}$, where $E_{\max,\mathrm{p}}$ is the maximum proton energy (assumed to be the same in all sources). An exponential cutoff is then applied at $E_{\max,i}$. 

In the present study, we consider five different scenarios, referred to as:
\begin{enumerate}
\item MC-high (mixed composition, with high $E_{\max}$)\vspace{3pt}

\item MC-4EeV (mixed composition with $E_{\max} = 4$~EeV)\vspace{3pt}

\item MC-15EeV (mixed composition with $E_{\max} = 15$~EeV)\vspace{3pt}

\item pure-p (pure proton model)\vspace{3pt}

\item pure-Fe (pure iron model)

\end{enumerate}

The ``MC-high model'' is the so-called mixed-composition model introduced in \citet{Allard:2005}, in which the UHECR source composition, below $E_{\max,\mathrm{p}}$, is assumed to be similar to that of low-energy Galactic cosmic-rays. A good fit of the UHECR spectrum data is obtained by assuming a spectral index $x = 2.3$ and a maximum energy $E_{\max} = Z\times 10^{20.5}$~eV. It follows from this high value of the maximum proton energy that the composition is dominated by protons at all energies. Although the propagated spectrum expected for this model \citep{Allard:2005,Allard:2007a,Allard:2007b} is compatible with the observations, the expected composition above $10^{19}$~eV does not appear to reproduce the Auger data concerning the average penetration depth of the induced atmospheric shower, $\langle X_{\max}\rangle$, nor its shower-to-shower fluctuation, RMS($X_{\max}$), as a function of energy. In principle, this model can however be reconciled with the data if the results concerning these observables are regarded as showing evidence of a change in the underlying hadronic physics rather than a change in UHECR composition.

The next two source models belong to the category of the so-called ``low $E_{\max}$ models'' described in \citealt{Allard:2008},  which refers to mixed-composition models in which the protons do not reach the highest energies. In these models, protons are accelerated by the sources only up to a maximum energy that is lower than the GZK-cutoff energy range. In such scenarios, the source composition above $10^{19}$~eV is gradually becoming heavier, with a dominant contribution of Fe nuclei above, say, 30~EeV. This appears conform to the evolution of the UHECR composition suggested by the Auger data.

In the case of the mixed-composition ``MC-4EeV model'', the adopted parameters are identical to those used in \citealt{Allard:2012}, where $E_{\max,i} = Z_{i} \times 4$~EeV and the spectral index required to reproduce the observed spectra is relatively hard, namely $x = 1.4$. The global abundance of heavy nuclei at the sources must be larger than in the MC-high model, to avoid an observable drop in the overall spectrum above $E_{\max,\mathrm{p}}$. However, the relative abundances of the heavy nuclei ($Z \ge 2$) at the source are assumed to be the same. This model was shown to reproduce relatively well the composition trend drawn from the $X_{\max}$ observations made by Auger.

An intermediate case is also considered, namely the ``MC-15EeV model'', in which the maximum energy of the protons is set equal to 15~EeV, so that $E_{\max,i} = Z_{i}\times 15$~EeV. While in the MC-4EeV model the maximum proton energy is such that the abundances of C, N, O and the other intermediate nuclei is very low above 50~EeV, these elements are still present with a significant abundance above 50~EeV in the source composition of the MC-15EeV model.

Finally, we explore two ``extreme'',  less realistic, but instructive models. The ``pure-p model'' corresponds to a scenario in which only protons are assumed to be accelerated at the source. This is considered for reference as an extreme case of light UHECR composition. The corresponding spectral index is $x = 2.6$, and the maximum energy is taken as $E_{\max} = 10^{20.5}$~EeV (or larger). From the point of view of the propagated composition at the highest energies, say above 50 EeV, the pure-p model is very similar to the MC-high model, which is dominated by protons at all observable energies. The general anisotropy features obtained in both cases are thus also very similar. Therefore, we do not show separately the results corresponding to the less realistic pure-p model in this paper.

\begin{figure*}[ht!]
\centering
\includegraphics[trim=0cm 0cm 0.5cm 0cm,clip,width=0.31\textwidth]{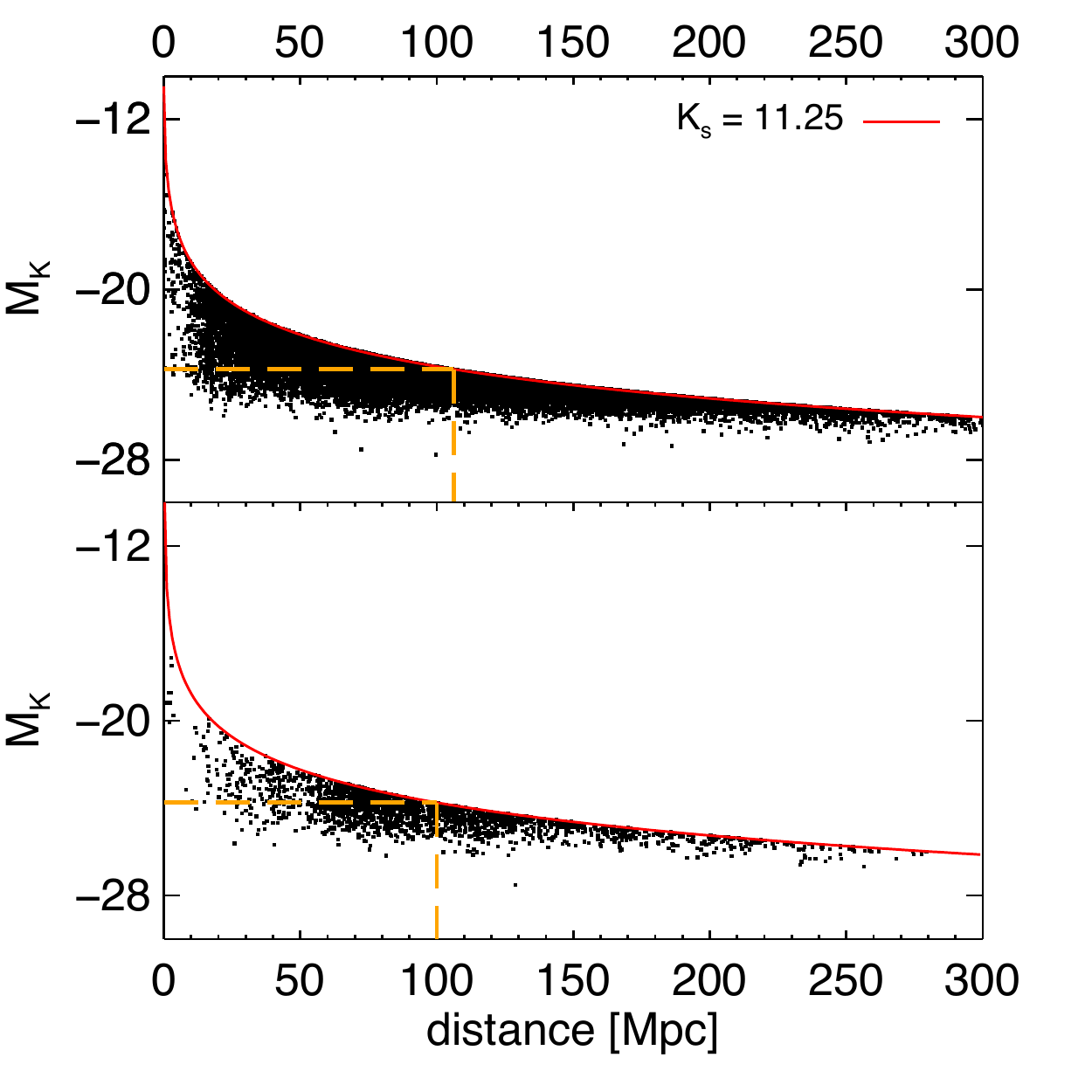}
\includegraphics[trim=0.25cm 0cm 0.25cm 0cm,clip,width=0.31\textwidth]{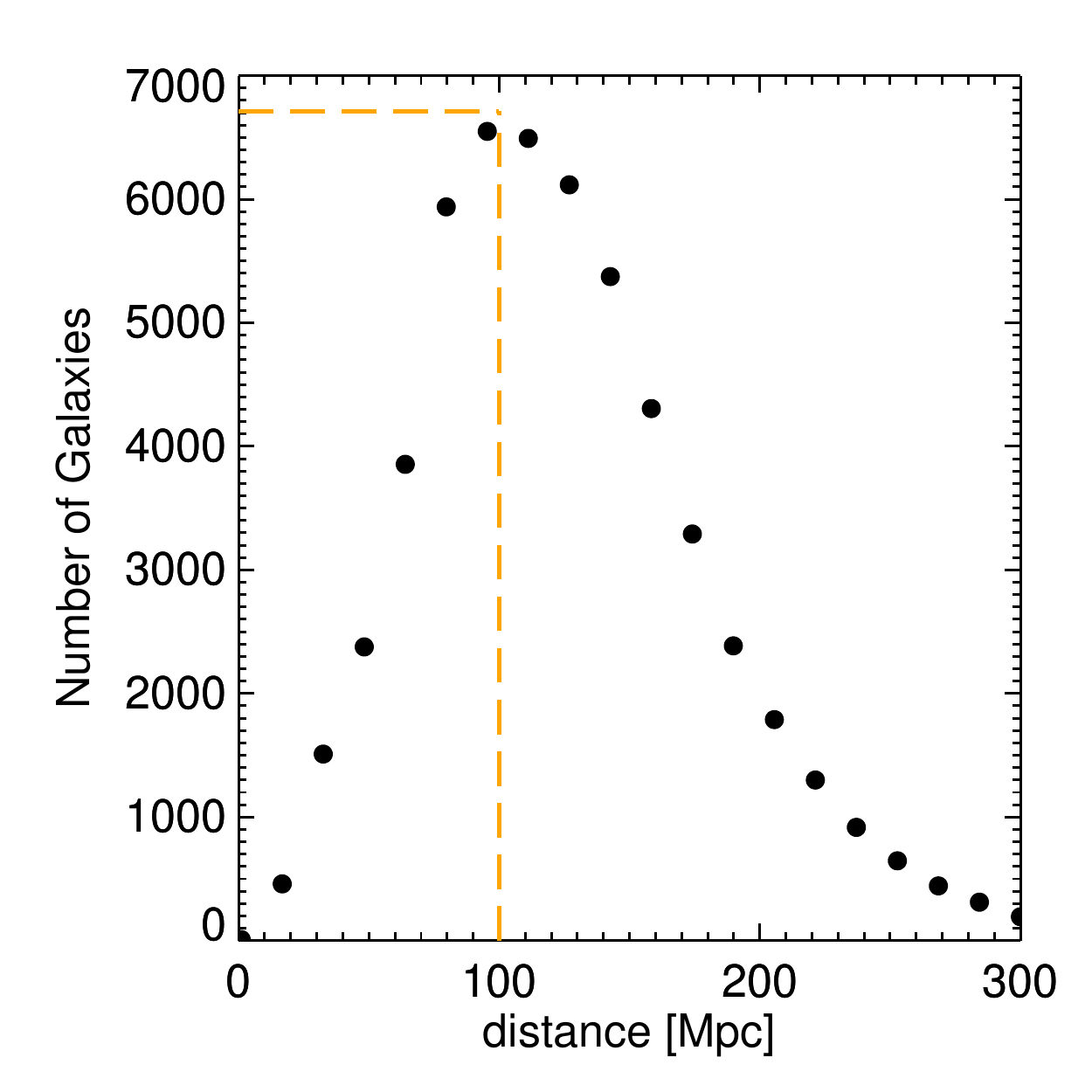}
\includegraphics[trim=0.25cm 0cm 0.25cm 0cm,clip,width=0.31\textwidth]{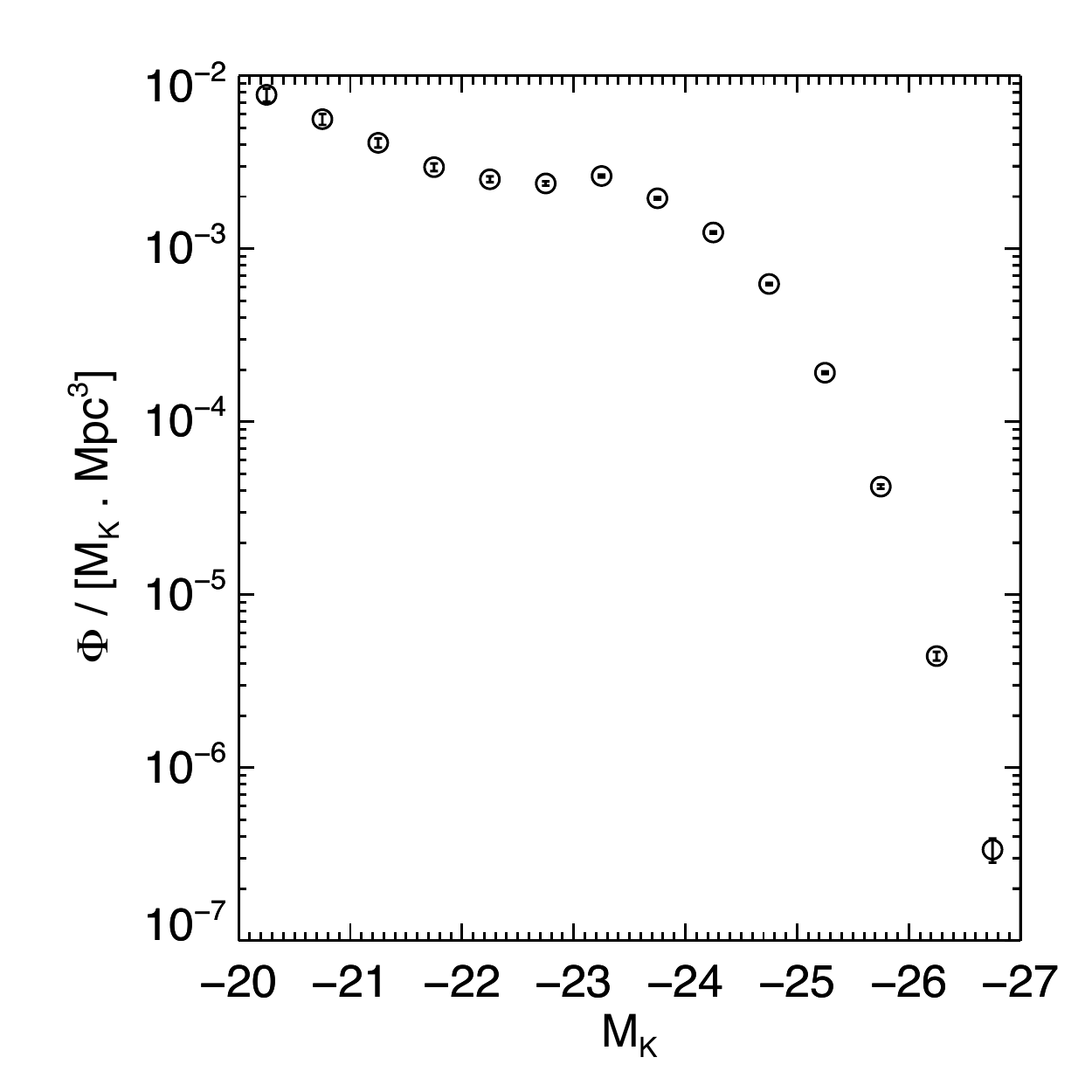}
\caption{Left: Absolute magnitude vs distance of galaxies with $K_s \leqslant 11.25$~mag. The top panel shows the galaxies in the original sample whereas the bottom panel shows the galaxies generated in the direction of the Galactic plane. The $K_s = 11.25$~mag limit is given by the red dashed line. To prevent a bias toward the faint galaxies at small distances, we use a volume-selected sample with maximum distance 100~Mpc corresponding to minimum absolute magnitude $M_K = -23.75$ (orange dashed lines). Middle: Size of a volume-limited sample as a function of the maximum distance. Right: Luminosity function of the original sample, used to produce the additional sample with the same luminosity distribution.}
\label{fig:2MRSSample}
\end{figure*}

In the ``pure-Fe model'', only Fe nuclei are assumed to be accelerated at the source. Although this is not a realistic scenario from the astrophysical point of view, it is used for reference as an extreme case of heavy-composition models. The observed UHECR spectrum can be well reproduced within the pure-Fe scenario (e.g. \citealt{Allard:2008}) using a source spectral index $x = 2.3$ and a maximum energy for Fe nuclei of $E_{\max} = 26 \times 10^{20.5}$~eV (or above), so that secondary protons, produced by photo-dissociation during propagation, can have energies up to $\sim 10^{20.5}$~eV. This model is characterized by a composition at the highest energies that is dominated by heavy nuclei ($A > 40$), but with the presence of a significant fraction (25-30\%) of secondary protons, which are much less deflected by Galactic magnetic fields. This feature has interesting consequences for the anisotropy patterns, similar to what would be obtained from a composite astrophysical scenario in which a few sources provide a subdominant component of protons at high energy, in addition to a dominant heavy component. Therefore, this scenario is also explore to provide some hint of what a more complex, but probably more realistic scenario would imply (see the discussion in Sect.~\ref{sec:pureFe}).


\subsection{Source distribution and density}
\label{sec:sourceDistrib}

The anisotropy patterns in the UHECR sky maps also depend on the distribution of the sources in the nearby universe and on their spatial density.

In the absence of any clear indication about the nature of the sources, the most natural choice is to assume that they are distributed in a similar way as ordinary matter. The distribution of matter is known to be non uniform in the nearby universe. In our simulations, the UHECR source distribution in tridimensional space (direction and distance) is derived from the distribution of galaxies in the 2MASS Redshift Survey catalog (2MRS, \citealt{Huchra:2012}). More specifically, we use the initial survey with cuts in the near infrared magnitude $K_\mathrm{s} \leqslant 11.25$~mag and in Galactic latitudes $|b| \ge 10\degr$. This catalog, which contains 20\,860 galaxies, is linked to the Extragalactic Distance Database (EDD, \citealt{Tully:2009}) to obtain the distance of nearby galaxies ($\sim$ 10--20~Mpc/h), for which the peculiar velocities dominate over the cosmic expansion, while the Hubble law in the linear regime is used to estimate the distances of more distant galaxies. Each galaxy in the resulting catalog is represented by a black dot in the distance-luminosity plane in Fig.~\ref{fig:2MRSSample} (left, top panel).

On the y-axis, $M_{\mathrm{K}}$ represents the absolute (intrinsic) magnitude in the K-band (near infrared). The red-line corresponds to the luminosity cut at $K_{\mathrm{s}}\le 11.25$~mag.

\begin{figure*}[t!]
\centering
\includegraphics[trim=0.25cm 0.5cm 0.25cm 0.5cm,clip,width=\textwidth]{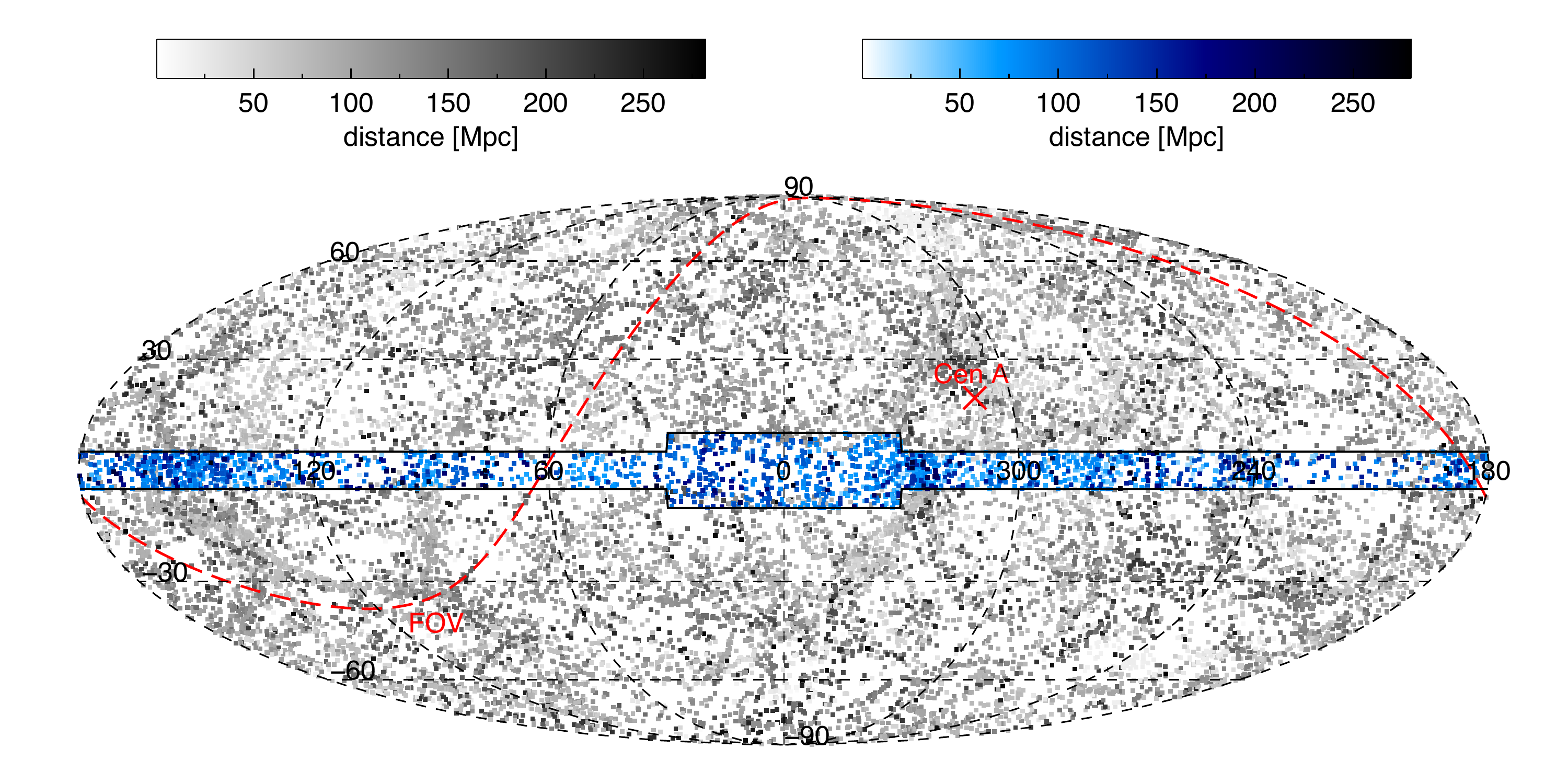}
\caption{Galaxies in the 2MRS catalog displayed in a Mollweide projection in Galactic coordinates. The black dots are galaxies in the original sample and the blue dots are the randomly generated galaxies in the Galactic plane. The solid line indicates the region that was populated. The red dashed line indicates the border of the field of view of the Pierre Auger Observatory. A red cross marks the direction of the Centaurus A radio-galaxy.}
\label{fig:2MRSSky}
\end{figure*}

To compensate for the missing sources in the Galactic plane, we follow the filling method described in \citealt{Crook:2007}, which consists in randomly populating the original sample to enhance the galaxy number density behind the Galactic plane, in a way that reflects the density observed just below and above it. As for the intrinsic luminosity of the additional galaxies, it is drawn according to the luminosity distribution function of the 2MRS sample. This luminosity function is derived from the catalog itself, and is shown in Fig.~\ref{fig:2MRSSample}, on the right. This procedure generates an additional set of $2\,094$ galaxies and ensures the continuity of the structures across the plane. These additional galaxies are shown in the lower left panel on Fig.~\ref{fig:2MRSSample}. The complete catalog is displayed in Fig.~\ref{fig:2MRSSky}.

A magnitude-limited survey is affected by radial-selection effects, since at each distance $D$, only galaxies brighter than an intrinsic magnitude $M_{0}(D)$ can be detected with apparent luminosity $K_{\mathrm{s}}$. The curve $M_{0}(D)$ corresponding to $K_\mathrm{s} \leqslant 11.25$~mag is represented by the above-mentioned red line in Fig.~\ref{fig:2MRSSample}. This curve can be used to build a volume-limited sample from the magnitude-limited survey. For instance, the dashed orange lines on the plot show that within a distance $D = 100$~Mpc (vertical line), all galaxies with intrinsic luminosity larger than absolute magnitude $M_K = -23.75$ (horizontal line) are present in the sample. Thus, the galaxies in the lower-left rectangle delimited by these lines make a complete sample of galaxies up to distance $D = 100$~Mpc (restricted to bright-enough galaxies). It contains 6\,720 galaxies, 1267 of which have been added by the above completing procedure.

Similarly, volume-limited samples can be obtained by selecting the galaxies in the lower-left corner of rectangles built in the same way on Fig.~\ref{fig:2MRSSample} (left) for different distances, with a corresponding cut at larger or lower luminosity. The resulting numbers of galaxies are shown in the middle plot of Fig.~\ref{fig:2MRSSample}, as a function of maximum distance. As can be seen, the maximum number of galaxies in the various volume-limited samples is obtained for a cut at 100~Mpc, which by coincidence turns out to correspond to the so-called GZK sphere, related to the indicative horizon of 100~Mpc corresponding to $\sim 90$~EeV protons or Fe nuclei.

In our simulations, we use this particular sample of galaxies as the \emph{seed catalog} within which the UHECR sources are randomly chosen, with different values of the source density, $n_{\mathrm{s}}$. As already indicated, the complete set of galaxies contains 6\,720 galaxies, which corresponds to a source density $n_\mathrm{s} = 1.6 \times 10^{-3}$~Mpc$^{-3}$. To explore a UHECR scenario with a source density of $10^{-5}$~Mpc$^{-3}$ (respectively $10^{-4}$~Mpc$^{-3}$), we thus simply select randomly 1 galaxy out of 160 (respectively 1 out of 16) in the catalog.

Note that we do not need to assume that the actual UHECR sources are necessarily among the 2MRS galaxies. All the anisotropy analyses that we perform are investigations of the intrinsic anisotropy of the simulated data. So the actual position of the sources in the sky is not relevant. Only the relative positions and the global angular/distance distribution is important. We thus simply assume that the overall distribution of the sources is similar to that of the galaxies.


\section{UHECR propagation and sky maps}
\label{sec:skymaps}


\subsection{General procedure}

Our main goal is to simulate realistic UHECR sky maps and to quantify their intrinsic anisotropies. For this, we compute the propagation of the UHECRs from their sources to the Earth and determine their energy, nuclear type and arrival direction taking into account the relevant energy loss processes, the possible change of nuclear species and the deflections in the extragalactic and Galactic magnetic fields.

We use the Monte-Carlo code presented in \cite{Allard:2005}, to compute the energy losses and photo-dissociation processes in the extragalactic photon backgrounds (\citealp[see][]{Allard:2008,Decerprit:2012b} for a more detailed description). We also compute the 3D geometrical trajectories as influenced by the magnetic fields using the fast integration method described in detail in \citet{Globus:2008}, where a comparison with a full numerical integration is given. This allows us to keep track of the time-dependence (i.e. redshift-dependence) of the energy losses without having to assume rectilinear transport.

The propagation is treated numerically in two separate steps. In the first step, we propagate the UHECR protons and nuclei in the extragalactic medium, following them in energy, mass, and geometrical spaces from their emission at a given source located at a distance D (and Galactic coordinates $l$ and $b$), at a redshift/time $z$. This provides us with the propagated flux of the UHECRs injected by the whole set of sources as they enter our Galaxy, characterized by their energy and mass distribution as well as the apparent arrival directions.

The second step takes into account the deflections by the Galactic magnetic fields and relates the UHECR arrival directions on a fictitious sphere representing the boundary of the Galaxy to the observed arrival directions on Earth. This is done by inverting the relation between the different directions in the sky and the directions of the particles at the entrance of the Galaxy, as derived from the back-propagation of negatively-charged nuclei in the Galactic magnetic field, as explained below. The resulting ``transfer function'' of the Galaxy can then be applied to the extragalactic UHECR sky map to produce the desired sky map on Earth.

Finally, we analyze the anisotropy of the simulated data set by searching for significant excesses in the angular 2-point correlation function. In the next subsections, we describe in more detail the ingredients of the procedure, and we give the results in the next section.


\subsection{Propagation in the extragalactic magnetic field}
\label{sec:EGMF}

The extragalactic magnetic field (EGMF) is poorly known, and its spatial distribution, intensity, coherence length, time evolution and origin are uncertain. Observations imply the presence of $\mu$G fields in the core of large galaxy clusters. However, the spatial extension of these large field regions and their volume filling factor in the universe are difficult to evaluate. Efforts have been made to model local magnetic fields using simulations of large-scale structure formation that include an MHD treatment of the magnetic field evolution (see the pioneering studies by \citealp{Dolag:2002, Sigl:2004}; or more recent calculations by \citealp{Das:2008, Ryu:2008, Ryu:2010, Donnert:2009}). Some of these simulations are constrained by the local density/velocity field to provide more realistic field configurations in the local universe. These simulations rely on different assumptions regarding the origin of the fields and the mechanisms involved in their growth. They are ultimately normalized to the values observed at the present epoch in the central regions of galaxy clusters (see the discussion in \citealt{Kotera:2008}). The outcome of the different simulations strongly differ. In particular, the volume filling factors for strong fields (above 1~nG, say) vary by several orders of magnitude from one simulation to the other. In contrast, an interesting simple alternative to complex hydrodynamical simulations, offering more freedom to test different models of magnetic field evolution with local density, has been proposed by \citet{Kotera:2008}.

In view of the above-mentioned uncertainties, we use a simplified approach, assuming that the universe is filled with a purely turbulent, homogeneous magnetic field. Admittedly, such a topology of the EGMF is not realistic, but since our main purpose is to study the effect on the UHECR sky maps of the magnetic deflection in the extragalactic space, we simply investigate the impact of a magnetic blurring upstream of the Galaxy for different typical values of these deflections. The smallest impact corresponds to no magnetic field, while the largest impact would be obtained for large EGMF intensities of the order of a few nG. Large localized magnetic fields, on the other hand, could be important if the volume filling factor is not too small, but this would mostly result in the apparition of fake secondary sources at the location of the magnetic cores, and thus produce a similar phenomenology, with only a larger apparent source density \citep{Kotera:2008}. Therefore, we simply simulate a uniform turbulent EGMF, with a method following that described in \citealt{Giacalone:1999}). We assume a Kolmogorov-like turbulence with a maximum scale $\lambda_{\max}=1 \rm Mpc$ and different magnetic field variance, $\sqrt{\langle \rm B^2 \rangle}$, ranging from 0.1 to 3~nG. 


\subsection{The Galactic magnetic field model}
\label{sec:GMF}

To simulate the transport of the UHECRs in the Galaxy, we implement a representative model of the Galactic magnetic field (GMF). We follow the modeling of \citet{Jansson:2012a,Jansson:2012b} who consider three types of magnetic structures: i) a coherent field with spatial scales on the order of a few kpc, ii) an isotropic turbulent field with spatial scales on the order of tens of pc, and iii) a {\it striated} field, which refers to an anisotropic turbulent field whose orientation is aligned with the large scale coherent field, but whose strength and sign vary on a small scale. 

\begin{figure}[t!]
\centerline{\includegraphics[trim=0cm 0cm 0cm 0.75cm,clip,width=7cm]{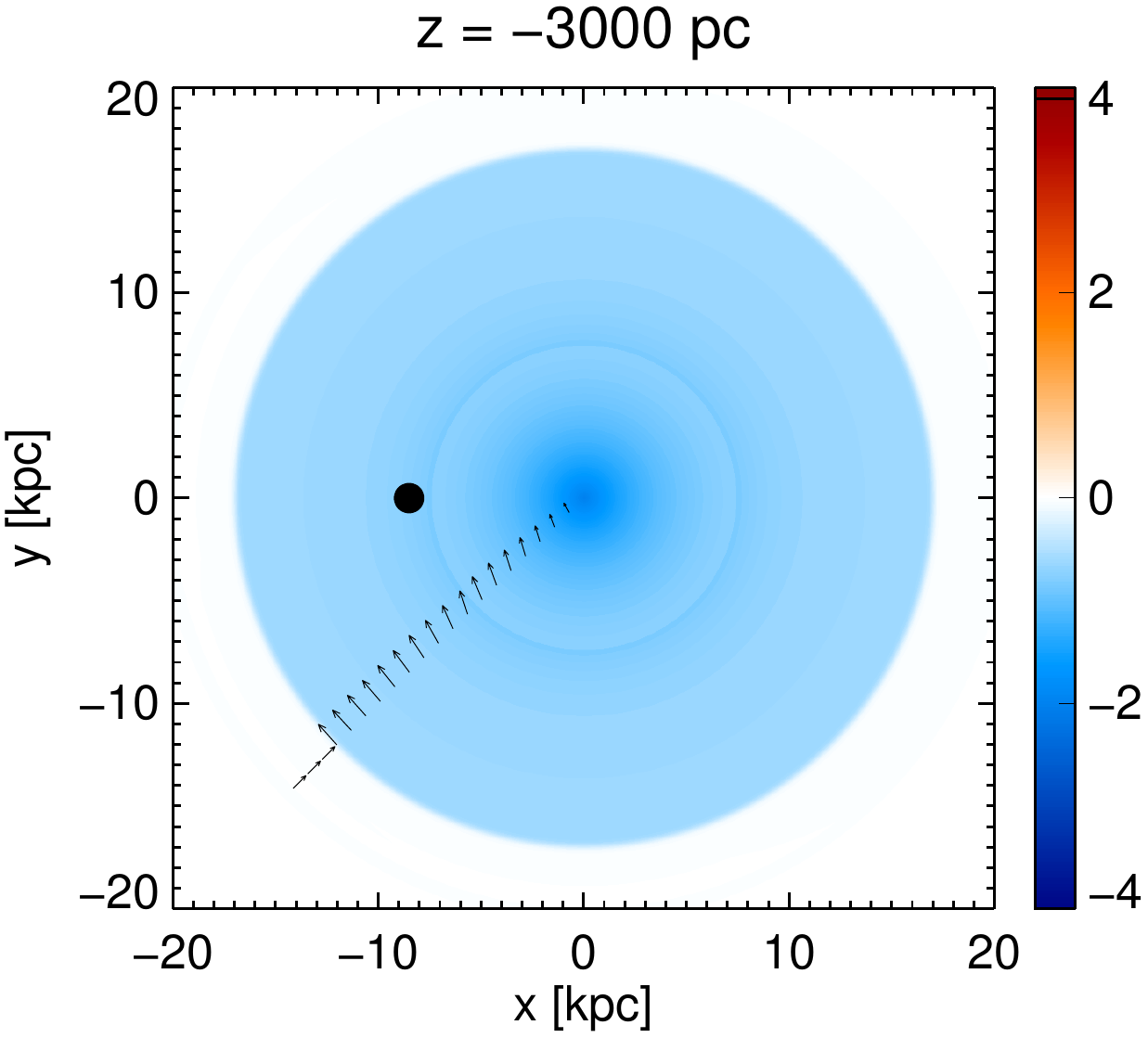}}
\bigskip
\centerline{\includegraphics[trim=0cm 0cm 0cm 0.75cm,clip,width=7cm]{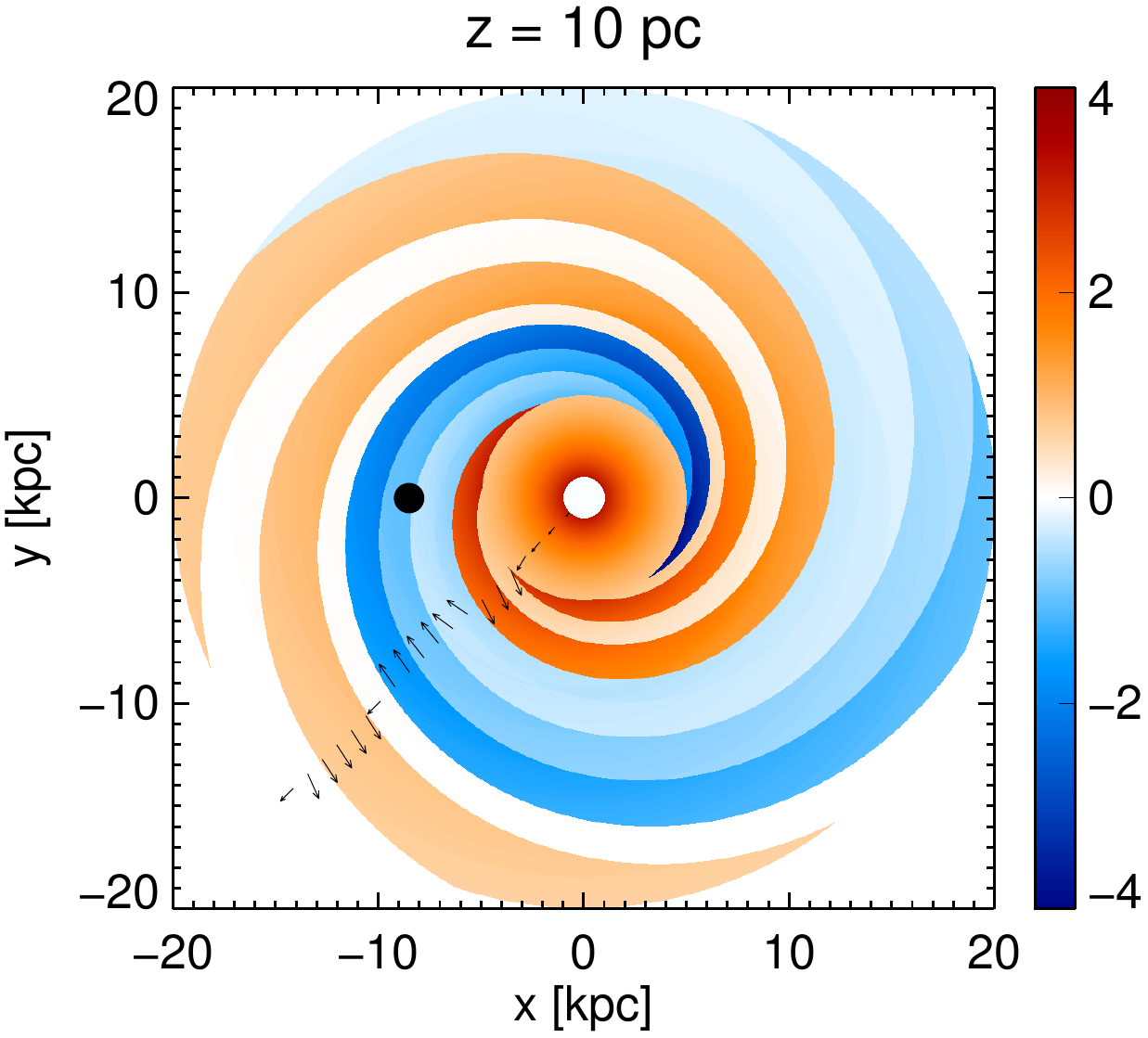}}
\bigskip
\centerline{\includegraphics[trim=0cm 0cm 0cm 0.75cm,clip,width=7cm]{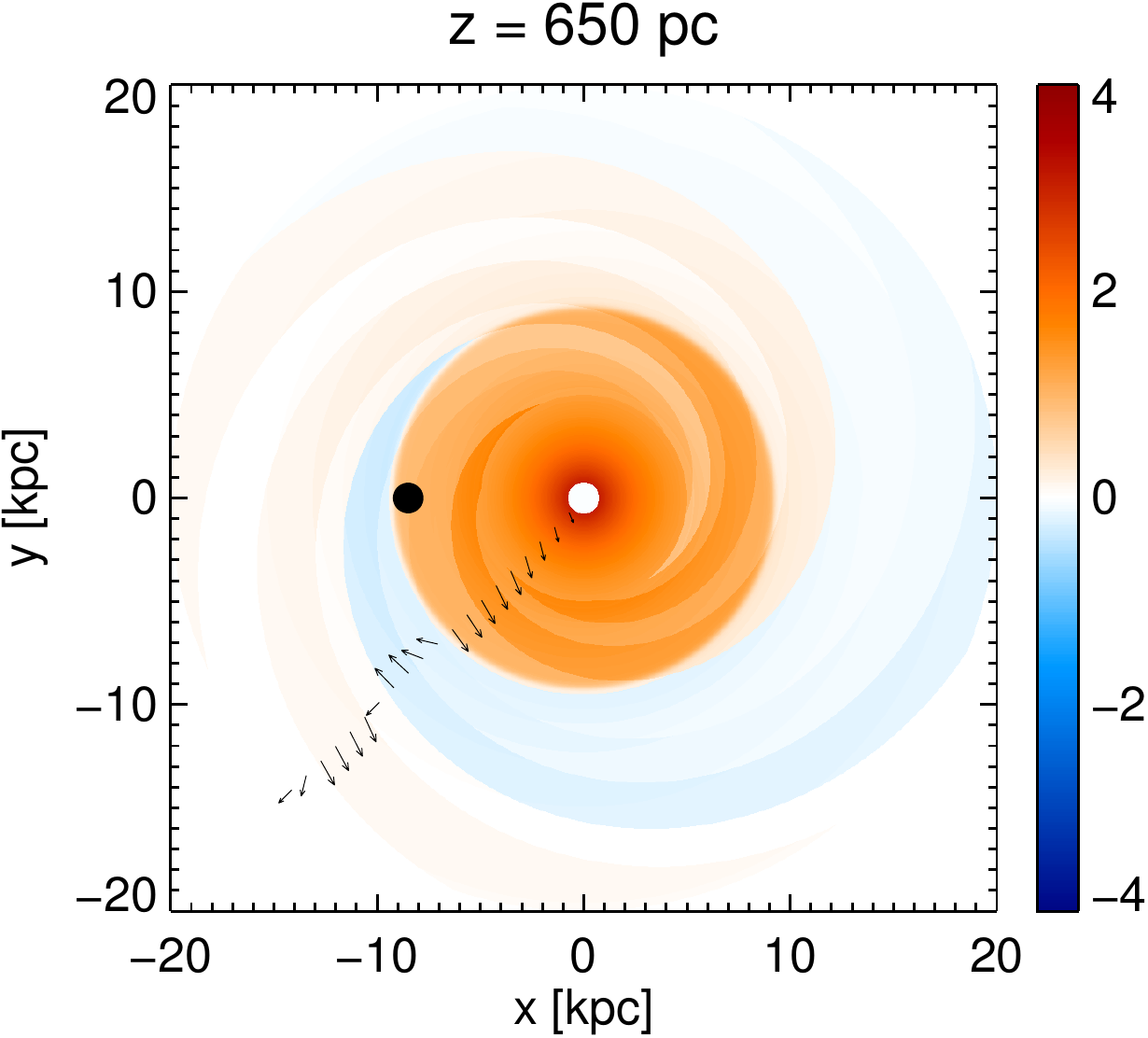}}
\caption{The regular component of the GMF model in the x-y plane (parallel to the Galactic plane) as seen from above. The slices are located at vertical height z = -3~kpc (top), z = 10~pc (middle) and z = 1~kpc (bottom). The Sun, represented by a black dot, is located at (-8.5,0,0)~kpc. The black arrows give the direction of the the field along the x = y line. The magnitude of the field is color coded and takes negative values for negative values of the azimuthal component of the field.}
\label{fig:GMFModel}
\end{figure}

The large scale coherent field is modeled as the superposition of three separate components: a disk component, an halo component and an out-of-plane halo component. The disk component originates from the \citet{Brown:2007} model where the magnetic field is concentrated in the plane and closely follows the spiral arms of the Galaxy. Several large scale reversals of the magnetic field occur along the Galactic radius. The disk field is symmetric with respect to the Galactic plane and transitions smoothly to a strictly azimuthal (toroidal) halo field at low vertical extent. This halo field decreases exponentially with scale height and takes different amplitudes below and above the plane. Finally, the out-of-plane halo component is inclined with respect to the Galactic plane, with a constant inclination far from the Galactic center and an almost perpendicular orientation closer to the Galactic axis. This so-called X-field is motivated by observations of X-shaped field structures in external galaxies \citep{Krause:2006,Krause:2007,Beck:2009}. The large scale regular field model of \citet{Jansson:2012a,Jansson:2012b} -- sum of the disk field, the toroidal halo field and the X-field -- is illustrated in Fig.~\ref{fig:GMFModel}.

For the random component of the GMF, we follow the modeling procedure of \citet{Giacalone:1999} and assume a Kolmogorov-like turbulence with a coherent length $\lambda_c = 100$~pc. Its intensity is given by the field variance, which we assume follows the magnitude of the regular component of the GMF, with a global enhancement factor of 3. In other words, the turbulent field is typically of rms magnitude $6\,\mu$G where the regular field is $2\,\mu$G. This scaling is assumed isotropic, so that magnetic turbulence associated with this random component is isotropic and spatially homogeneous on small scales.

The last component of the GMF model is the striated field, which is included after the recent works of \citet{Jaffe:2010,Jaffe:2011}. Here, the field is either parallel or antiparallel to the regular field with a coherence length of 100~pc, and its magnitude follows the magnitude of the regular component, according to \citet{Jansson:2012a,Jansson:2012b}. 

\subsection{Particle deflections in the Galactic magnetic field}
\label{sec:particleDeflections}

To build the UHECR sky maps, we need to connect the arrival direction of the particles into the Galaxy, as resulting from the extragalactic propagation, with the direction in which they are eventually observed on Earth. This is done in a statistical way applying the following procedure.


\subsubsection{Trajectories and global deflections}

First, we back-propagate a very large number of protons away from the Earth, until they reach a sphere of radius 50~kpc centered on the Galactic center, loosely considered as the ``boundary'' of the Galaxy, beyond which the influence of the GMF is negligible. More specifically, we propagate antiprotons with fixed energies between $10^{17.5}$~eV and $10^{20.5}$~eV, by steps of $\Delta \log(E/[{\rm eV}]) = 0.1$, starting from the Earth in different directions. For each energies, the starting directions are regularly distributed over the celestial sphere using an HEALPix grid \citep{Gorski:2005} with resolution parameter N$_{\rm side}$ = 1024, which corresponds to 12,582,912 directions, or a pixel size of $\sim 3.5$~arcmin. The spatial transport of the particles is then computed by simply integrating the equation of the trajectory governed by the Lorentz force (\citealp[see][]{Harari:1999} for a discussion on cosmic ray propagation in the GMF).

\begin{figure}[t!]
\centering
\includegraphics[trim=0.75cm 0cm 0.25cm 0cm,clip,width=\columnwidth]{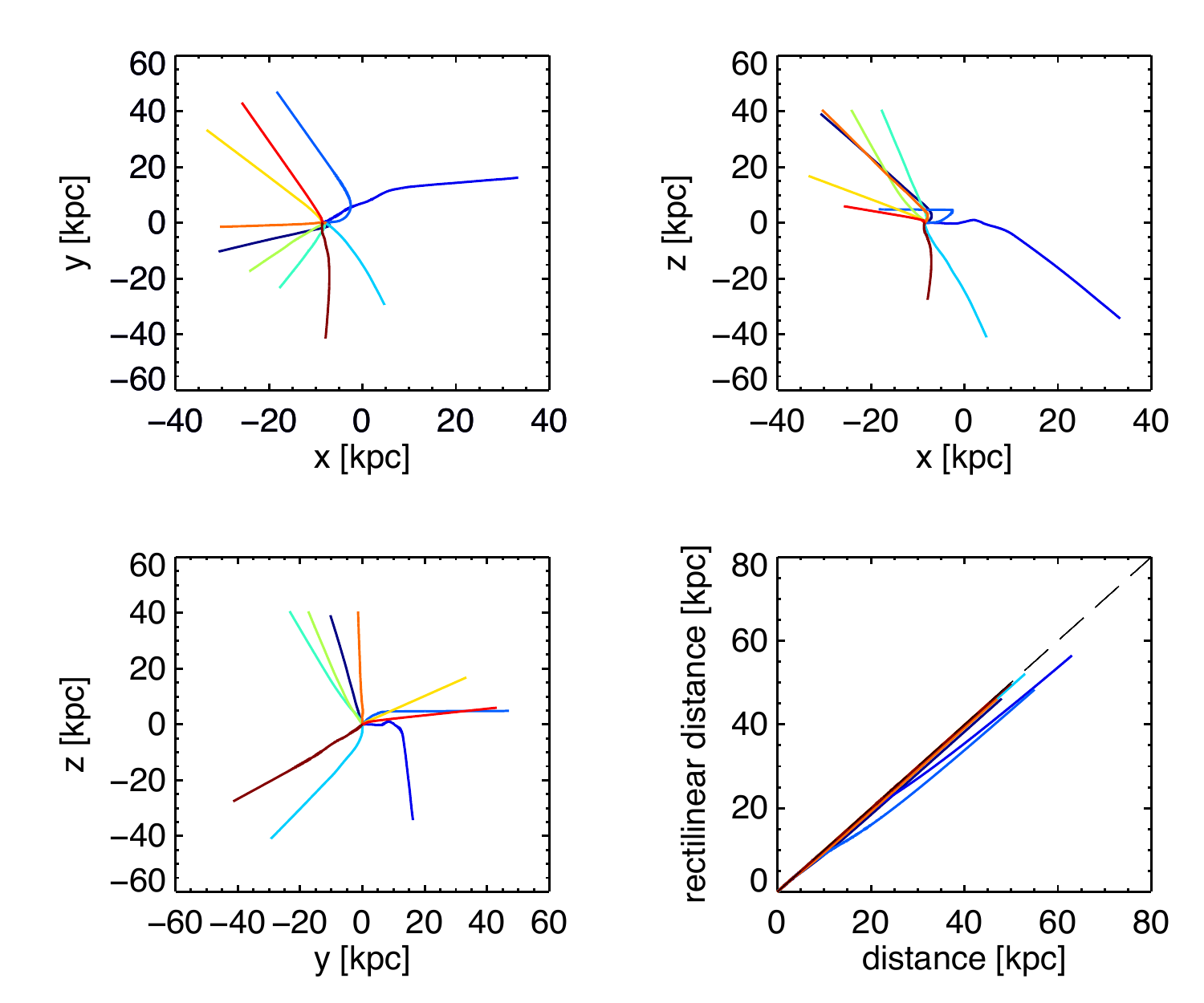}
\caption{Top and lower left panels: projection of 10 trajectories of 5~EeV protons in the x-y, x-z and y-z plane. Lower right panel: rectilinear distance as a function of the curvilinear distance for the same trajectories.}
\label{fig:trajectoryGMF}
\end{figure}

Figure~\ref{fig:trajectoryGMF} shows a set of ten trajectories of back-propagated 5~EeV (anti)protons bended by the GMF. The distance traveled by the particles (along their trajectory, but measured in the Galactic frame) is always larger than the rectilinear distance from Earth, because of the deflections. This is shown on the lower right panel of Fig.~\ref{fig:trajectoryGMF}. However, even when the deflections are relatively large, the difference does not exceed about~10\%. Whatever the starting direction, the (anti)protons are found not to be confined by the GMF, and the residence time inside the Galaxy remains negligible compared to the energy loss time, so we can neglect their interactions with the local photon fields. This justifies that we only consider the Lorentz force when computing the particle transport in the Galaxy and, as a consequence, all UHECRs with the same magnetic rigidity behave in the same way. The trajectories of 5~EeV protons are thus also those of 30~EeV carbon nuclei,  40~EeV oxygen nuclei, 130~EeV iron nuclei, or any nucleus with a 5~EV rigidity.

\begin{figure}[t!]
\centering
\includegraphics[trim=1.5cm 0cm 0.25cm 0cm,clip,width=0.48\linewidth]{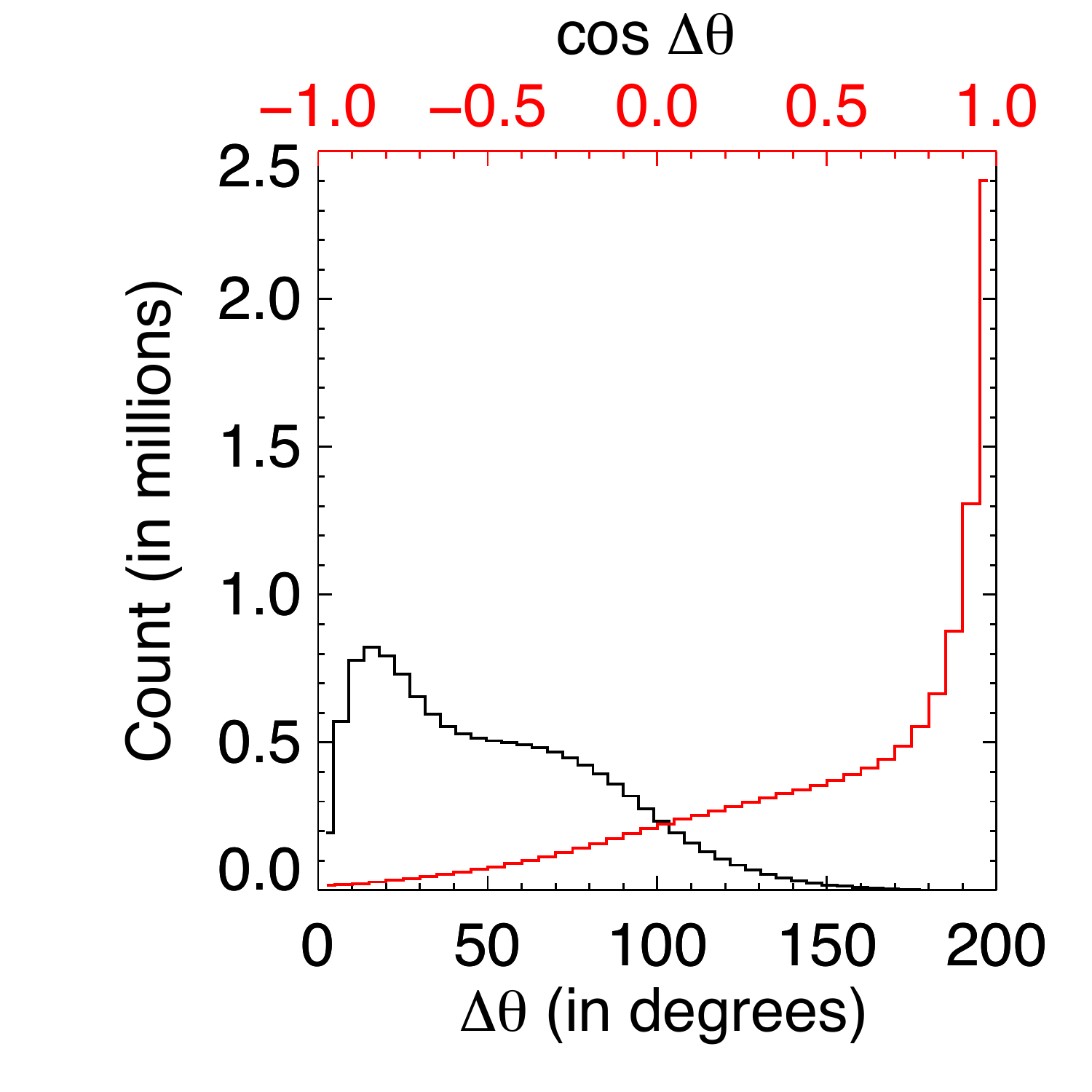}
\includegraphics[trim=1.25cm 0cm 0.5cm 0cm,clip,width=0.48\linewidth]{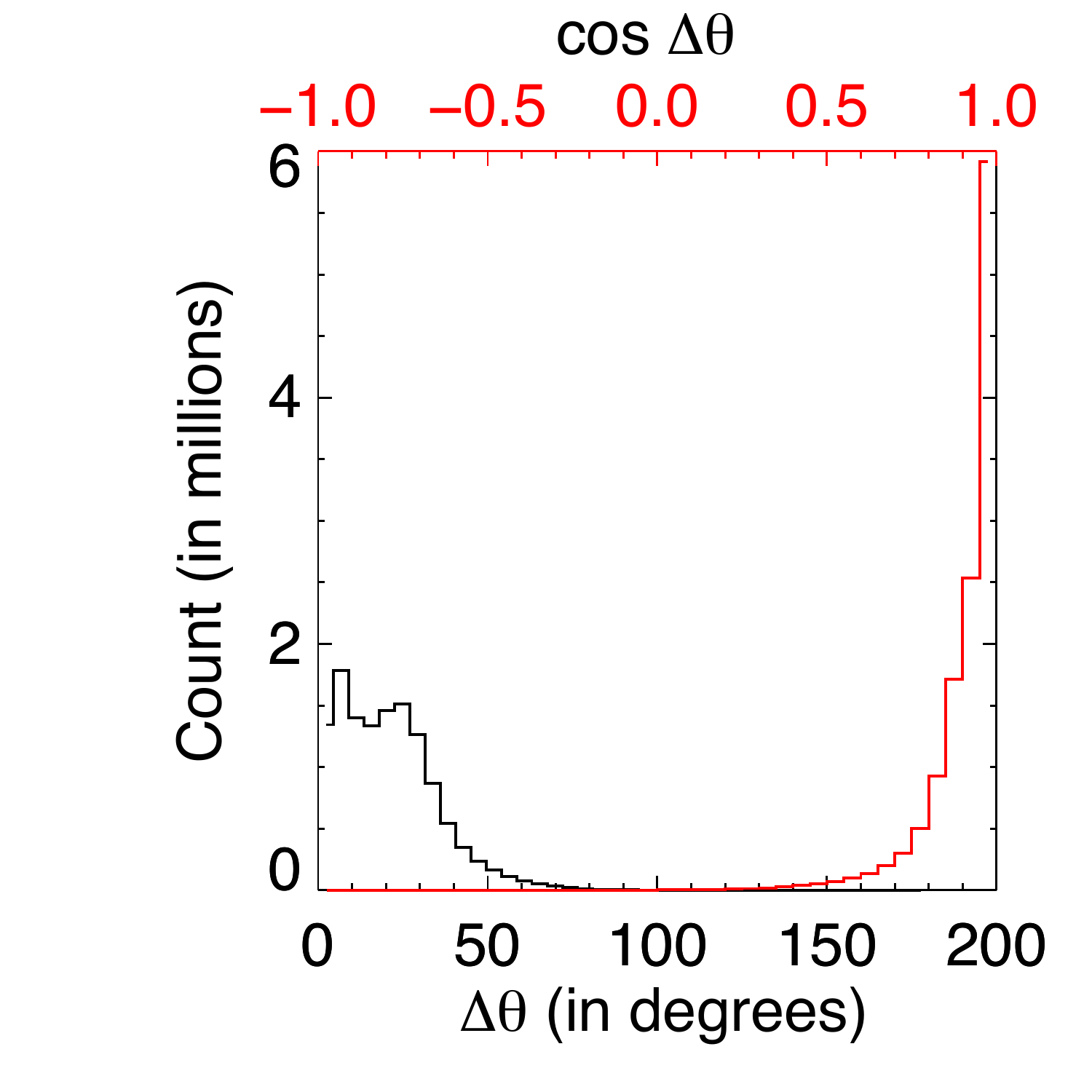}
\includegraphics[trim=1.5cm 0cm 0.25cm 0cm,clip,width=0.48\linewidth]{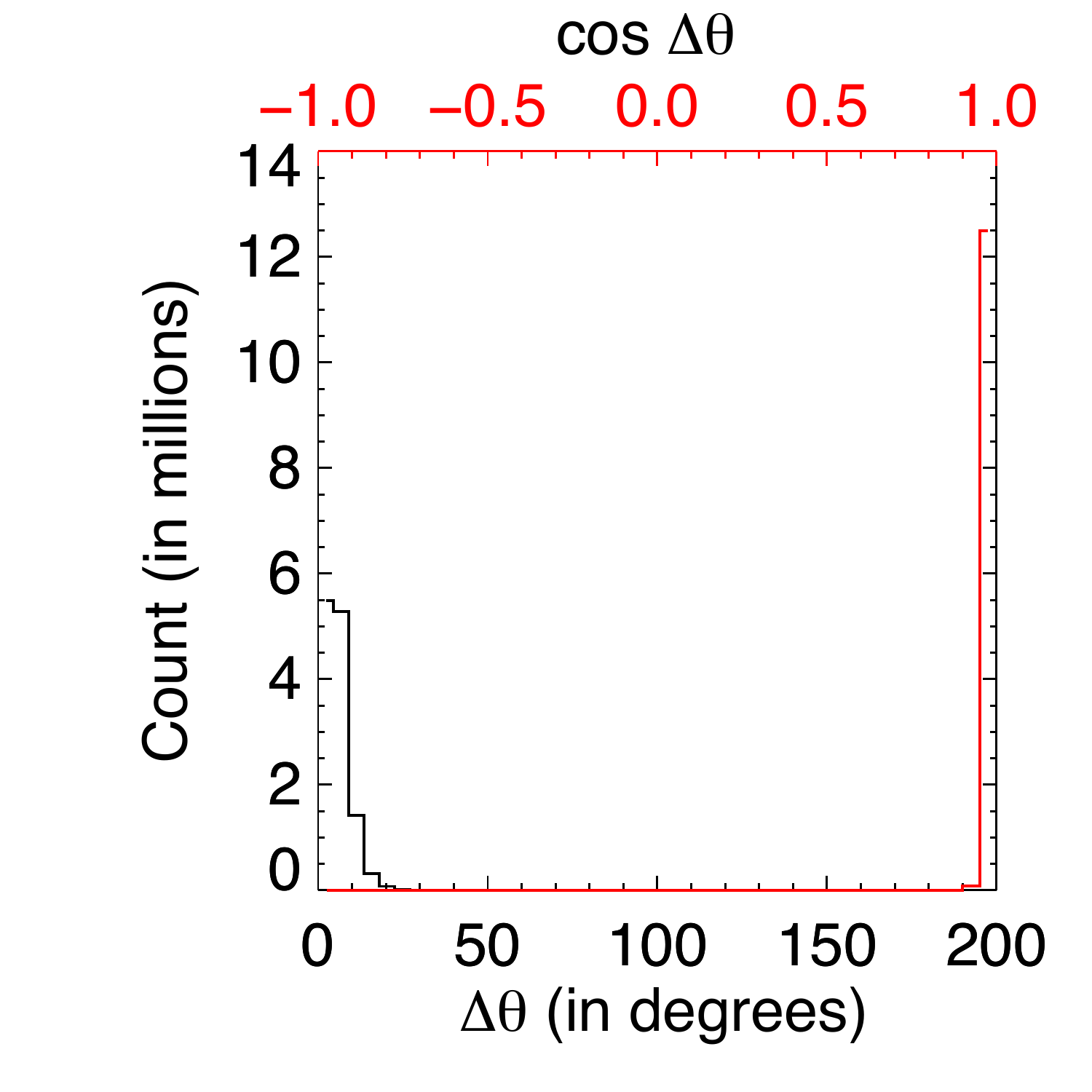}
\includegraphics[trim=1.25cm 0cm 0.5cm 0cm,clip,width=0.48\linewidth]{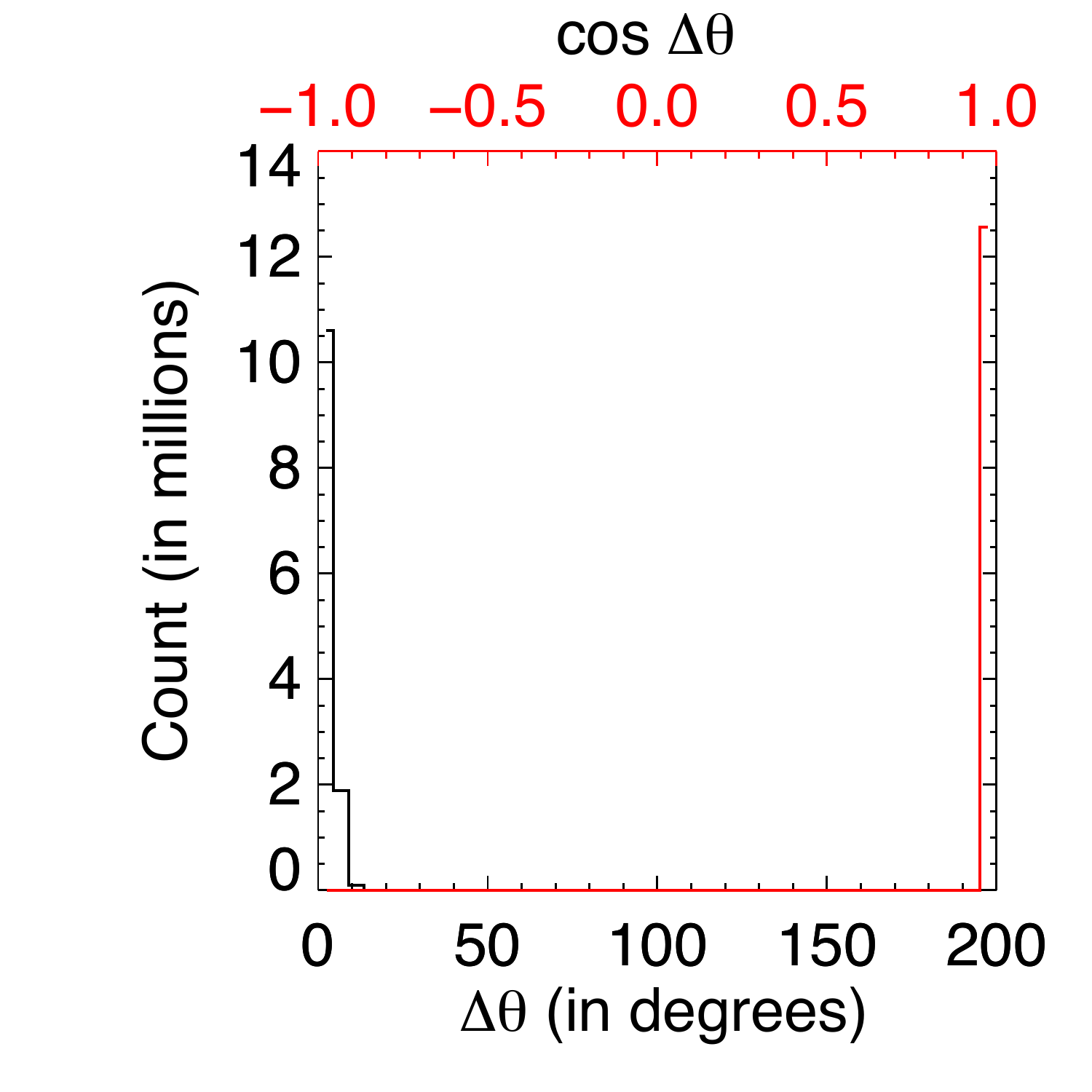}
\caption{Histogram of the deflection angles of the 12\,582\,912 UHECRs back-propagated from evenly distributed directions with a rigidity of 5~EV (top left), 16~EV (top right), 63~EV (bottom left), and 130~EV (bottom right). Both the distributions of the deflection angle (in red) and of its cosine (in blue) are represented.}
\label{fig:deflectionHistogram}
\end{figure}

The above back-propagation gives us, for each rigidity, a one-to-one relation between the $\sim 12.6$ million starting directions on the celestial sphere and the direction in which the corresponding (anti-)UHECR leaves the Galaxy. In Fig.~\ref{fig:deflectionHistogram}, we show the histogram of the resulting angular deflections for all these UHECRs for four different rigidities: $10^{0.7} \simeq 5$~EV (relevant for $\simeq 130$~EeV Fe nuclei), $10^{1.2} \simeq 16$~EV (relevant for $\simeq 95$~EeV C nuclei), $10^{1.8} \simeq 63$~EV and $10^{2.1} \simeq 130$~EV.

\begin{figure}[t!]
\centering
\includegraphics[trim=1cm 0cm 0cm 0cm,clip,width=7cm]{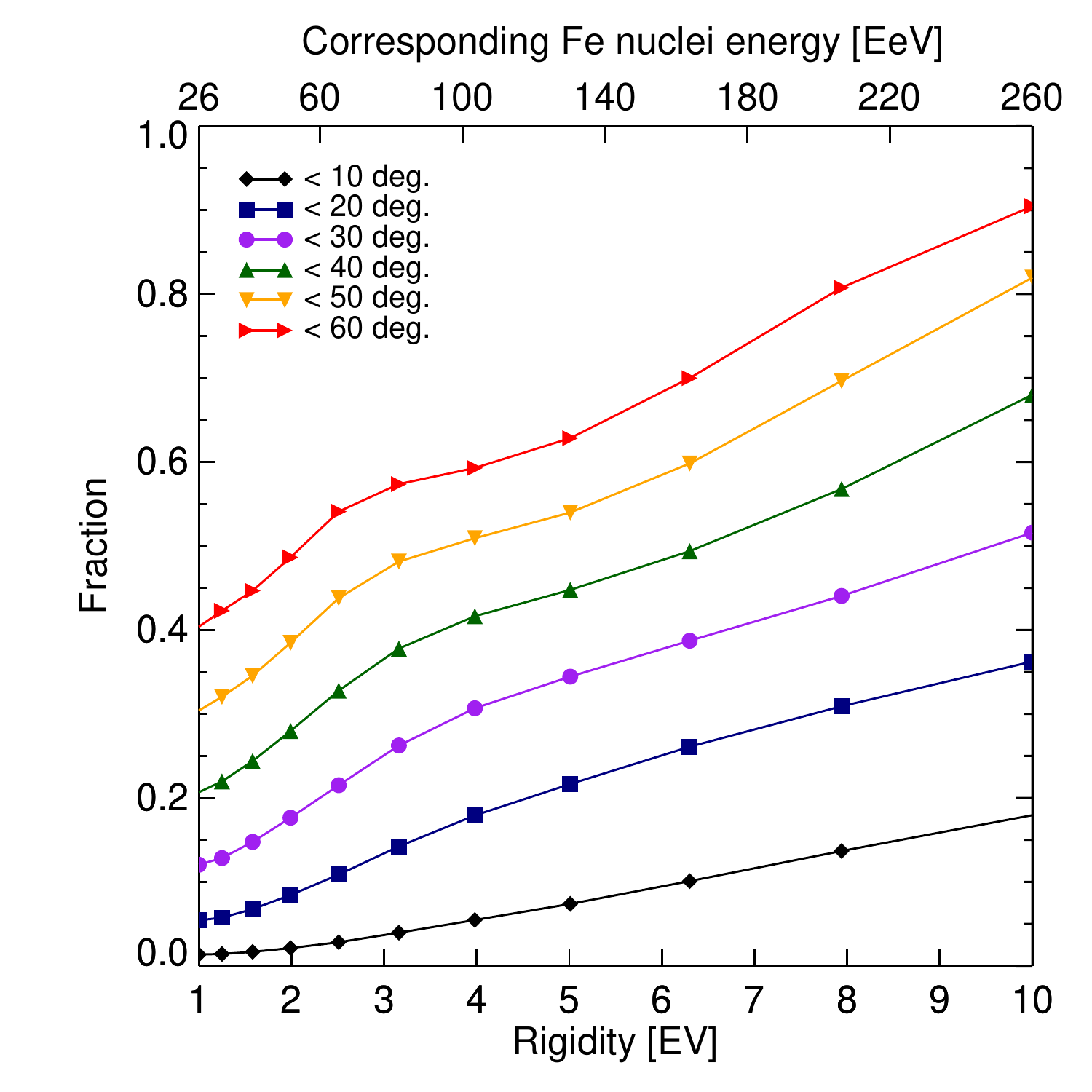}
\caption{Fraction of the UHECRs experiencing deflection angles lower than 10, 20, $\ldots$ 60 degrees (as indicated) across the GMF, as a function of rigidity (lower axis) or corresponding Fe nuclei energy (top axis).}
\label{fig:deflectionFractionIron}
\end{figure}

As expected, the deflections are much larger at low rigidities, and UHECRs with intermediate rigidities experience a wide range of deflections, depending on the arrival directions. The difference between the top left panel and the bottom right panel illustrates the difference between a proton primary and an iron nucleus primary at the highest energies. Obviously, direct pointing astronomy seems inaccessible if the UHECRs are dominated by Fe nuclei. However, we recall here that direct source pointing should not be the only goal of UHECR astronomy, and the study of anisotropy patterns can provide important information about the UHECR sources. In addition, a small fraction of protons (or low rigidity nuclei) may lead to the apparition of small angular scale multiplets if the statistics is large enough to allow the detection of a few of them. Moreover, even in the most unfavorable case where all UHECRs above $\sim 80$~EeV are Fe nuclei, a significant fraction of them are found to experience deflections smaller than the typical angular separation between sources. To illustrate this point in a more quantitative way, we show on Fig.~\ref{fig:deflectionFractionIron} the fraction of arrival directions corresponding to UHECR deflections smaller than 10, 20, 30, 40, 50 and 60 degrees, as a function of rigidity (translated into Fe nuclei energy on the top x-axis). Finally, we note that some knowledge of the regular component of the GMF can in principle be used to correct for the non-random part of the UHECR deflections.

\begin{figure}[ht!]
\centering
\includegraphics[trim=1cm 0cm 0cm 0cm,clip,width=7cm]{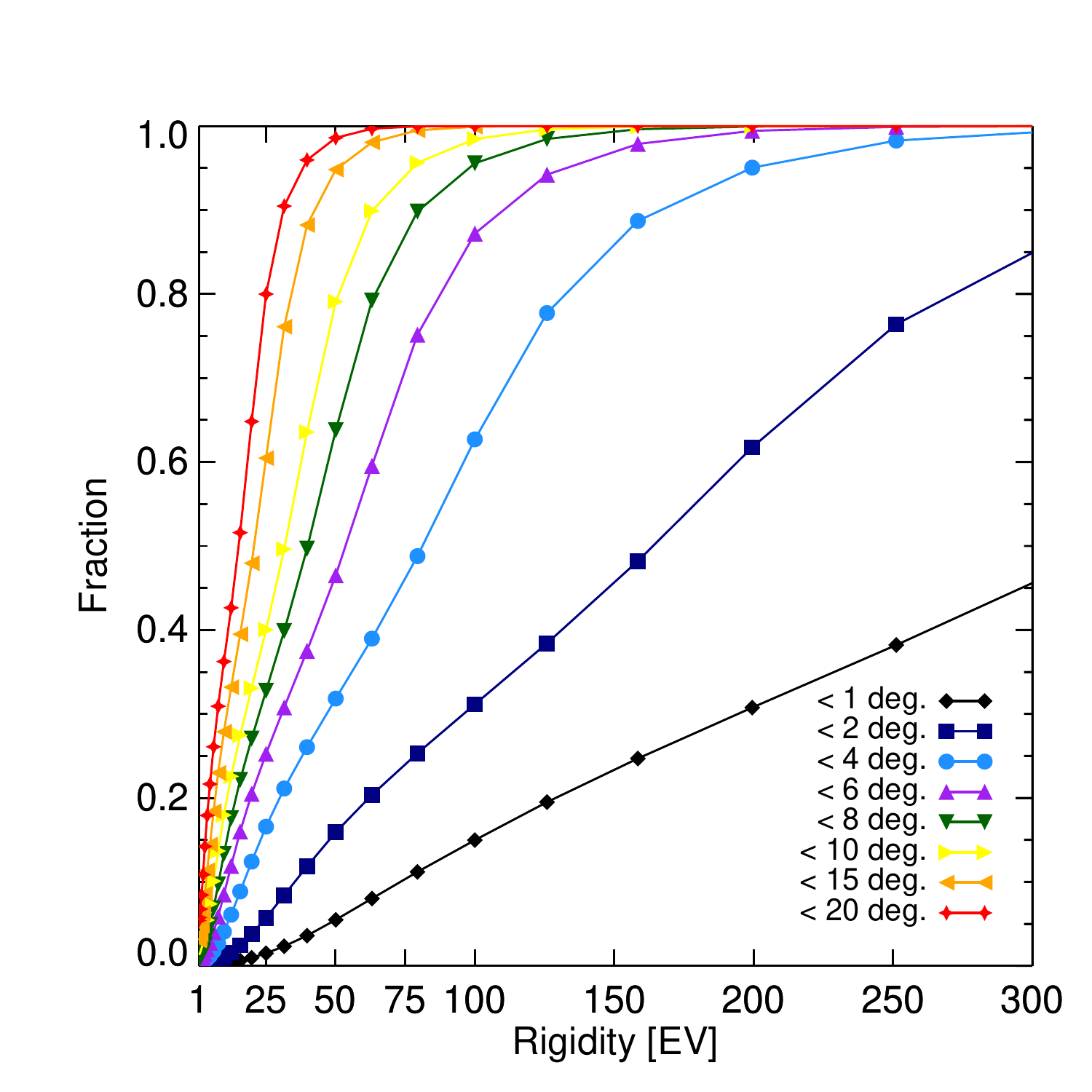}
\caption{Fraction of the UHECRs experiencing deflection angles lower than 1, 2, $\ldots$ 20  degrees (as indicated) across the GMF, as a function of rigidity (lower axis), with obvious conversion into proton energy.}
\label{fig:deflectionFractionProton}
\end{figure}

For comparison, we show in Fig.~\ref{fig:deflectionFractionProton} the deflection fractions of protons with energies between 1~and 300~EeV. At 50~EeV, 20\% of the protons are deflected by more than 10 degrees. This fraction drops to less 3\% above 100~EeV. 

While the distributions illustrated by Figs.~\ref{fig:deflectionHistogram},~\ref{fig:deflectionFractionIron} and~\ref{fig:deflectionFractionProton} mix all the arrival directions together, more information about the UHECR deflection patterns across the GMF can be obtained by looking separately at different pixels in the sky map.


\subsubsection{Backward and forward deflection maps}
\label{sec:deflectionMaps}

Starting from the above one-to-one relation between the starting directions of the back-propagated particles and their directions out of the Galaxy, we then use a coarser sampling of the celestial sphere, choosing a HEALPix resolution parameter $N_{\mathrm{side}} = 64$. This defines 49,152 pixels of slightly less than $1\deg^{2}$ evenly distributed over the sky, each of which contains 256 of the original directions on the fainter grid. Thus each direction on the sky (with a resolution of $\sim 1\deg$) is now linked with 256 directions at the boundary of the Galaxy, which are in effect the arrival directions of UHECRs with the rigidity under consideration that would be observed on Earth in that direction (with the assumed GMF).

\begin{figure}[hp!]
\centering
\includegraphics[width=\columnwidth]{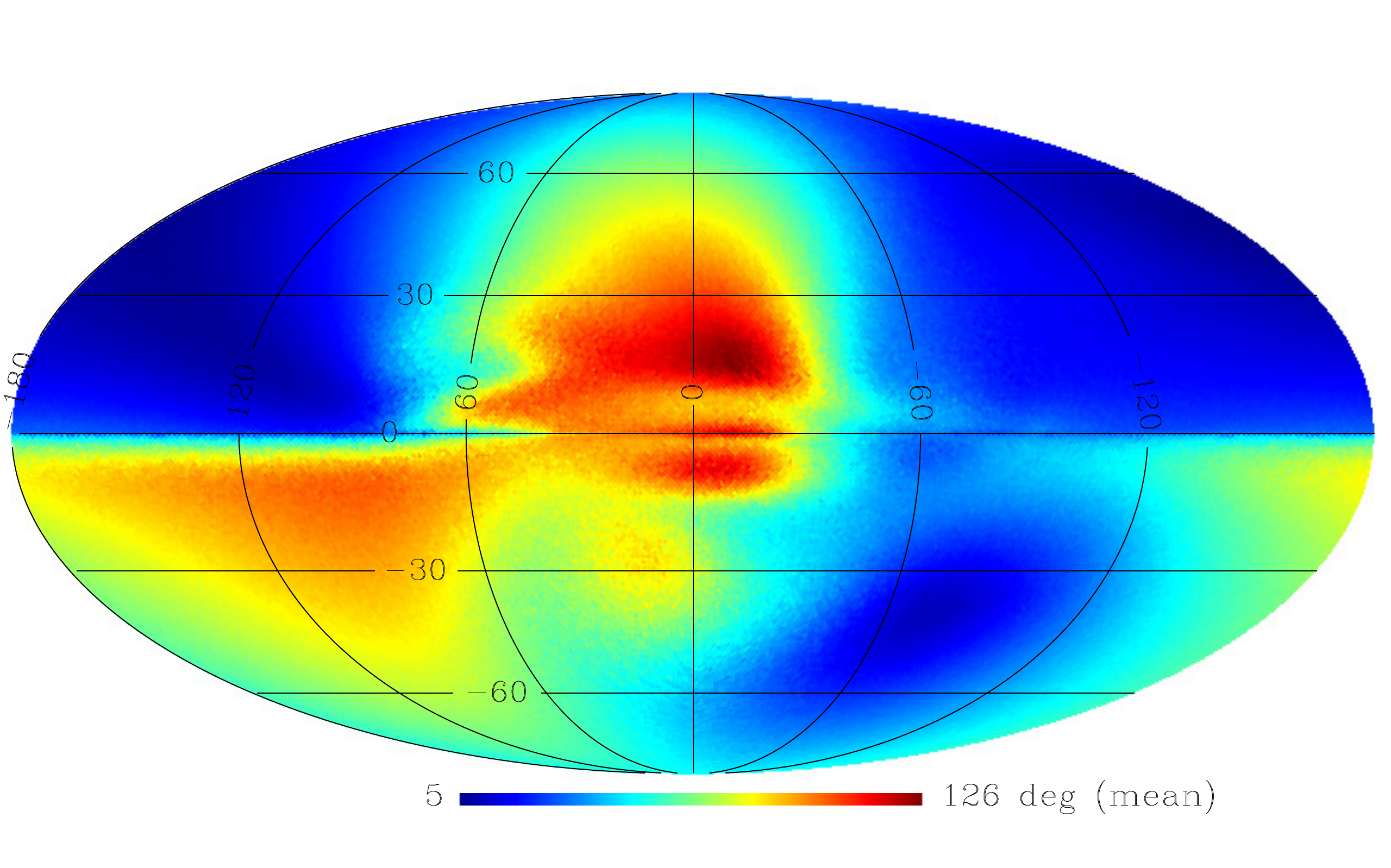}
\includegraphics[width=\columnwidth]{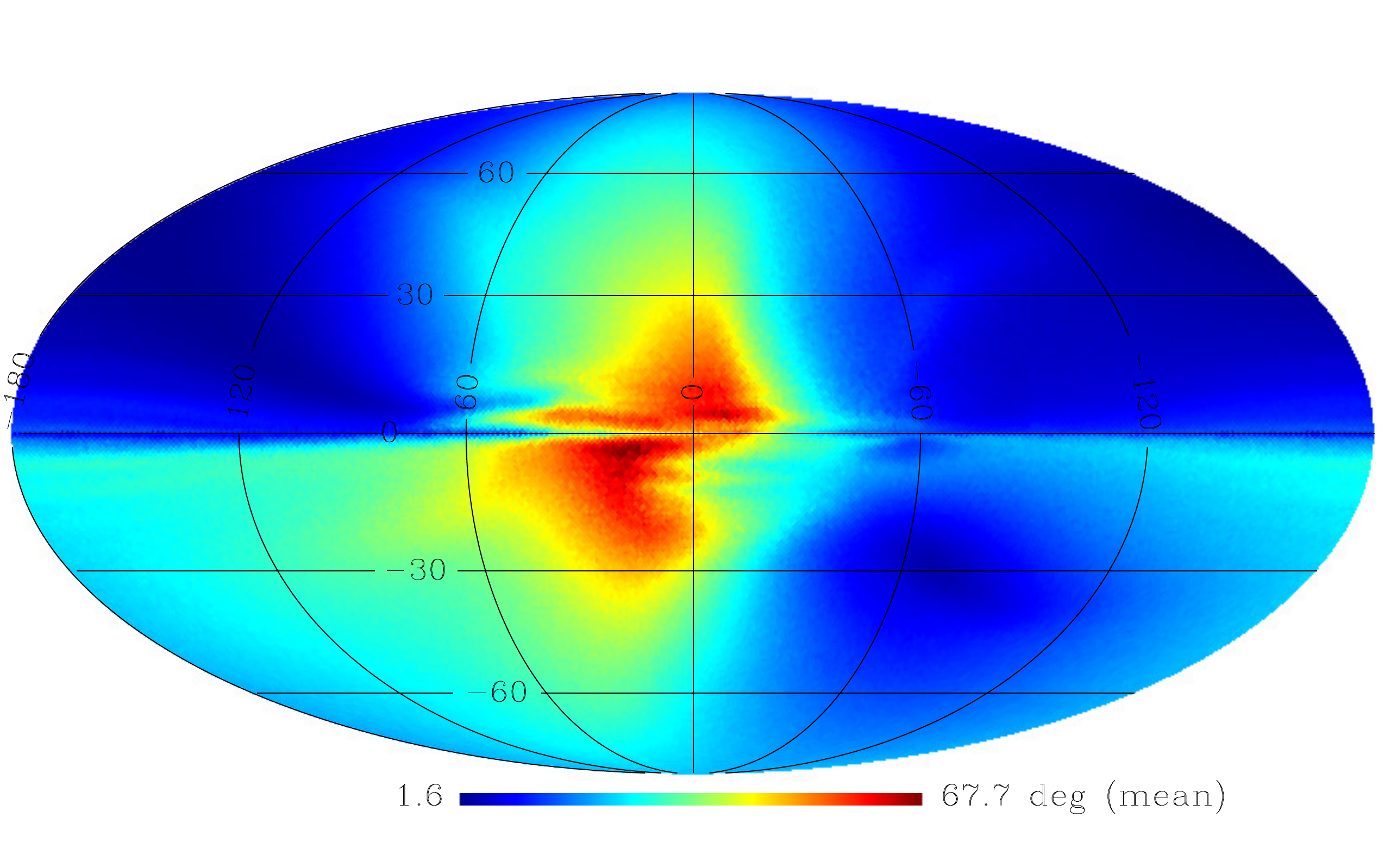}
\includegraphics[width=\columnwidth]{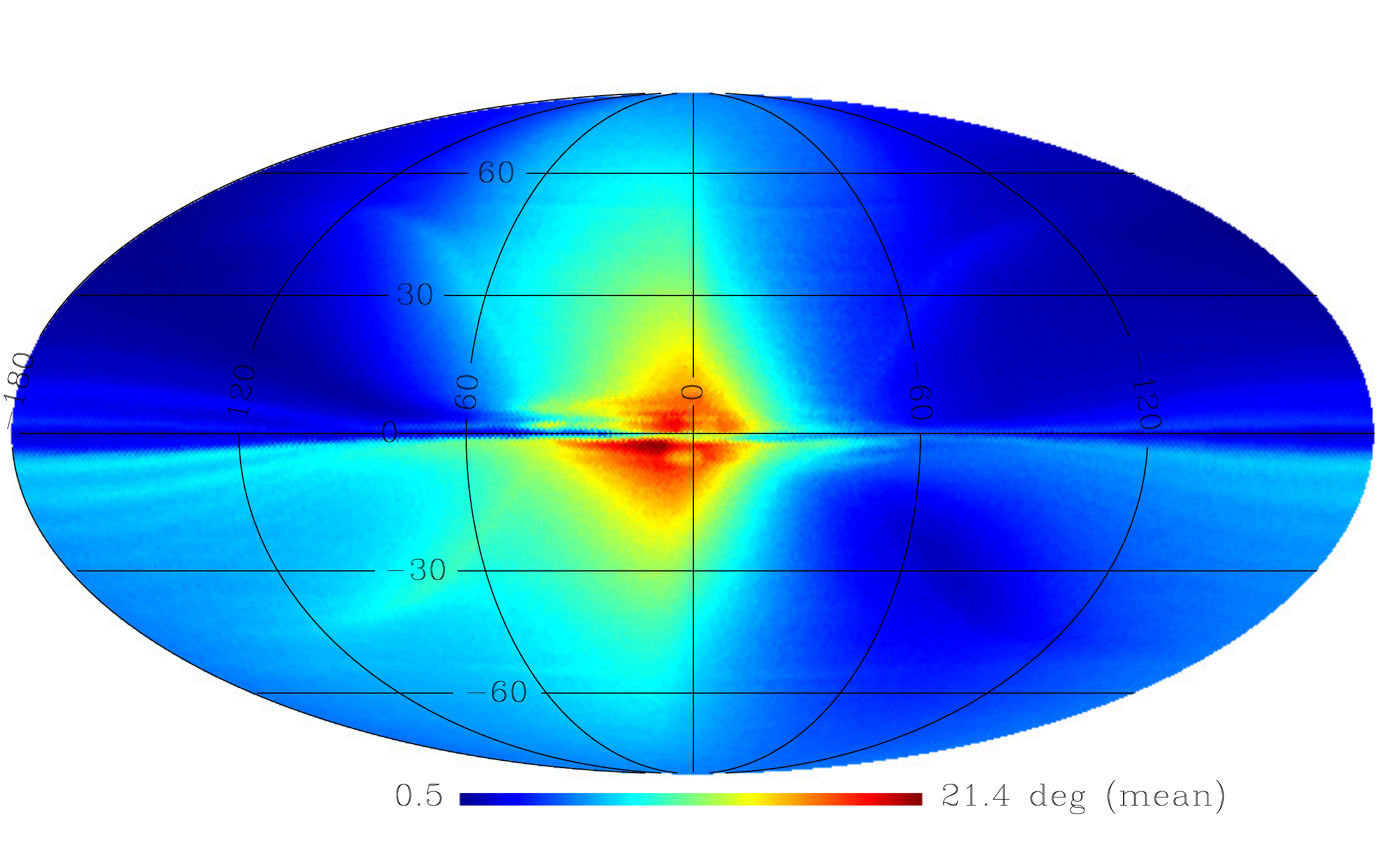}
\includegraphics[width=\columnwidth]{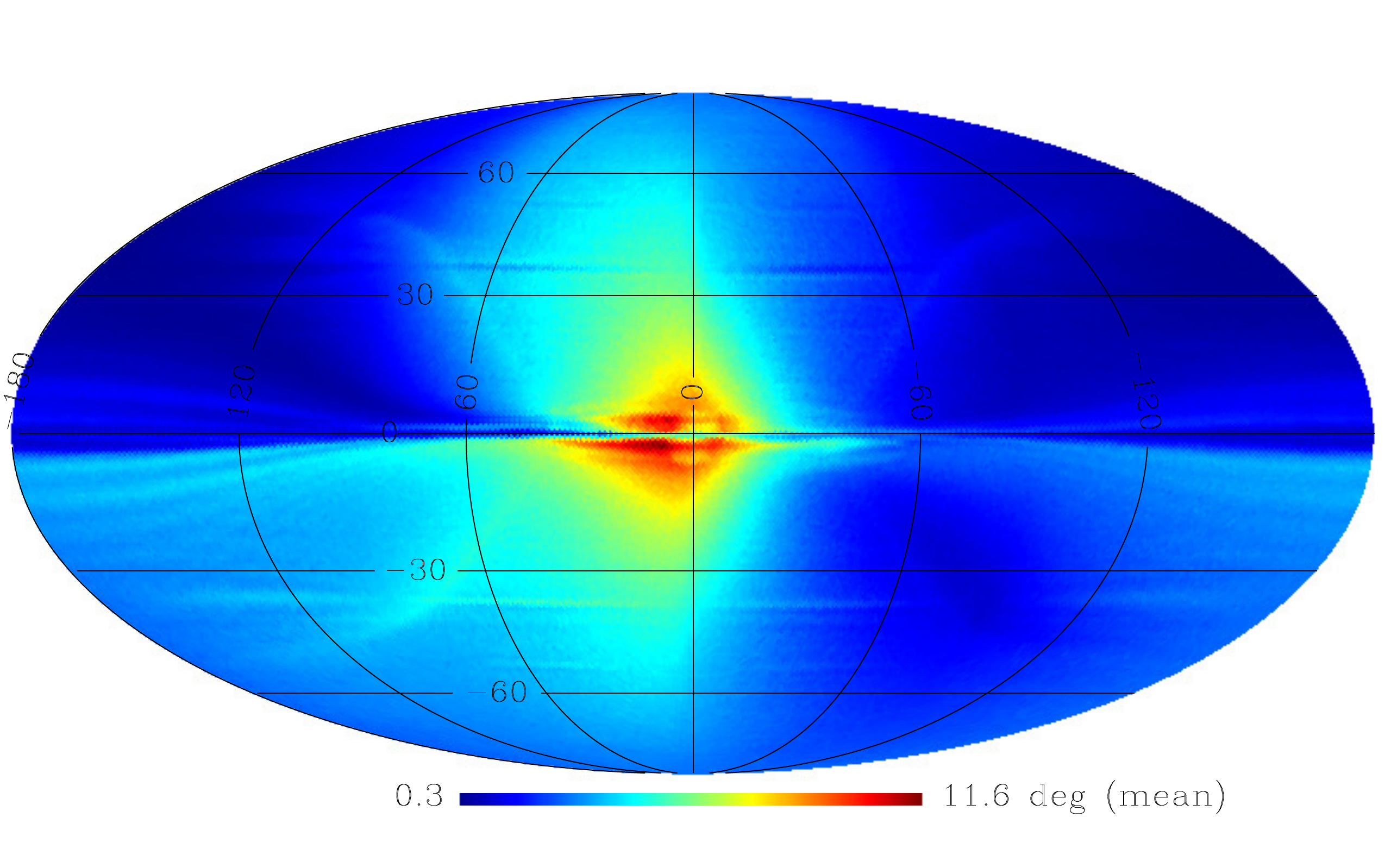}
\caption{Backward deflection maps, displayed in Mollweide projection in Galactic coordinates, showing the mean deflection angle between the 256 incoming directions at the entrance of the Galaxy and the observed direction, for the same rigidity as in Fig.~\ref{fig:deflectionHistogram}: 5~EV, 16~EV, 63~EV, and 130~EV (from top to bottom).}
\label{fig:deflectionMapBackwardMean}
\end{figure}

For each pixel in the coarse grid (observed direction), we computed the average angular deflection, i.e. the mean of the 256 angles between the incoming directions at the entrance of the Galaxy and the observed direction. The result is shown in Fig.~\ref{fig:deflectionMapBackwardMean}, where we plot the map of these mean deflections in color code for the same four rigidities as above. Note that, although the patterns show similar shapes, associated with the structure of the regular component of the GMF, the color code spans different ranges in each map, to follow the global reduction of the deflection with increasing rigidity.

It is interesting to note that the range of deflections is generally quite large over the celestial sphere. UHECRs observed in some directions are on average much less deflected than in others. Particles observed in a large circle of $\sim 15^{\circ}$ radius around the Galactic center are notably much more likely to have been deflected by a large amount than particles observed towards anti-center longitudes, especially in the Northern (Galactic) hemisphere. This strong contrast between observing directions is partly reflected in the global distribution of the angular deflections of Fig.~\ref{fig:deflectionHistogram}, where the top panels show two wide, but distinct peaks. But most importantly, it can in principle be exploited to perform refined anisotropy analyses attributing different weights to different regions, based on some prior knowledge of the relative deflection amplitude. This is not attempted here.

The above deflection maps may however be misleading, since they give information about the average deflection of the UHECRs \emph{observed} in different directions, but not about the UHECRs coming from sources \emph{located} in these directions. For the same reason, the results of the back-propagation of charged particle cannot be exploited to build simulated sky maps until an inversion is done to relate the arrival directions of cosmic rays at the entrance of the Galaxy (from their particular extragalactic sources) to the actual directions in which they are observed on Earth. We may call a ``Galactic pixel'' a pixel in the sky map where a back-propagated particle goes out of the Galaxy, or where a forward-propagated particle enters the Galaxy. Likewise, an ``Earth pixel'' is a pixel in the sky map where a back-propagated particle starts its trajectory, or where a forward-propagated particle is eventually observed.

\begin{figure}[hp!]
\centering
\includegraphics[width=\columnwidth]{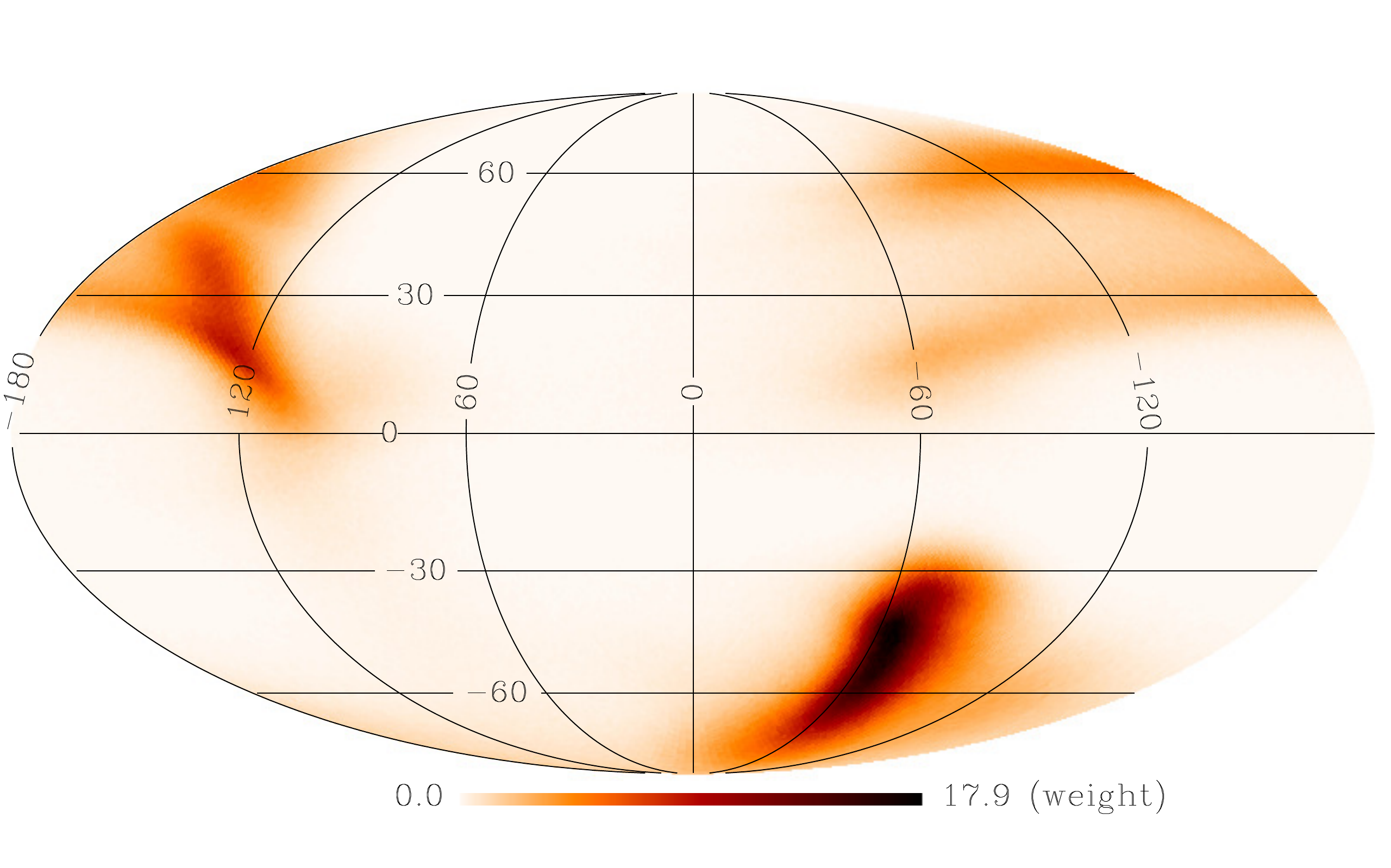}
\includegraphics[width=\columnwidth]{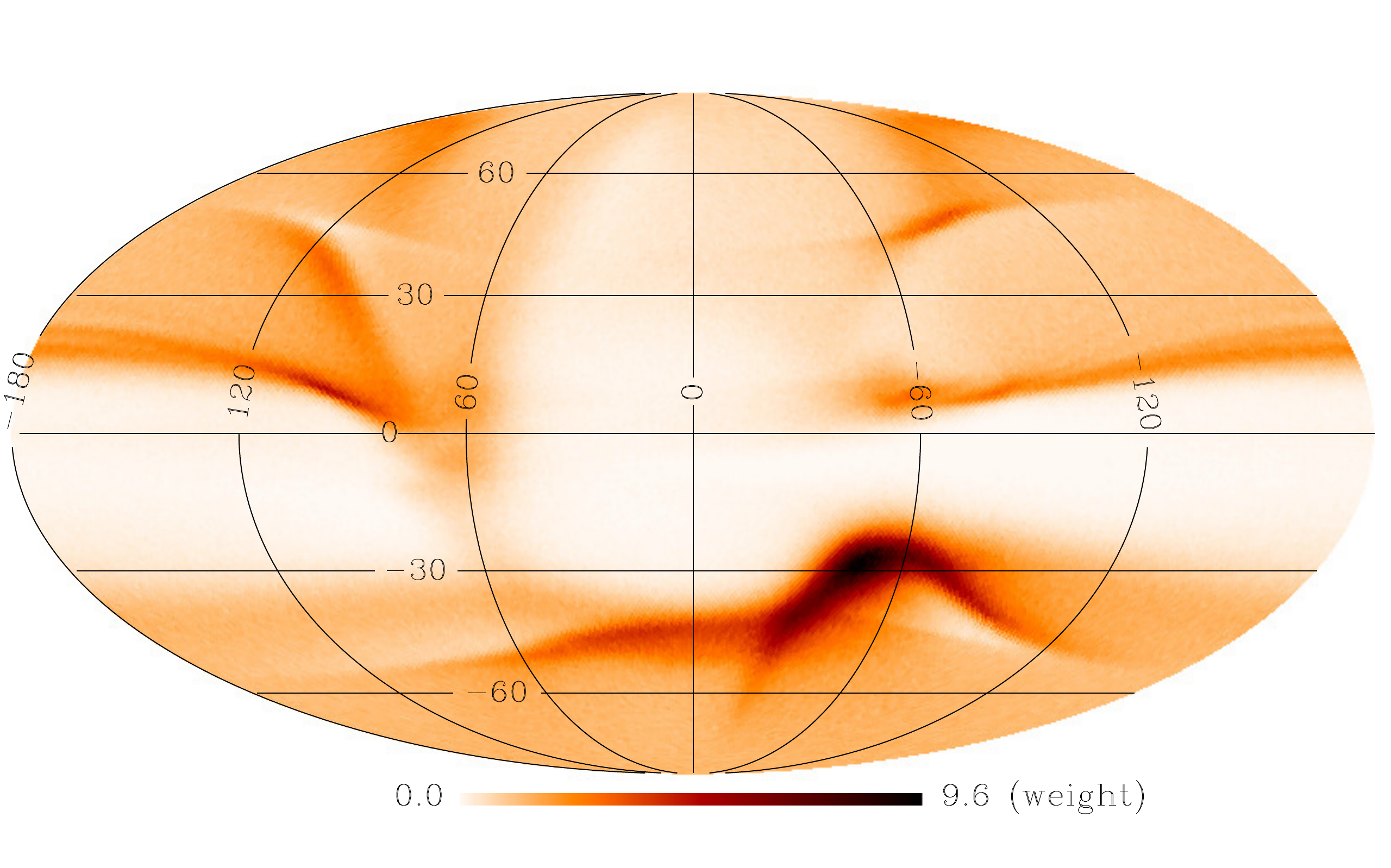}
\includegraphics[width=\columnwidth]{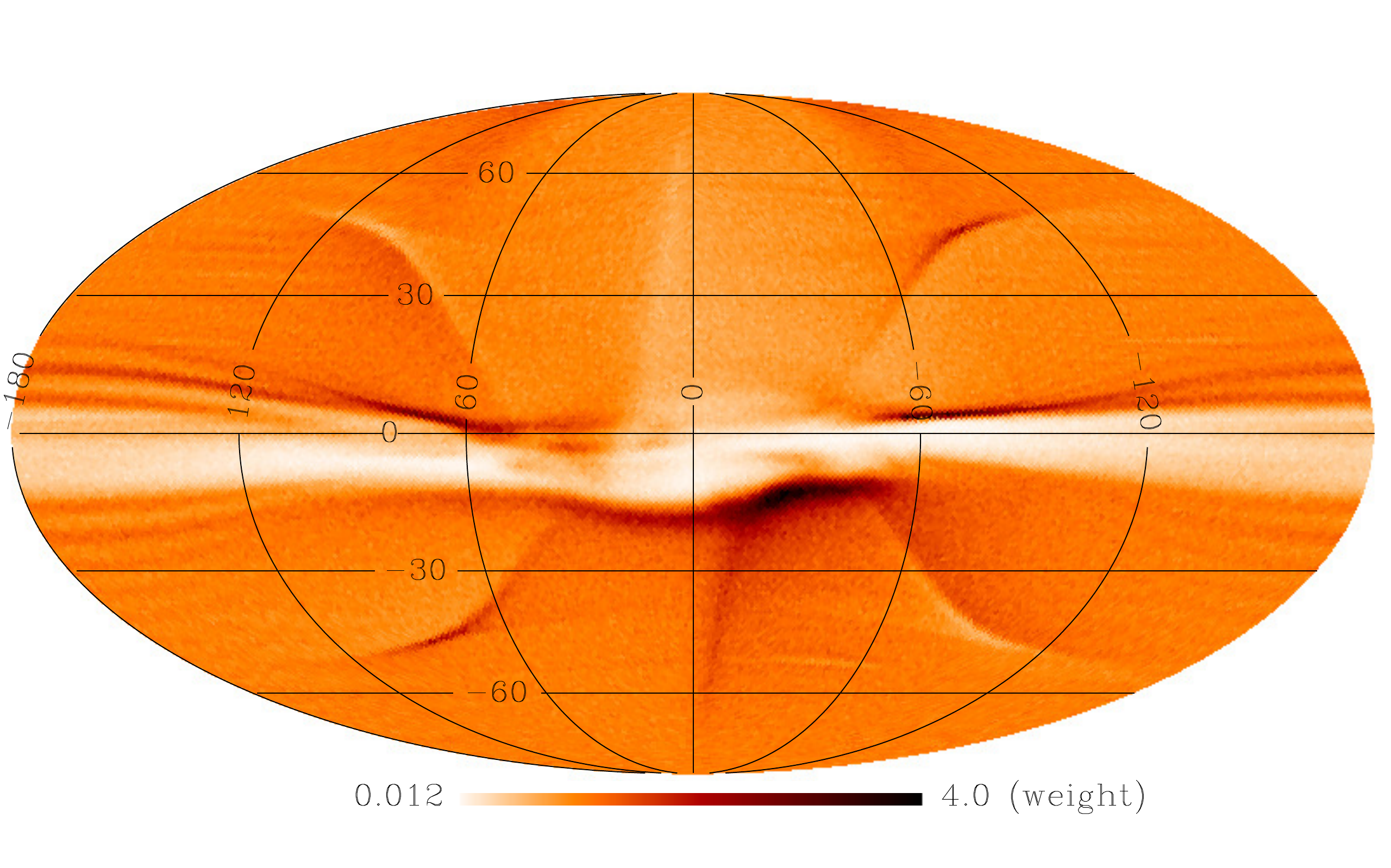}
\includegraphics[width=\columnwidth]{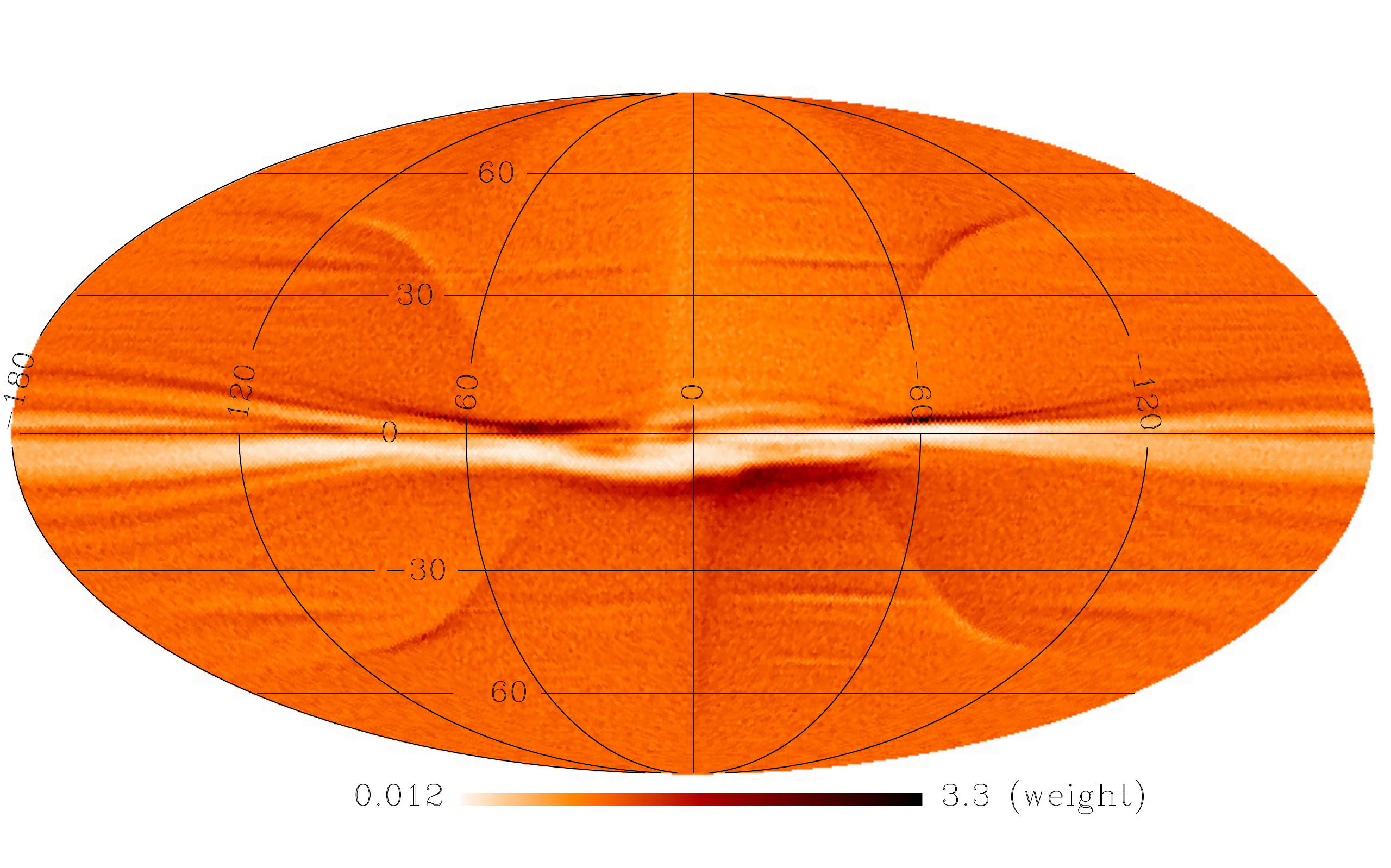}
\caption{Magnification maps, in Mollweide projection and Galactic coordinates, showing the magnification factor of the overall intensity of a UHECR source as a function of its position in the sky, as resulting from the deflections of the particles by the GMF. Four different rigidities are shown: 5~EV, 16~EV, 63~EV, and 130~EV (from top to bottom).}
\label{fig:magnification}
\end{figure}

The inversion is done by simply keeping track, for each UHECR incoming direction, i.e. for each Galactic pixel, of the different Earth pixels in the direction of which back-propagated particles were initially sent to exit the Galaxy in that pixel. The number of Earth pixels related to a given Galactic pixel is not known \emph{a priori}. On average, 256 pixels on the fine grid are associated with the Galactic pixels on the coarse grid (this is also, of course, the number of Earth pixels that would be associated with each Galactic pixel if there were no deflection at all). However, some directions turn out to be more likely to be exited into by back-propagated particles than others, because the GMF can focus back-propagated particles from a wide range of directions into a smaller solid angle, or conversely. This is illustrated in Fig.~\ref{fig:magnification}, which shows the magnification factors for each Galactic pixel, defined as the number of Earth pixels associated with that pixel divided by 256. For instance, a magnification factor of 2 indicates that a source in that direction will contribute twice as much flux to the UHECRs observed on Earth as if there were no deflection (or as the average of what could be expected if it were anywhere in the sky).

There is a different magnification map for each particle rigidity and, as can be seen, the level of magnification can vary by large amounts even between two relatively nearby source directions. This is the well-known phenomenon of the so-called caustic curves, which are singular mathematical lines where the wave front of a radial stream of particles bent by the (regular-only) magnetic field would be tangent to the line of sight. Strongly contrasted caustics are found to appear only for intermediate rigidities in the energy range of interest. Indeed, for low rigidities, the particles are strongly deflected along any direction and, in the limit of very large deflections, an essentially isotropic flux is produced and the magnification tends to one in all directions. Conversely, at very large rigidity, the deflections become very small in any direction, and the particle transport across the GMF tends to the trivial one-to-one relation between Earth pixels and identical Galactic pixels. At intermediate rigidities, complex structures can be observed, with regions of large magnification neighboring regions of large demagnification. 

Quantitatively, one can see from the top panel of Fig.~\ref{fig:magnification} that, for some source locations, the magnification factor can reach values as high as 10 to 18 for Fe nuclei around 130~EeV, while we appear to be essentially blind to other regions of the sky. The same is still true at a rigidity of 16~EV, with magnification factors up to 9.6. The magnification factors do not exceed the value of 3 or 4 at rigidities larger than 60~EV, but regions of the sky, notably just below the Galactic plane, appear to be strongly demagnified, down to almost complete invisibility, even for protons which are, yet, very little deflected in this energy range.

\begin{figure}[hp!]
\centering
\includegraphics[width=\columnwidth]{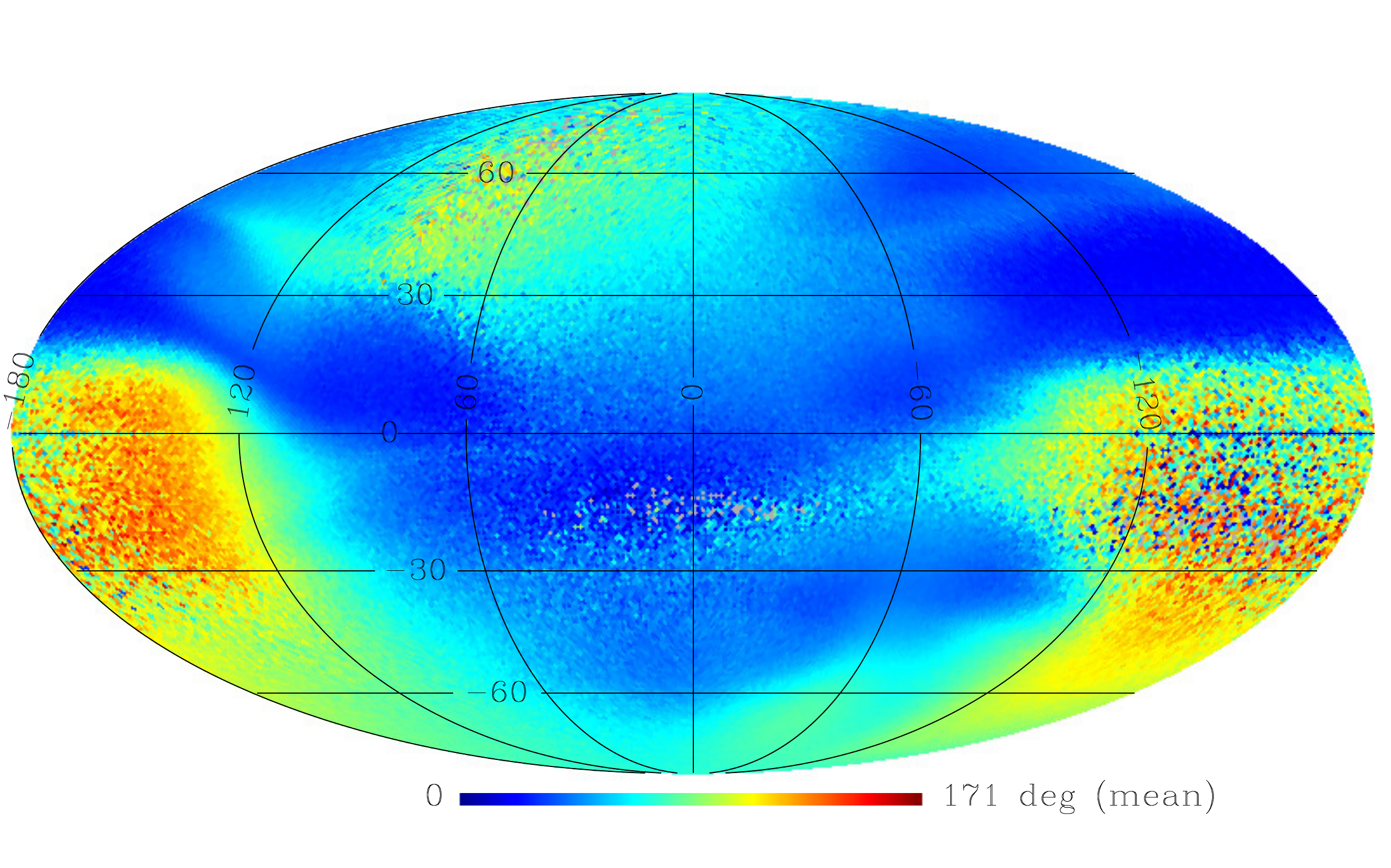}
\includegraphics[width=\columnwidth]{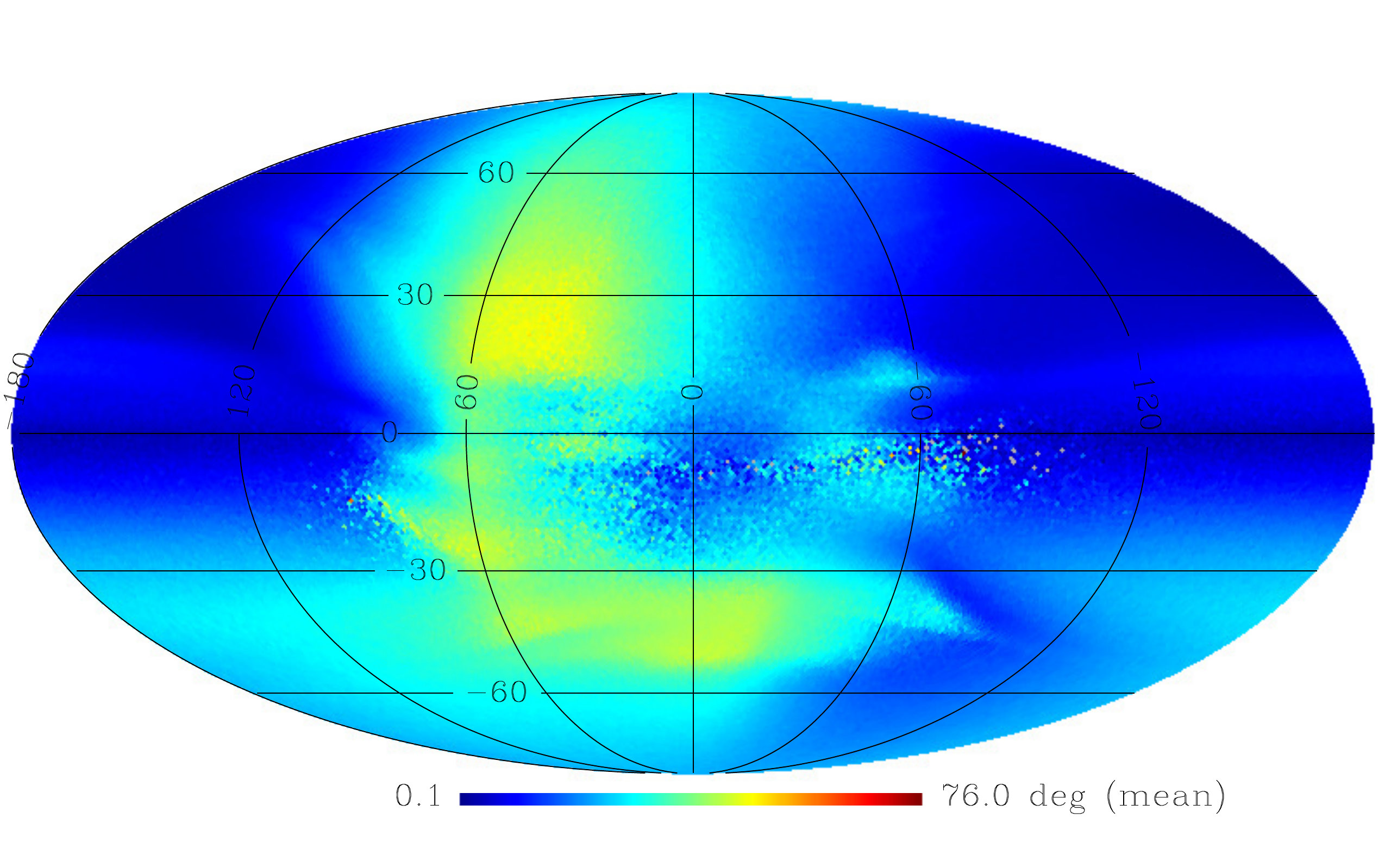}
\includegraphics[width=\columnwidth]{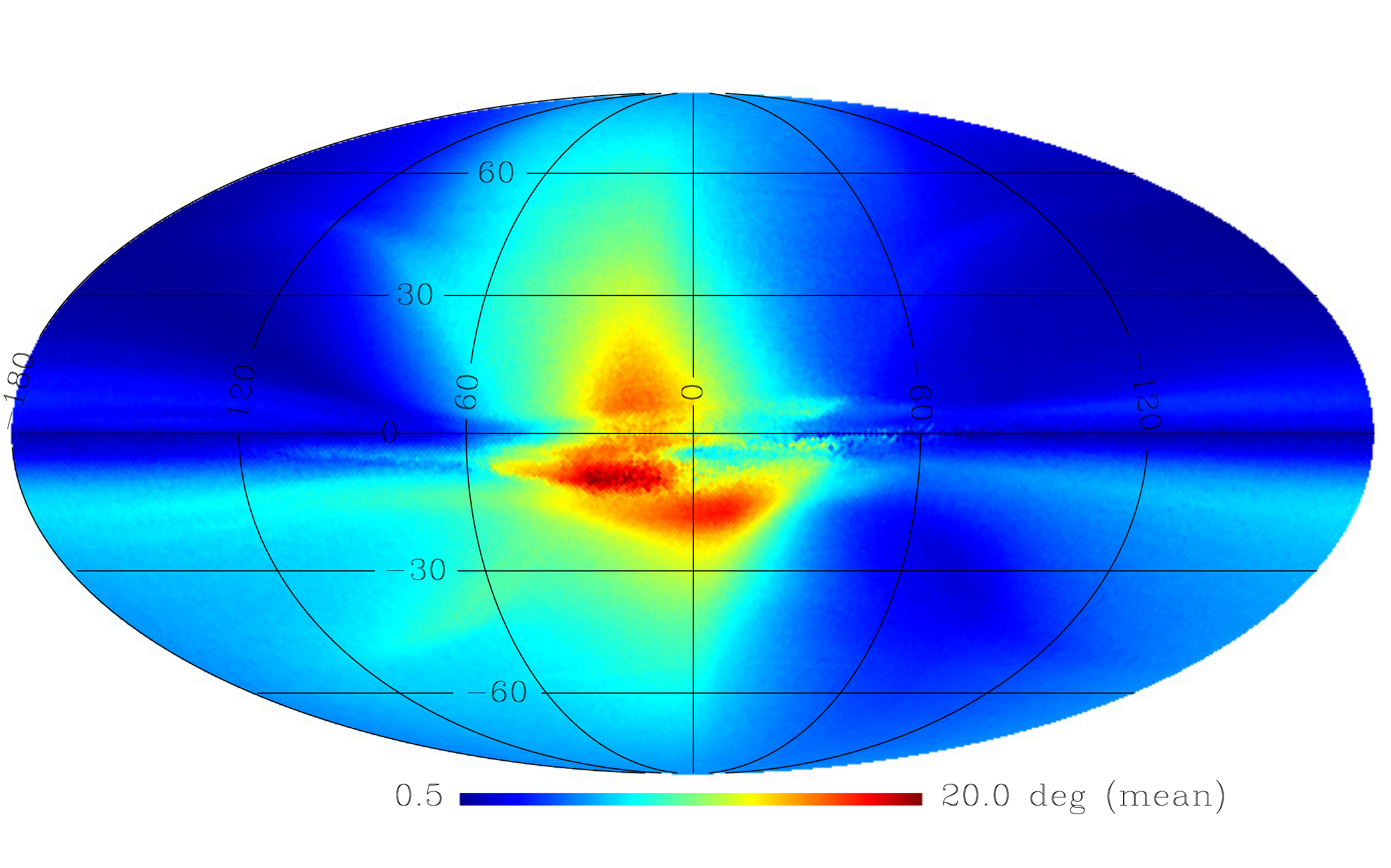}
\includegraphics[width=\columnwidth]{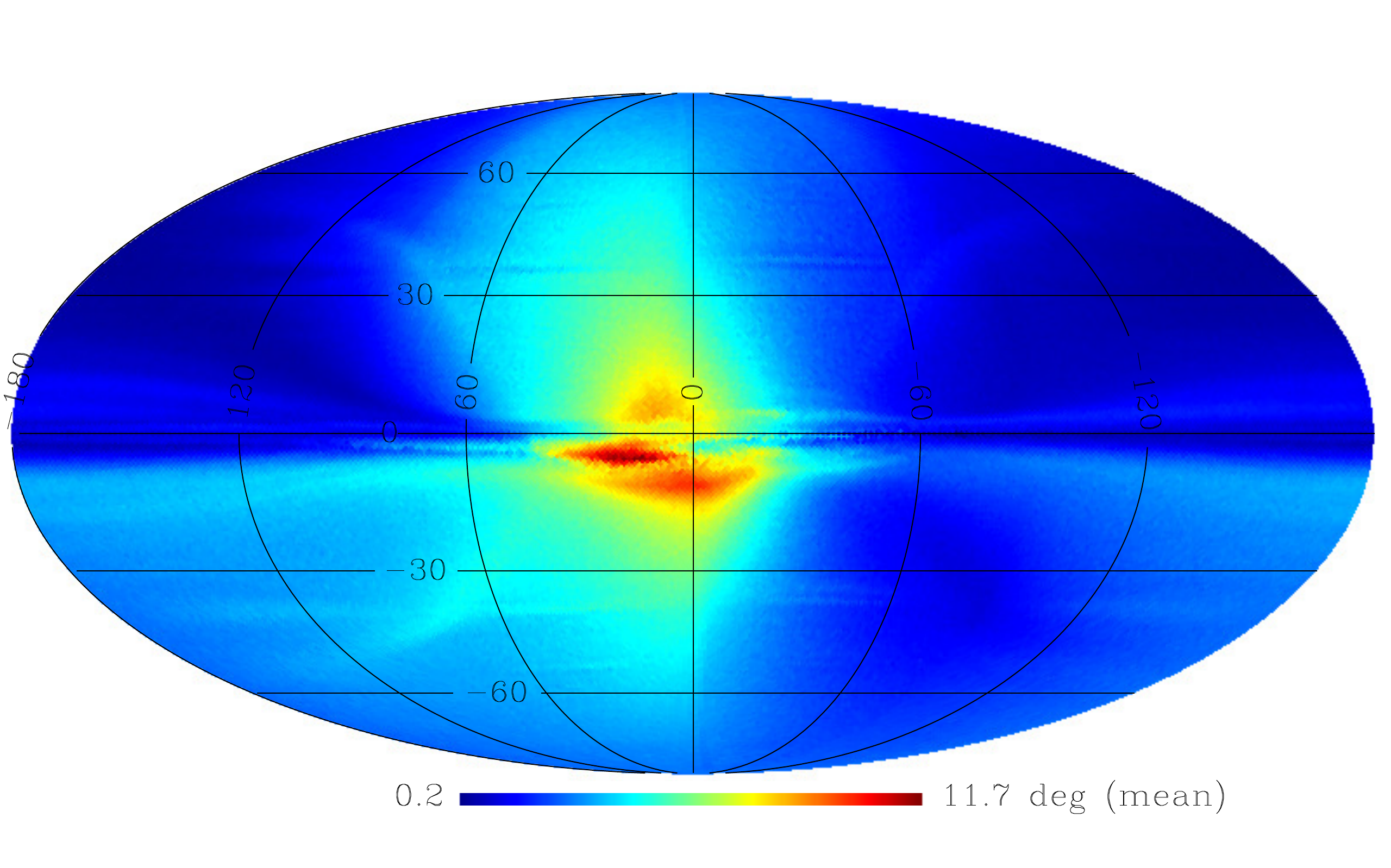}
\caption{Forward deflection maps, displayed in Mollweide projection in Galactic coordinates, showing the mean deflection experienced by the UHECRs coming from a given direction at the entrance of the Galaxy and the corresponding arrival directions at Earth, for the same rigidity as in Fig.~\ref{fig:deflectionHistogram}: 5~EV, 16~EV, 63~EV, and 130~EV (from top to bottom).}
\label{fig:deflectionMapForwardMean}
\end{figure}

It is worth noting that there is no contradiction with the so-called Liouville theorem, which indicates that an incoming isotropic flux must remain isotropic, whatever the intervening magnetic field may be. This simply means that dead zones (or angles) from which cosmic rays are deflected away and never reach the Earth are exactly compensated by focused zones from where incoming cosmic rays are deflected into the apparent direction of the dead zone. Here, the sources are discrete and some positions can be partially or completely hidden to us by the GMF, while others can be magnified to larger apparent luminosities. In principle, this can modify the source statistics by potentially large amounts, depending on the range of rigidities of the UHECR particles. In addition, since the magnification of a source in a given direction depends so much on the rigidity, the magnification factor can be very different for different types of particles at a given energy. In particular, a given source location can result in a large magnification (or demagnification) of the Fe component in a specific energy range, while the proton component will be unaffected. This may result in noticeable modifications of the composition and spectrum of individual sources or of separate regions of the sky. These interesting effects and the associated constraints that may be derived from them will be studied in a statistical way in a forthcoming paper. 

In addition to the magnification factors, the inverted (i.e. forward) relation between Galactic pixels and Earth pixels allows to determine where the UHECRs entering the Galaxy in a given direction will be observed on Earth. From the set of Earth pixels associated with a given Galactic pixel, it is straightforward to compute the average deflection experienced by the UHECRs coming from a source in that direction, as a function of rigidity. This is shown synthetically in Fig.~\ref{fig:deflectionMapForwardMean} in the form of so-called ``forward deflection maps'', where we plot the mean of the set of deflections of UHECRs not observed in a given direction (as in the ``backward deflection maps'' of Fig.~\ref{fig:deflectionMapBackwardMean}), but initially coming from that direction. This mean is calculated from the set of Earth pixels associated with that direction, which contains on average 256 pixels but can be much more or much less numerous depending on the magnification factor. Some particularly blind source directions turned out to be associated with no Earth pixels at all, and it was thus impossible to derive a mean deflection. The corresponding pixels have been represented in gray on Fig.~\ref{fig:deflectionMapForwardMean}.

As mentioned above for the backward deflection maps, an important feature of the forward deflection maps is the strong contrast between the observational situation for different source directions. Even at the lowest rigidity represented here (top panel), where deflections are on average very large, some significant parts of the sky give rise to much smaller deflections, represented in blue on the plots. The prior knowledge of these regions (which derives directly from the knowledge of the GMF) should help deriving meaningful constraints from the distribution of UHECR arrival directions and anisotropy patterns.

Finally, it is interesting to note that large magnification factors are not necessarily associated with large particle deflections, but rather large angular gradients of particle deflections. Large deflections usually correspond to either randomization, in which case the magnification tends to be close to unity, or to obscuration, in which case the magnification factor can become much lower than 1, or even tend to zero (e.g. towards the Galactic center, behind which a source has very low probably to be visible at UHE). On the other hand, large magnification usually require an ordered variation of the deflections occurring over a range of nearby directions, which can coherently extend the solid angle ``feeding'' a given direction.

Of course, the exact patterns observed on these various maps (backward, forward and magnification maps) depend on the GMF model used in the propagation code, which is unlikely to be correct all across the Galaxy. However, we may hope that the assumed GMF model is sufficiently representative of the actual GMF for the above results to give a reasonable idea both of the typical average deflections and standard deviation values and of their range of variations over the sky.


\subsection{Generation of UHECR sky maps}

\subsubsection{General procedure}

The final step of the propagation procedure is the building of the simulated sky maps, each of which represents a particular set of UHECR events distributed over the sky, as could be observed by a given experiment. For this, we simply put together the above elements.

First, we select an astrophysical scenario, i.e. we choose one of the five generic composition/spectrum models described in Sect.~\ref{sec:sourceCompo} and summarized in Table~\ref{table:compo}, and assume a given source density, $n_{\mathrm{s}}$. We then generate a particular realization of this scenario, by selecting the location of the sources through a random draw in the source catalog with density $n_{\mathrm{s}}$, as explained in Sect.~\ref{sec:sourceDistrib}. A total of 500 different realizations are simulated for each astrophysical scenario, in order to explore the so-called \emph{cosmic variance} of the models, i.e. the range of sky map properties that can be expected within a given scenario, depending on the contingent distribution of the actual sources currently active in the local universe.

For each realization, we build different sky maps, depending on the intended observatory (determined by its coverage map), and event statistics (determined by the total exposure of the experiment). In the current paper, we choose either the partial sky coverage of the Pierre Auger Observatory, or the almost uniform sky coverage of JEM-EUSO with an exposure of 300~000 km$^{2}$~sr~yr, as discussed in Sect.~\ref{sec:introduction}.

The UHECR particles are generated one by one, with their own energy and nuclear type, according to the source spectrum and source composition of the scenario under investigation. Each particle is propagated with the Monte Carlo code described above in the EGMF. We then apply the magnification factor appropriate to the resulting direction of the UHECR as it enters the Galaxy, as derived in Sect.~\ref{sec:deflectionMaps}. For this, we normalize the magnification map to the maximum magnification factor at all energies, and apply a standard acceptance/rejection method. To determine the observed direction of the UHECRs on Earth, we choose randomly between the various Earth pixels associated with the incoming direction (i.e. Galactic pixel). The next step consists in applying the coverage map of the experiment, i.e. accepting/rejecting the events according to the normalized exposure in the relevant arrival direction. In the case of the JEM-EUSO-like detector, we apply an additional acceptance/rejection procedure to account for the detection efficiency as a function of energy. For this, we use the efficiency curve computed for JEM-EUSO, as given in \cite{JEM-EUSO:2013a}. Finally, we apply an error on the energy and direction to reflect the experimental uncertainty on the reconstructed shower parameters. For JEM-EUSO, we use a simplified and conservative Gaussian energy resolution uncertainty of $30\%$ for all the UHECR events. The angular resolution is also assumed to be Gaussian with a width of 2 degrees, but given the patterns and amplitudes of the deflections for the different models, changing this parameter did not appear to have any significant impact on the results.

The above procedure is applied to each UHECR, one after the other, until the intended statistics is collected for the detector under consideration. The resulting sky map is the final output of the simulation, from which a systematic search for anisotropy can be performed, as discussed in Sect.~\ref{sec:anisotropySearches}.


\subsubsection{Sky map statistics}

It is important to note that the total number of events to be observed by a given observatory is not known a priori. The main source of uncertainty on the UHECR flux resides in the so-called \emph{energy scale} of the measured UHECR spectrum. Both the Auger spectrum and the HiRes/TA spectrum suffer from systematic uncertainties on the reconstructed energy of the UHECR events. While a joint working group has proposed to build a fiducial energy spectrum by rescaling the Auger energy scale upward and the TA energy scale downward (\citealp{Dawson:2013},~\citetalias{Matthews:2013}), the exact flux remains uncertain. In this paper, we considered separately the two assumptions on the energy scale (i.e. we did not apply any rescaling), which result in two different assumptions for the UHECR flux as a function of energy, and thus two different statistics expected above some energy threshold, for a given choice of the total exposure.

Another source of uncertainty in the expected number of events is the absence of a clear knowledge of the shape of the UHECR spectrum, which requires a larger statistics to be known precisely and may be different in different regions of the sky. As a matter of fact, detecting anisotropies in the UHECR arrival directions above a given energy is equivalent to detecting a different energy spectrum in different directions. The results discussed below show that significant anisotropies should be expected at the highest energies, whatever the assumed astrophysical scenario, so we cannot use the current knowledge of the spectrum in a limited region of the sky (even barring its imperfection) to predict the number of events that an all-sky coverage experiment should detect. The uncertainty associated with the current poor knowledge of the shape of the spectrum is however much smaller than that associated with the energy scale. For our present purpose, we simply assume that either the Auger flux or the HiRes/TA flux hold over the whole sky, and derive a fiducial spectrum by averaging the spectra obtained by fitting the currently available data with our different models. This fiducial spectrum is then used to determine the expected numbers of events for a given detector. Two reference spectra are thus built, one for each choice of the energy scale. From these, we determined the following statistics for the JEM-EUSO-like detector with the quoted total exposure. In the case of the Auger energy scale assumption, we expect respectively 1100, 250 and 100 events above 50~EeV, 80~EeV and 100~EeV (implementing the energy detection efficiency of JEM-EUSO \citep{JEM-EUSO:2013a}. In the case of the HiRes/TA energy scale assumption, we expect respectively 2100, 580 and 260 events above the same energies. Although model-dependent, these numbers represent the best-guess limits on the number of events that can be extrapolated from the current knowledge of the spectra measured with low statistics, partial sky-coverage ground observatories, within the framework of the astrophysical scenarios investigated here.

Finally, in addition to these fiducial statistics for a future observatory similar to the JEM-EUSO project, we also consider the current Auger statistics as a reference point to select astrophysical models which appear compatible with the current data, as far as anisotropy is concerned (see below). For this, we apply the Auger coverage map and accumulate 84 UHECRs above 55~EeV, which corresponds to the statistics gathered by the Pierre Auger Observatory on June, 2011 (according to~\citetalias{Kampert:2011}, in which the last search for small scale anisotropy above 55~EeV with Auger is reported).

\subsubsection{Reading the sky maps}

\begin{figure}[t!]
\centering
\includegraphics[trim=2.75cm 3.25cm 1.25cm 2cm,clip,width=\columnwidth]{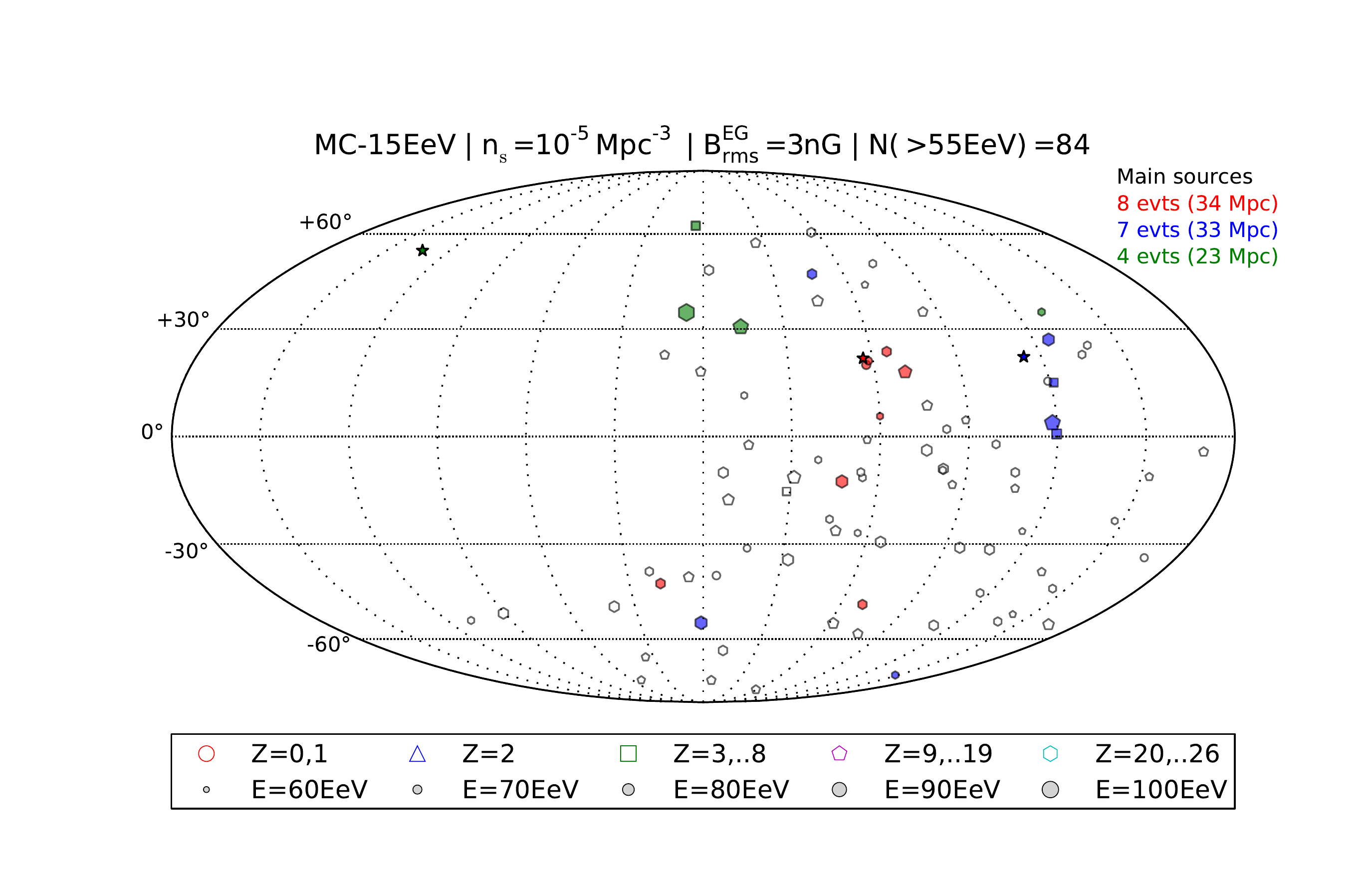}
\includegraphics[trim=2.75cm 3.25cm 1.25cm 1.75cm,clip,width=\columnwidth]{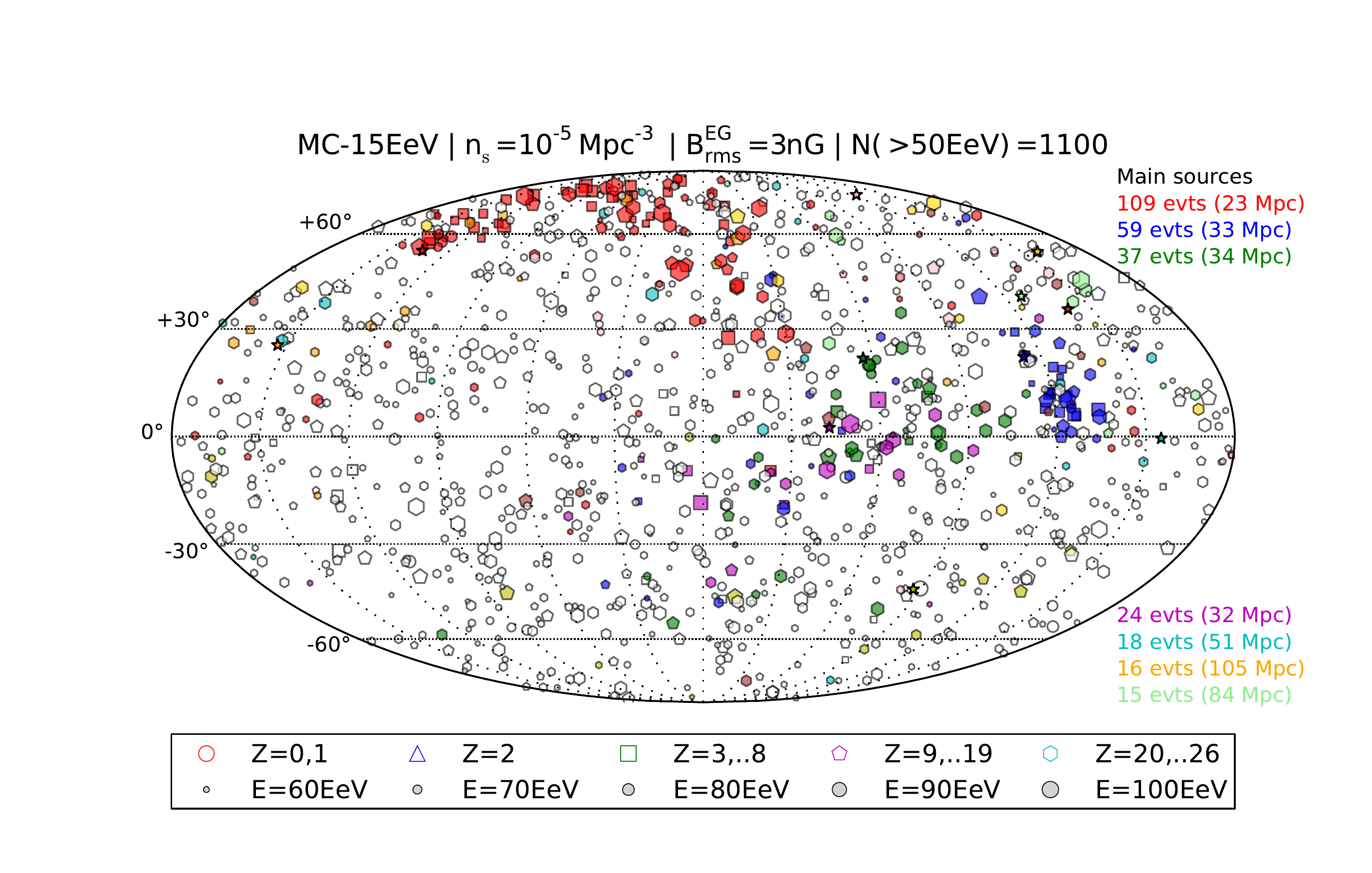}
\includegraphics[trim=2.75cm 3.25cm 1.25cm 1.75cm,clip,width=\columnwidth]{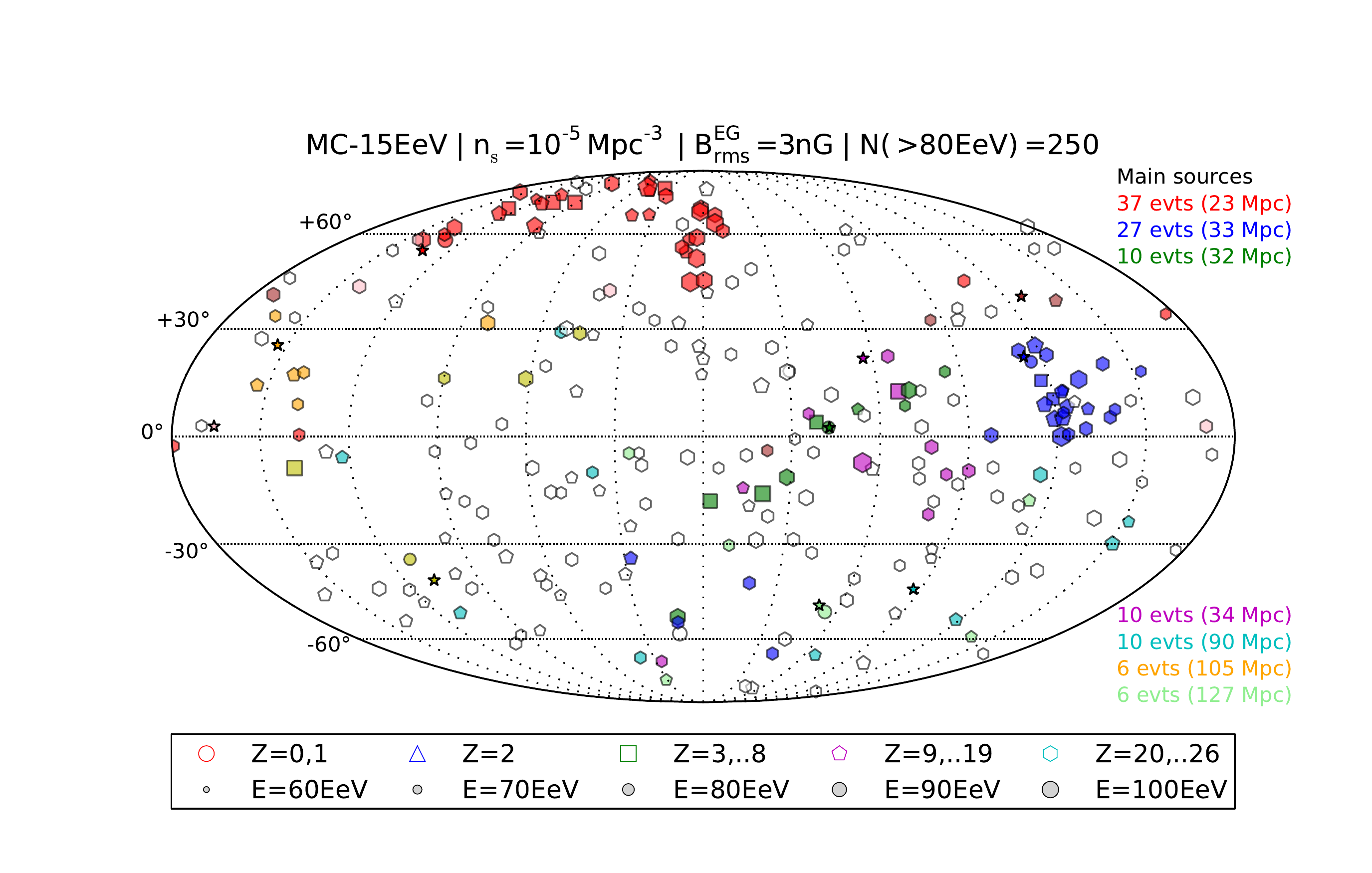}
\includegraphics[trim=2.75cm 1cm 1.25cm 1.75cm,clip=true,width=\columnwidth]{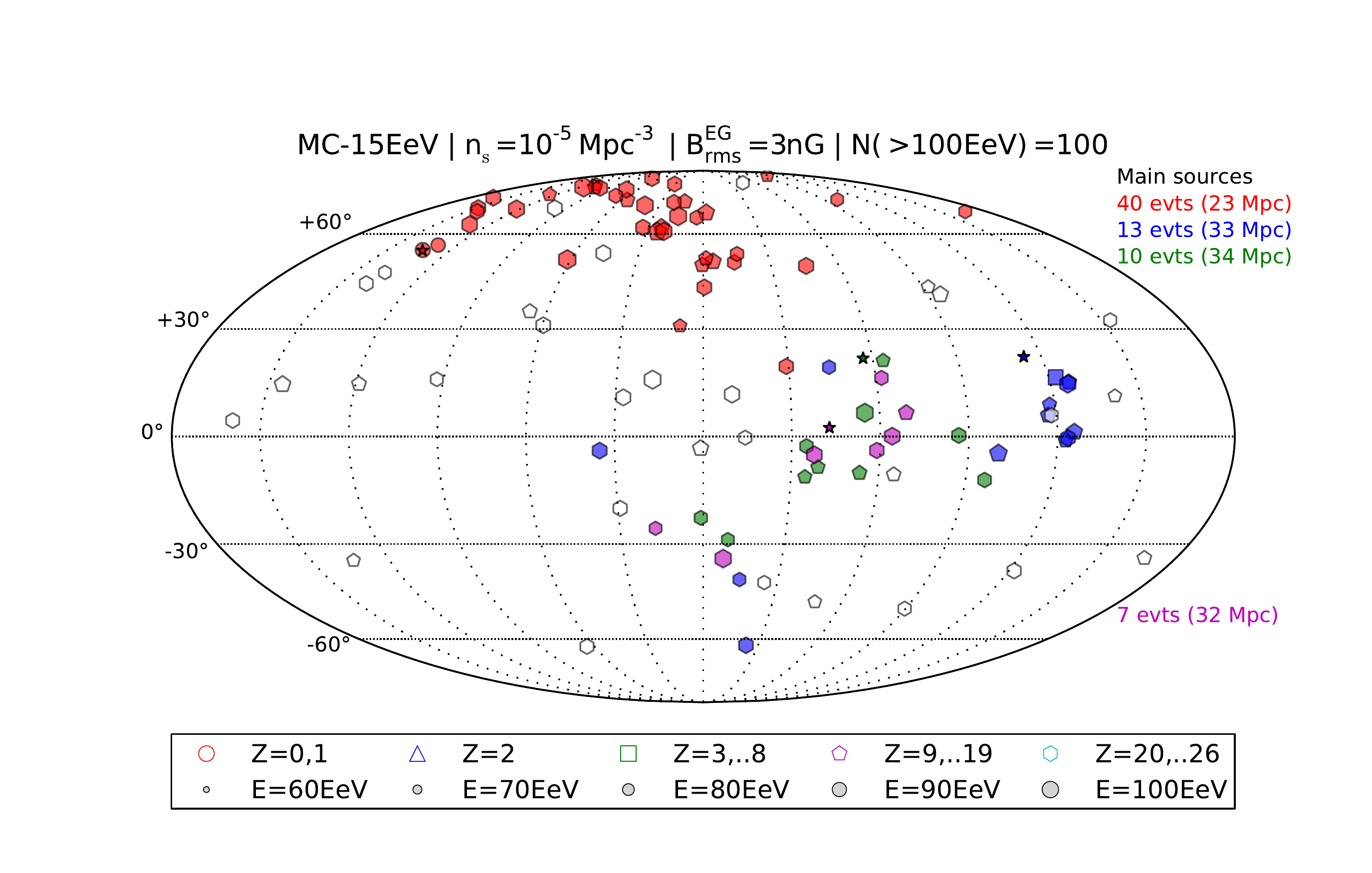}
\caption{Examples of sky maps corresponding to the MC-15EeV model (see text), simulated for the current statistics of Auger (top panel) and for the expected statistics that JEM-EUSO would gather with a total exposure of $300,000\,\mathrm{km}^{2}\,\mathrm{sr}\,\mathrm{yr}$, assuming the flux normalization given by the Auger energy scale (see text). The second, third and fourth maps are drawn with a (reconstructed) energy threshold of 50~EeV, 80~EeV and 100~EeV respectively.}
\label{fig:skyMap1}
\end{figure}

An example of a set of sky maps is shown in Fig.~\ref{fig:skyMap1}. This is the result of a typical simulation, corresponding to a particular realization of a mixed-composition model with a source density of $n_{\mathrm{s}} = 10^{-5}\,\mathrm{Mpc}^{-3}$ and a maximum proton energy of $E_{\mathrm{p},\max} = 15$~EeV (MC-15EeV model). The map on the top panel is the Auger-like reference map, showing the arrival direction of 84 UHECR events above 55~EeV. The map is shown in Galactic coordinates, and the wide region without events on the left and upper right parts of the map are regions of the sky inaccessible to the Auger detector.

The symbols and color codes obey the following rules:
\begin{itemize}
\item the shape of the symbols representing the events give an indication of the mass of the associated UHECR: polygons with larger numbers of sides correspond to heavier nuclei, as indicated on the map, and protons are shown as circles;
\item the size of the symbols is proportional to the particle energy: larger symbols correspond to higher energy particles;
\item events shown with the same color correspond to UHECRs coming from the same source. However, only the most intense sources (by decreasing multiplicity and provided that they contribute at least 3 events and 1\% of the total flux) are shown with a separate color. All the other events are shown in black;
\item the colored stars correspond to the real location of the sources of the events sharing the same color.
\end{itemize}

As can be seen in the top panel of Fig.~\ref{fig:skyMap1}, in this particular example only three sources have a multiplicity larger than 4 in the map corresponding to the Auger statistics of reference. The most intense one is shown in red, with 8 events out of the 84 events recorded. The second source is in blue, with 7 events, and the third source is in green, with 4 events in the field of view. It is interesting to note that the source of these 4 events color-coded in green is actually far away in the non-observable part of the sky. The distance to the color-highlighted sources is also indicated on the map. In this case, the ``green source'' is the closest, most luminous source in the sky, at 23~Mpc. The four events observed from this source in the Southern hemisphere sky consist of one low-mass (square symbol), one intermediate mass (pentagon) and two heavy (hexagon) nuclei, deflected in the Auger field of view by the GMF. The rightmost event has the smallest energy, as indicated by its smaller size.

No obvious clustering of events is visible on the map, which is compatible with the absence of any clear small-scale anisotropy in the Auger data. In this model, this is mostly due to the low value of the maximum proton energy assumed, namely 15~EeV, which results in the dominant presence of heavy nuclei in the energy range under consideration, as can be checked directly on the map (polygonal symbols). Two protons (circles) can nevertheless be seen in red, very close to their actual source, represented by the red star. With this statistics, such a doublet cannot be used to pin point a source on the sky, as the arrival direction coincidence could be a simple chance coincidence. As a matter of fact, a few other doublets can be observed on the sky map, including an almost perfect (but random) coincidence of two heavy nuclei (hexagons) coming from different sources. Three other isolated protons can be seen on the map. They may be close to their actual source, but the low statistics does not allow to identify these sources. Larger statistics are needed to assess the presence of real multiplets with a reliable statistical significance.

The three following panels on the same Fig.~\ref{fig:skyMap1} show the sky maps obtained from the same realization of this astrophysical model (same source composition/spectrum/density scenario \emph{and} same sources), assuming a uniform full-sky coverage with the JEM-EUSO-like exposure and detection efficiency, with energy thresholds at 50~EeV, 80~EeV, and 100~EeV respectively. The three maps illustrate the GZK horizon effect, by which the distant sources contributing to the observed flux become less and less numerous as the energy increases. Even though the nearby, high-luminosity sources are also present, and even dominant at lower energies, the contribution of a very large number of sources distributed more or less uniformly makes it difficult to isolate the brightest sources. Even though a standard test of anisotropy should easily reveal the presence of a significant excess in the maps (see below), it may not be as easy to derive meaningful astrophysical information from the map drawn with an energy cutoff of 50~EeV, as from the map drawn with an energy cutoff of 100~EeV (bottom panel).

At the highest energies, the dominant source appears to account for 40 of the 100 events (in agreement with the general results about UHECR source statistics reported in \citealt{Blaksley:2013}). The corresponding source (red-colored events) is so-to-day self-isolated on the sky, because the more distant sources are cut off by the GZK effect. One should not be misled by the color code, however. In practice, of course, we will have no way, a priori, to distinguish the events coming from a given source. The green and purple sources appear much more mixed together, and could not be so easily isolated. This was to be expected anyway, since their angular separation is of the same order as the typical deflection of the individual events (here dominated by heavy nuclei), and their distance is roughly similar (33--34~Mpc), so that their apparent luminosity is almost identical and the GZK effect operates in the same way for both of them. This is a standard case of source confusion. The possibility to isolate (to a large extent) the dominant source in the sky at 100~EeV is nevertheless a common feature of most of the models and realizations that we have generated.

While the sky maps are sometimes useful to guide the eye, the best way to obtain definite and objective information about the UHECR sky is usually to perform unambiguous anisotropy analysis. This is what we did to analyze in a systematic way the thousands of sky maps generated by the above procedure, as explained in the next Section.

\section{Results and discussion}
\label{sec:results}

Our goal is to determine whether a future experiment with a total exposure of 300,000~$\mathrm{km}^{2}\,\mathrm{sr}\,\mathrm{yr}$ could observe significant anisotropies in the arrival directions of UHECRs. This depends not only on the general astrophysical model assumed, but also on its particular realization, i.e. on the specific distribution of the sources which would happen, within this model, to contribute to the observed flux of cosmic rays in our specific location in the universe, during the time of observation. Indeed, even for a given assumption about the source composition, spectrum and density, a relativity large range of situations could be encountered, exhibiting either very strong, moderate or low anisotropies depending on this particular source distribution.

In the present study, we explore the global anisotropy properties associated with the different scenarios in a statistical manner. To this end, we simulate 500 different realizations of each scenario and determine the probability that a given realization, chosen randomly, gives rise to a significant anisotropy.

\subsection{Intrinsic anisotropy searches}
\label{sec:anisotropySearches}

There are many ways to search for anisotropies in a data set, and for any given sky map, various tests can be performed, including searches for specific correlations with known astrophysical sources or for any type of potentially meaningful pattern noticed \emph{a posteriori} on a partial data set, and which can then be tested on subsequent, independent data sets as the sky map is being built over time. It is thus not possible here to do an exhaustive search of anisotropies, and most of the tests based on correlations with other catalogs of sources would anyway be mostly irrelevant in the case of our simulated sky maps, not mentioning the fact that the knowledge of the magnetic field is still incomplete at the moment. We thus chose to limit our studies to intrinsic anisotropies, as can be revealed by the analysis of the angular auto-correlation of the UHECR arrival directions, through localized excesses in the number of events or through the commonly used two-point correlation function.

The results shown here are in this respect conservative, in the sense that they do not exploit the whole arsenal of analysis tools available, but consider only the most standard ones, which are largely independent of the assumptions made on the GMF (see the related comment in Sect.~\ref{sec:particleDeflections}). In addition, these tests do not depend much on the actual choice of the sources, but rather on their overall properties and density in the nearby universe. Likewise, in our anisotropy searches, we did not make any use of the potential correlation between the arrival direction of the UHECRs and their energy, which should display, on average, some coherent patterns in the case of UHECRs originating from the same sources. Such correlations, which would in principle open new possibilities to identify sources or constrain their number and locations, will be studied separately.

Here, we present the results obtained with one of the most widely used statistical tests to measure the departure of a given data set from isotropy, based on the so-called two-point correlation function. This function characterizes the auto-correlation of the UHECR arrival directions by simply giving the number of pairs of UHECR events separated by less than a given angle on the celestial sphere, as a function of that angle. The test consists in comparing the observed numbers of pairs of events at any angular scale with the numbers of pairs present at the same angular scale in datasets of the same size built randomly from an isotropic distribution.

To ensure a good statistical power of this test, we computed $10^{6}$ different realizations of an isotropic sky for each number of events considered (corresponding to different exposures and energy thresholds), implementing the appropriate coverage map, either that of Auger or the uniform full-sky coverage relevant to JEM-EUSO. Each of these realizations has its own two-point correlation function, and the whole set of realizations gives us not only the average number of pairs of events at each angular scale, but also the distribution of the numbers of pairs that can be expected for an isotropic sky. From this distribution, we can determine the probability that a given sky map could be built from an underlying isotropic distribution, by simply counting the fraction of isotropic realizations that lie further away from the average distribution than the data set under study.

We also performed on the whole set of sky maps a standard blind search test for localized excesses in some directions of the sky, as a function of energy as well as angular window size, and analyzed the corresponding significance of the detected anisotropy using the Li \& Ma statistics \citep{Li:1983}. The results obtained were on average very similar to those derived from the two-point correlation function. Even though some particular sky maps appeared to show larger departures from isotropic expectations with one test rather than with the other, no additional information could be derived globally once averaged over the different realizations. Therefore, we only show here the results obtained with the auto-correlation function.

Obviously, scenarios in which Fe nuclei dominate are less likely to generate strong anisotropies, notably on small angular scales, than scenarios in which protons dominate. The deflection angle of individual particles is however not the only important parameter. The source density also has a strong influence on the possibility to detect significant anisotropies. In the case of a low source density, say $n_{\mathrm{s}} = 10^{-6}\,\mathrm{Mpc}^{-3}$, very few sources contribute to the observed flux, and multiplets with a large multiplicity are bound to be detected by observatories with an increased exposure, making it much more likely that clusters of events incompatible with an isotropic flux are observed. Conversely, a very large source density will result in much smaller multiplicities, even for the most luminous sources, and in addition a smaller angular separation between sources Êon average, making source confusion much more likely. To be definite, in the following we loosely refer to the case where $n_{\mathrm{s}} = 10^{-6}\,\mathrm{Mpc}^{-3}$ as the low-density case, to the case where $n_{\mathrm{s}} = 10^{-5}\,\mathrm{Mpc}^{-3}$ as the intermediate-density case, and to the case where $n_{\mathrm{s}} = 10^{-4}\,\mathrm{Mpc}^{-3}$ as the high-density case.


\subsection{Mixed-composition models with low $E_{\max}$}
\label{sec:MCModels}

The initial hope behind the intense observational efforts developed in the last decades to detect UHECRs was to identify their sources by direct pointing, in an energy range where the deflections by intervening magnetic fields would be small enough to let us identify tight clusters of events associated with individual sources, right behind their arrival directions. Such an ideal situation did not occur so far. A reason for this may be that the deflections are larger than initially anticipated, possibly because a dominant fraction of the highest energy cosmic rays are not protons, but heavier nuclei, with a smaller rigidity. If this is the case, then the absence of any clear anisotropy detected by the current detectors is quite easy to understand. The key question is now whether a new generation of detectors, increasing the exposure by a factor of ten or so, could change the situation significantly. In this section, we study the case of mixed composition models, in which the maximum energy of the protons in the sources is lower than the GZK energy scale, so that the highest energy particles are dominated by heavy nuclei, and most particularly Fe nuclei.


\subsubsection{Large source density: $n_{\mathrm{s}} = 10^{-4}\,\mathrm{Mpc}^{-3}$}

\begin{figure}[ht!]
\centering
\includegraphics[trim=0.75cm 0.25cm 1.5cm 0.5cm,clip,width=\columnwidth]{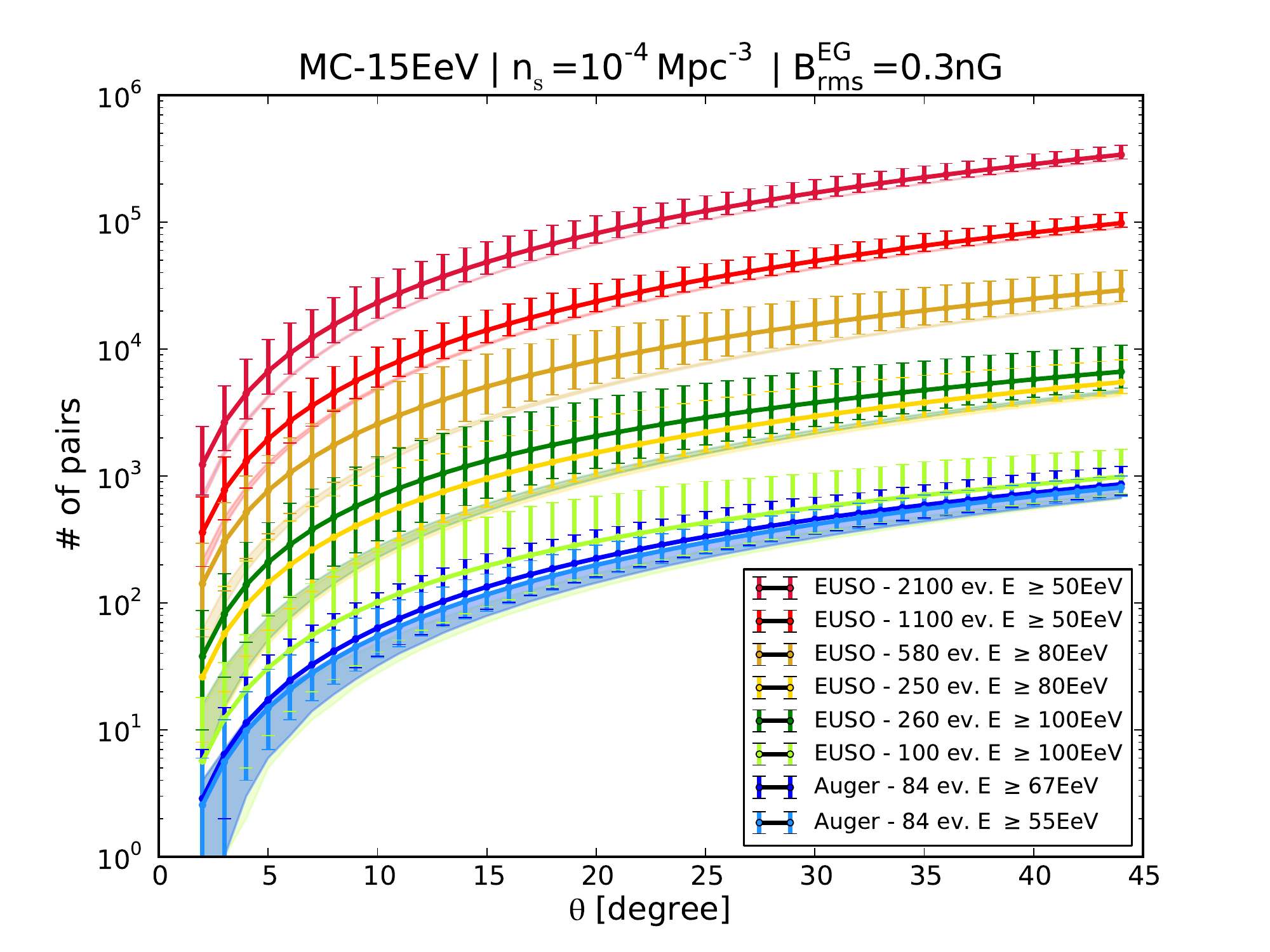}
\caption{Two-point correlation functions for the MC-15EeV model, with a source density $n_{\mathrm{s}} = 10^{-4}\,\mathrm{Mpc}^{-3}$ and an intensity of the random EGMF of 0.3~nG (r.m.s.). The different statistics for each experiment are shown in different colors, as indicated (light and dark tones of the same color are respectively for the Auger and TA energy scales, see text). The error bars contain 90\% of the 500 realizations simulated. Isotropic samples with the same statistics are given as a shaded area of the same color (the envelop contains 90\% of the isotropic samples).}
\label{fig:2ptMC15-4}
\end{figure}

In Fig.~\ref{fig:2ptMC15-4}, we show the two-point correlation functions obtained with the mixed-composition, low proton-$E_{\max}$ model, MC-15EeV, assuming a source density of $10^{-4}\,\mathrm{Mpc}^{-3}$. Curves with different colors correspond to different statistics, and for each angular scale, the error bars contain 90\% of the 500 realizations of that particular astrophysical scenario. These curves are interesting only in comparison with the isotropic expectations, which are shown by the shaded areas of the same color. These areas contain 90\% of the million isotropic realizations.

The two lowest curves, in blue, correspond to sky maps simulated with the Auger statistics. As can be seen, a large fraction of the realizations are compatible with isotropy, and the average two-point correlation function for this model lies only marginally away from the isotropic expectations, especially if the Auger energy scale (light blue) is assumed. This confirms that the MC-15EeV scenario is fully compatible with the current data, as far as the anisotropies based on the auto-correlation of UHECRs are concerned (the same holds for the blind search of localized excesses, at any angular scale).

The other six curves, however, show that with the statistics of JEM-EUSO, this scenario would produce significant anisotropies for most realizations, if not all. This is true for the sky maps built with a cutoff at 50~EeV (red curves), 80~EeV (beige curves) or 100~EeV (green curves), whatever the assumption on the energy scale (lighter tone for the Auger scale, darker tone for the TA scale): the lower limit of the error bars hardly touches the shaded areas corresponding to the isotropic expectations, for most of the angular scales.

\begin{figure}[ht!]
\centering
\includegraphics[trim=0.75cm 0.25cm 1.5cm 0.5cm,clip,width=0.98\columnwidth]{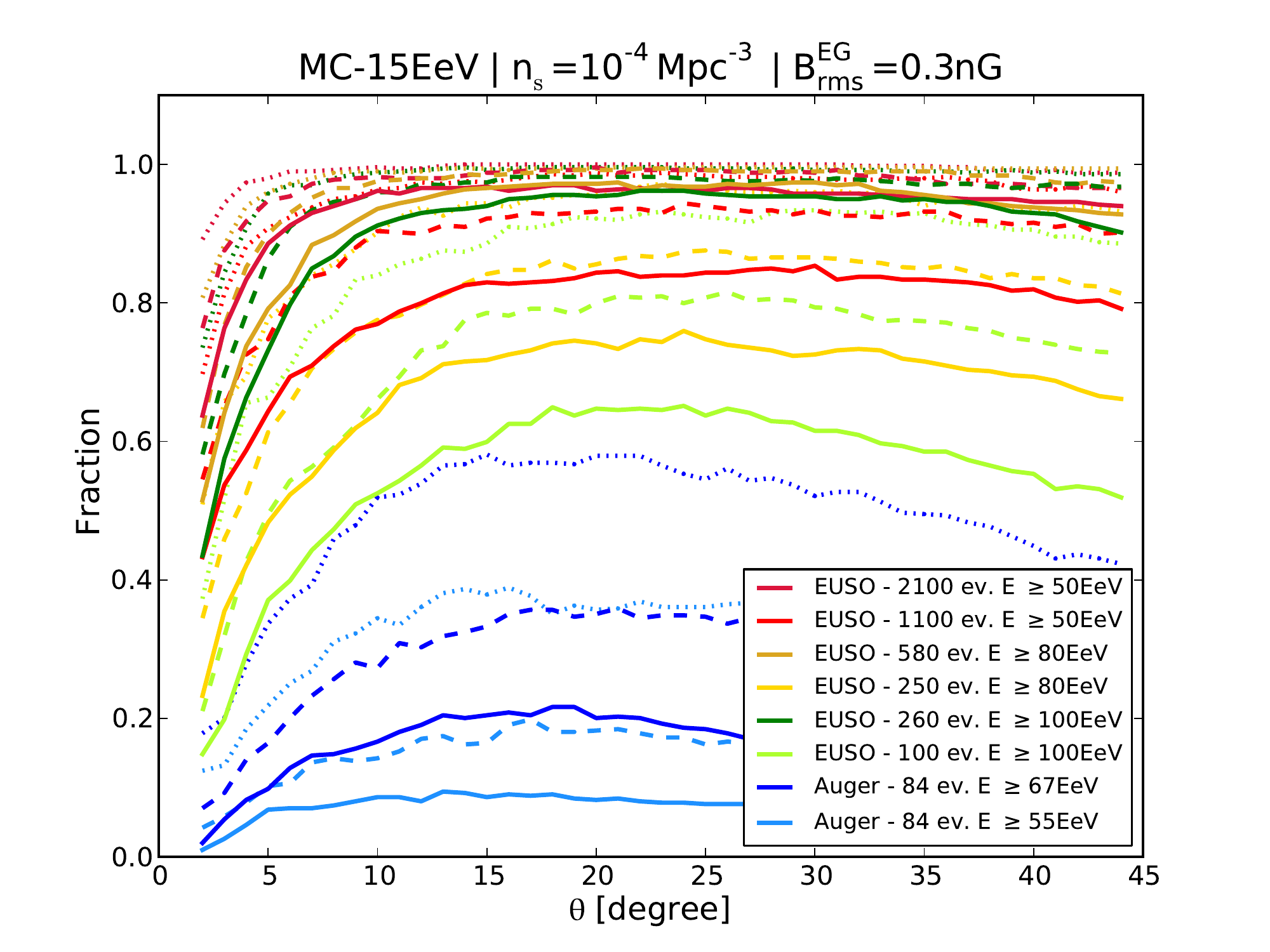}
\caption{Fraction of the 500 realizations of the same scenario as in Fig.~\ref{fig:2ptMC15-4}, which give an anisotropic signal with a significance larger than $2 \sigma$ (dotted lines), $3 \sigma$ (dashed lines), or $4 \sigma$ (plain line), for different statistics (as indicated), as a function of angular scale.}
\label{fig:2ptFractionMC15-4}
\end{figure}

Another way to look at these results is proposed in Fig.~\ref{fig:2ptFractionMC15-4}, where we show the fraction of the realizations of the same scenario which are further away than 2$\sigma$ (dotted lines), 3$\sigma$ (dashed lines) or 4$\sigma$ (plain lines) from the average isotropic expectation. Note that, since the distributions are not necessarily gaussian, what we mean by n-$\sigma$ is actually not the number of standard deviations away from the average of the distribution, but the distance away from the average that corresponds, respectively, to 95.5\%, 99.7\% and 99.994\% of the isotropic realizations. In other words, the non-gaussian tails are taken into account to provide real probabilities rather than standard deviations.

The various curves in Fig.~\ref{fig:2ptFractionMC15-4} confirm, in a more quantitative way, that the scenario under investigation is globally compatible with the Auger constraints on anisotropy, since less than 20\% of the realizations show an anisotropy with a significance of $3\sigma$, and less than 40\% of the realizations show an anisotropy as weak as 2$\sigma$ (for the two-point correlation function). Assuming the HiRes/TA energy scale instead of that of Auger increases the fraction of realizations showing some anisotropy, because the events considered are more energetic and thus less deflected, and because the corresponding horizon distance is somewhat reduced. However, more than 50\% of the realizations are still found not to display any significant auto-correlation.

The situation with the JEM-EUSO statistics is very different, since more than 80\% of the realizations display at least a $3\sigma$ anisotropy at all energies. If one assumes the HiRes/TA energy scale (dark colors), one even finds that 90\% of the realizations can be declared anisotropic with a $4\sigma$ significance (i.e. a $6.3\,10^{-5}$ chance probability). Even with the Auger energy scale, more than 50\% of the realizations show a $4\sigma$-significance anisotropy at 100~EeV, despite the lower statistics. This fraction increases to $\sim 70$\% at 80~EeV, and up to more than 80\% at 50~EeV. By contrast, the Pierre Auger Observatory is expected to find a $4\sigma$ anisotropy in less than 10\% of the realizations of such a scenario.

As can be seen, the angular scale where most realizations display the most significant anisotropy in this astrophysical scenario is relatively large, around 15--25 degrees. This is not surprising, given the predominance of high-$Z$ nuclei among the UHECRs.

\begin{figure}[ht!]
\centering
\includegraphics[trim=0.75cm 0.25cm 1.5cm 0.5cm,clip,width=0.98\columnwidth]{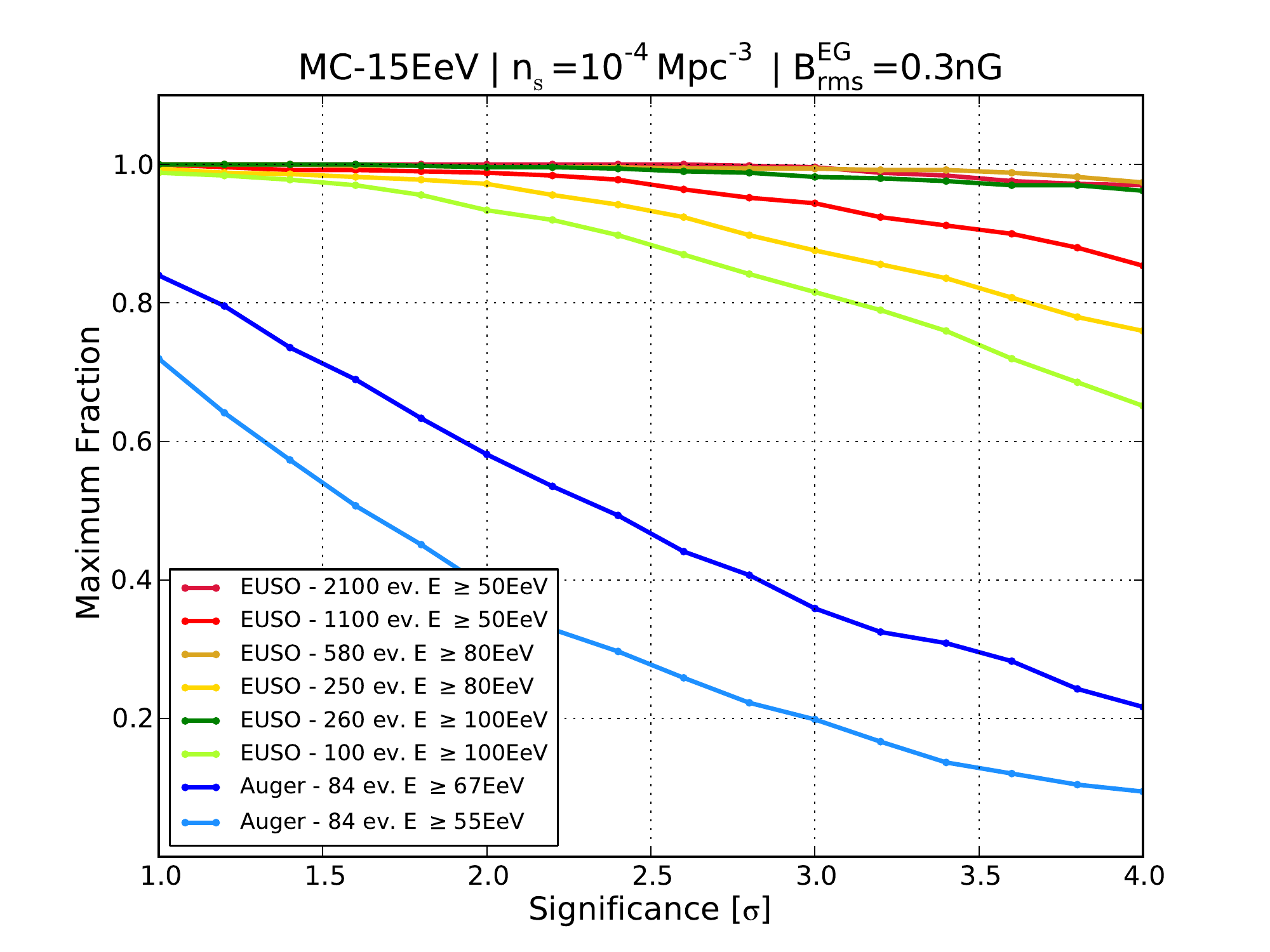}
\caption{Fraction of the 500 realizations of the same scenario as in Fig.~\ref{fig:2ptMC15-4}, which give an anisotropic signal with a significance larger than the significance indicated in abscissa, for different statistics (as indicated).}
\label{fig:2ptMaxMC15-4}
\end{figure}

In Fig.~\ref{fig:2ptMaxMC15-4}, we propose yet another way to look at the results, by showing the fraction of realizations that display an anisotropy stronger than a given significance, as a function of that significance (translated into a number of sigmas, as explained above). 
Since the significance depends on the angular scale, we chose for this plot the angular scale giving the maximum significance. In principle, one should penalize the resulting probability for searching at that particular angular scale (i.e. we should marginalize over angular scales). However, as clearly shown by Fig.~\ref{fig:2ptFractionMC15-4}, the angular scale where the maximum significance occurs is essentially always the same for the model under study (which is true also for the other models), and the curves are very flat over a large range of angles. The departure from anisotropy could thus be searched for \emph{a priori} in this angular range, by fixing the angular scale before any trial, say at 20$^{\circ}$, in which case no penalty factor should be applied. In conclusion, for each given astrophysical scenario tested, the penalization to be applied to Fig.~\ref{fig:2ptMaxMC15-4} and to similar plots for other scenarios should remain limited.

The curves in Fig.~\ref{fig:2ptMaxMC15-4} confirm the previous conclusion: while the Auger statistics is too low to allow the detection of any significant anisotropy in this MC-15EeV scenario, except in some ``lucky'' realizations representing only a small fraction of the possible skies, increasing the exposure up to that of JEM-EUSO will almost certainly lead to the detection of significant anisotropies. This is particularly true if the HiRes/TA energy scale holds, as shown by the three darker lines, almost superimposed close to the 100\% probability line. The JEM-EUSO statistics thus appear appropriate to efficiently constrain such a scenario.

It is worth noting that this scenario is one of the worse possible scenarios as far as the detection of anisotropies is concerned, since it implies that the high-energy particles are mostly heavy nuclei (thus experiencing large deflections), and the source density is large. Nevertheless, a detector able to reach the statistics of JEM-EUSO would be sufficient to detect significant anisotropies in essentially all the realizations of this scenario.

\begin{figure}[ht!]
\centering
\includegraphics[trim=0.75cm 0.25cm 1.5cm 0.5cm,clip,width=0.98\columnwidth]{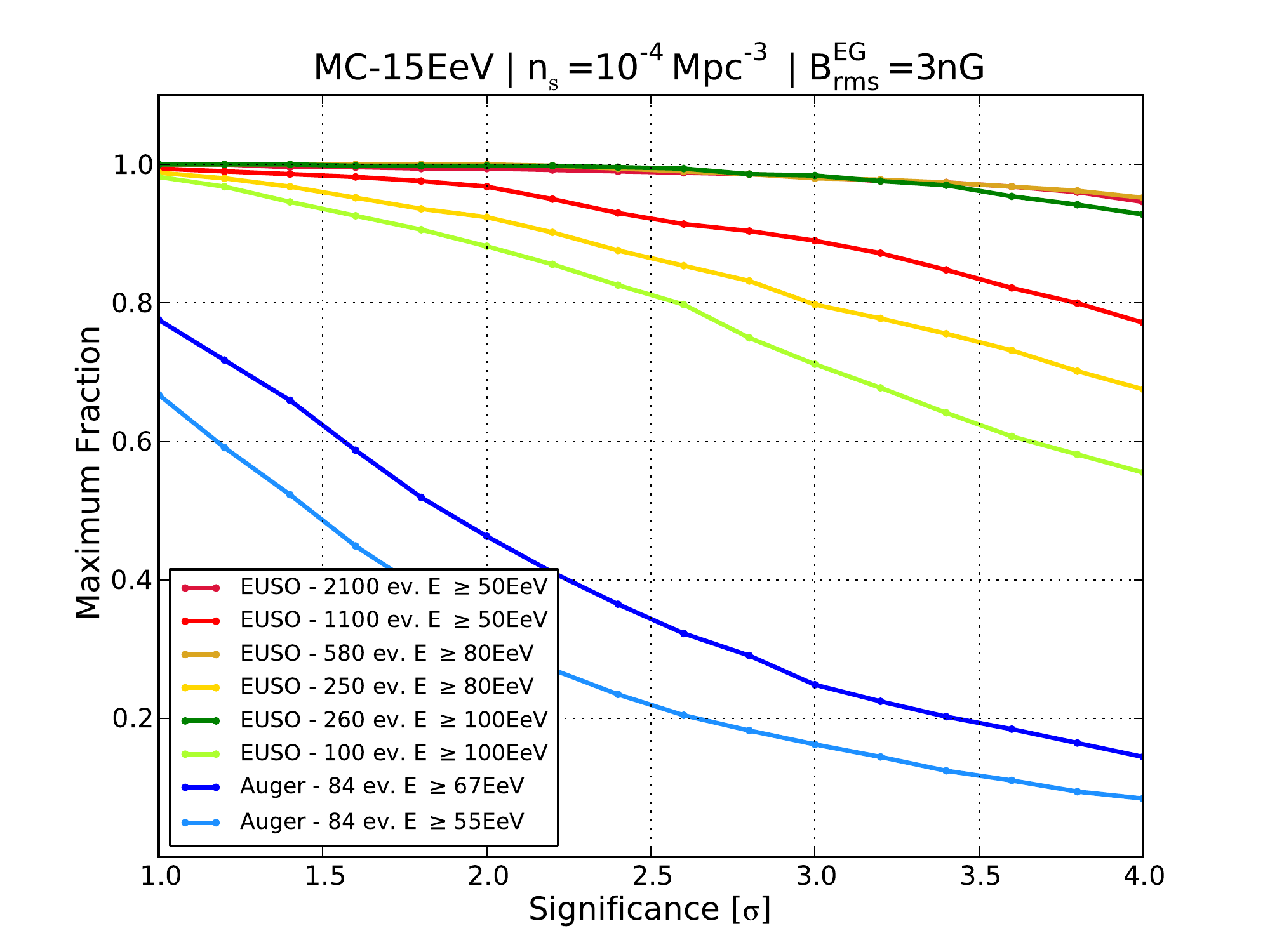}
\caption{Same as Fig.~\ref{fig:2ptMaxMC15-4}, with an extragalactic magnetic field of 3~nG instead of 0.3~nG.}
\label{fig:2ptMaxMC15-4-highEGMF}
\end{figure}

We also studied the effect of the extragalactic magnetic field. As discussed in Sect.~\ref{sec:EGMF}, the most extreme situation may be given by a uniformly distributed, random magnetic field with a root-mean-square amplitude of 3~nG. Figure~\ref{fig:2ptMaxMC15-4-highEGMF} shows how the results of Fig~\ref{fig:2ptMaxMC15-4} are modified in this case.
As expected, the fraction of realizations that display significant anisotropy are somewhat reduced, compared to the case with a lower magnetic field. This is due to the larger dispersion of the UHECR arrival directions at the entrance of the Galaxy, which in turn results in a more widely spread distribution of arrival directions on Earth. However, the conclusions remain essentially the same. More than 90\% of the realizations of the scenario under study display an anisotropy with a significance larger than 4$\sigma$ in the case of the HiRes/TA energy scale, and almost 80\% of them do so above 50~EeV in the case of the Auger energy scale.


\subsubsection{Low source density: $n_{\mathrm{s}} = 10^{-6}\,\mathrm{Mpc}^{-3}$.}

\begin{figure*}[hp!]
\centering
\includegraphics[trim=0.75cm 0cm 1.5cm 0cm,clip,width=\columnwidth]{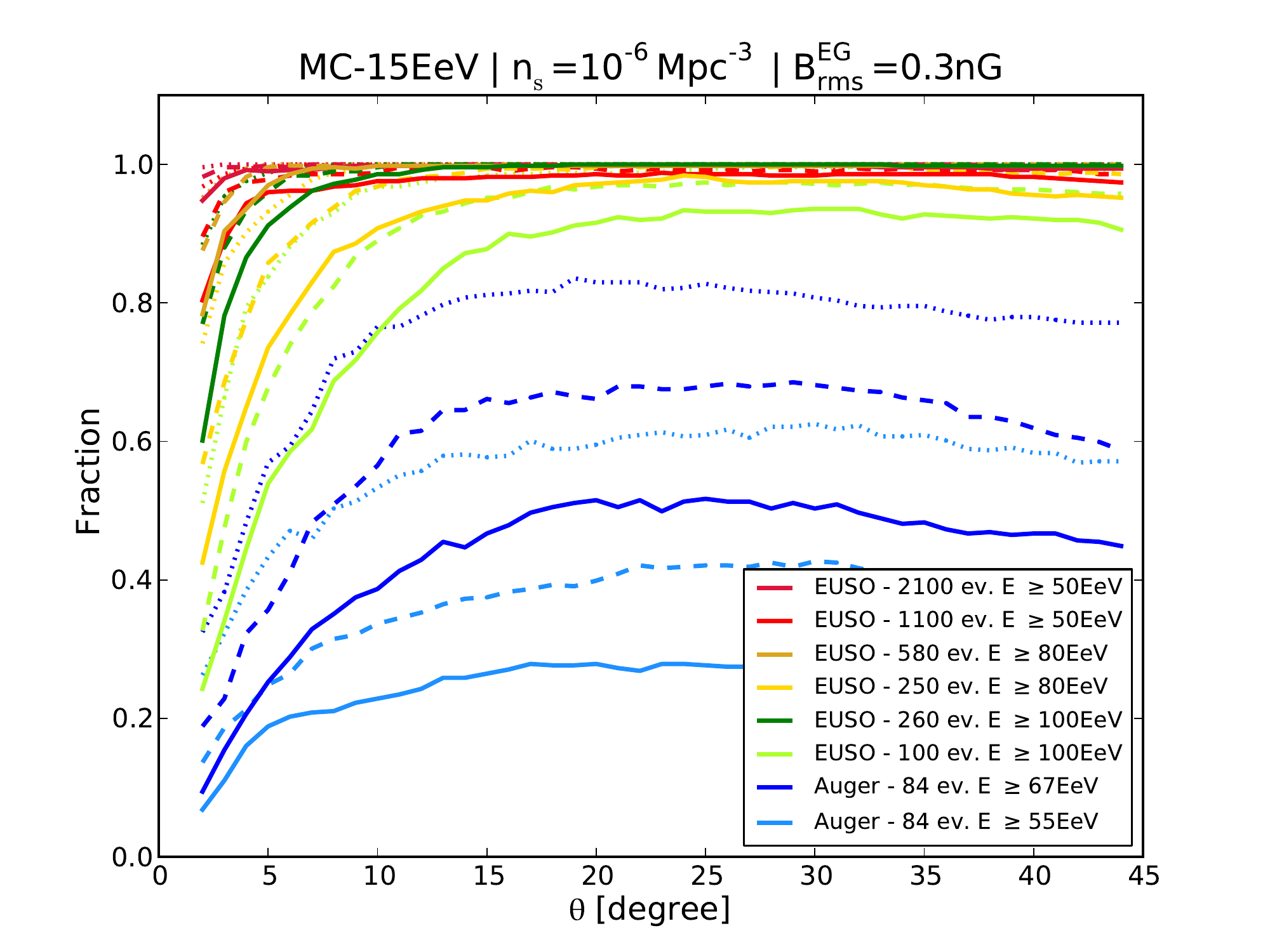}
\includegraphics[trim=0.75cm 0cm 1.5cm 0cm,clip,width=\columnwidth]{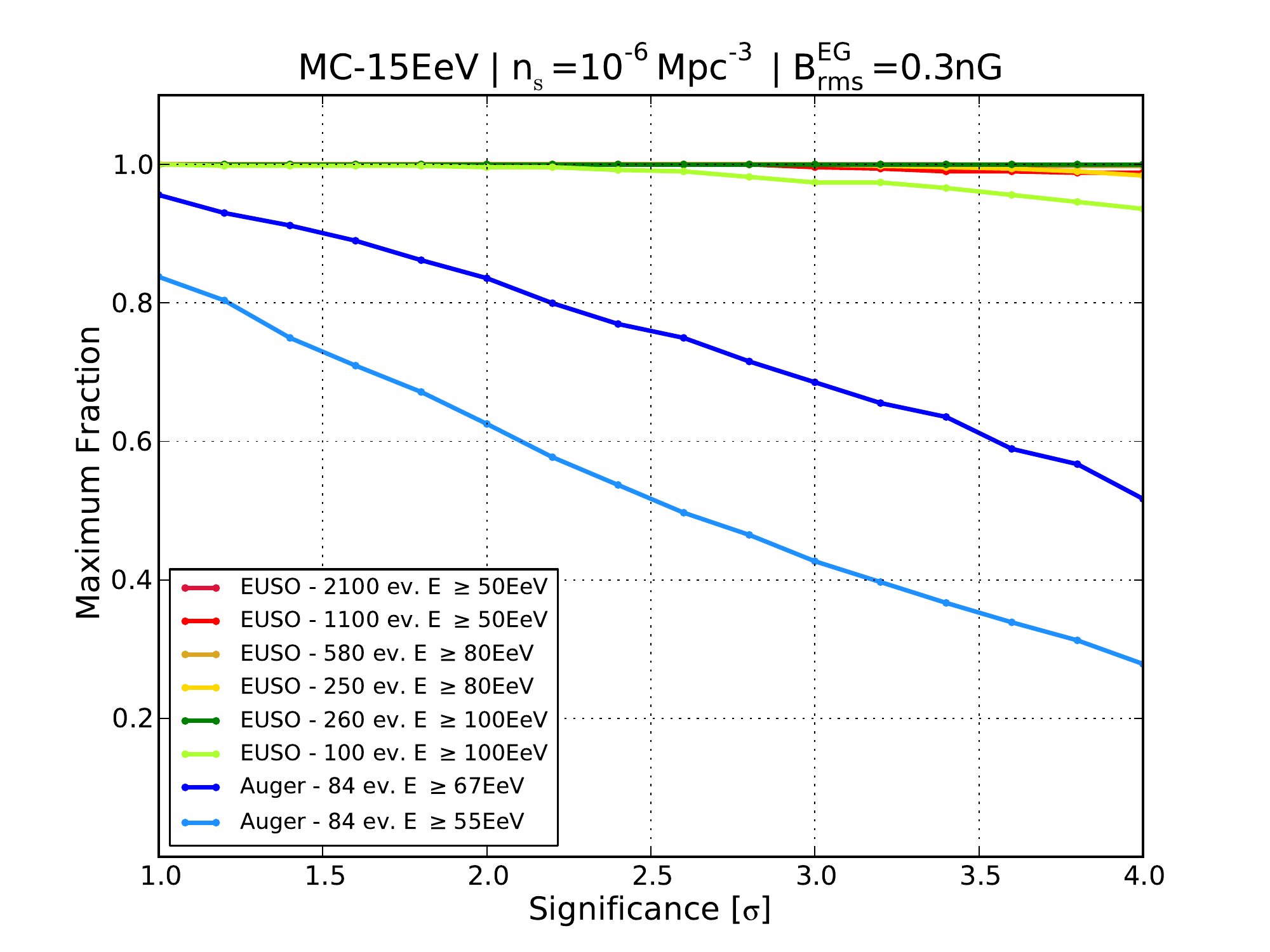}
\caption{Same as Fig.~\ref{fig:2ptFractionMC15-4} (left) and Fig.~\ref{fig:2ptMaxMC15-4} (right), but with the source density $n_{\mathrm{s}} = 10^{-6}\,\mathrm{Mpc}^{-3}$.}
\label{fig:2ptFractionMC15-6}

\includegraphics[trim=0.75cm 0cm 1.5cm 0cm,clip,width=\columnwidth]{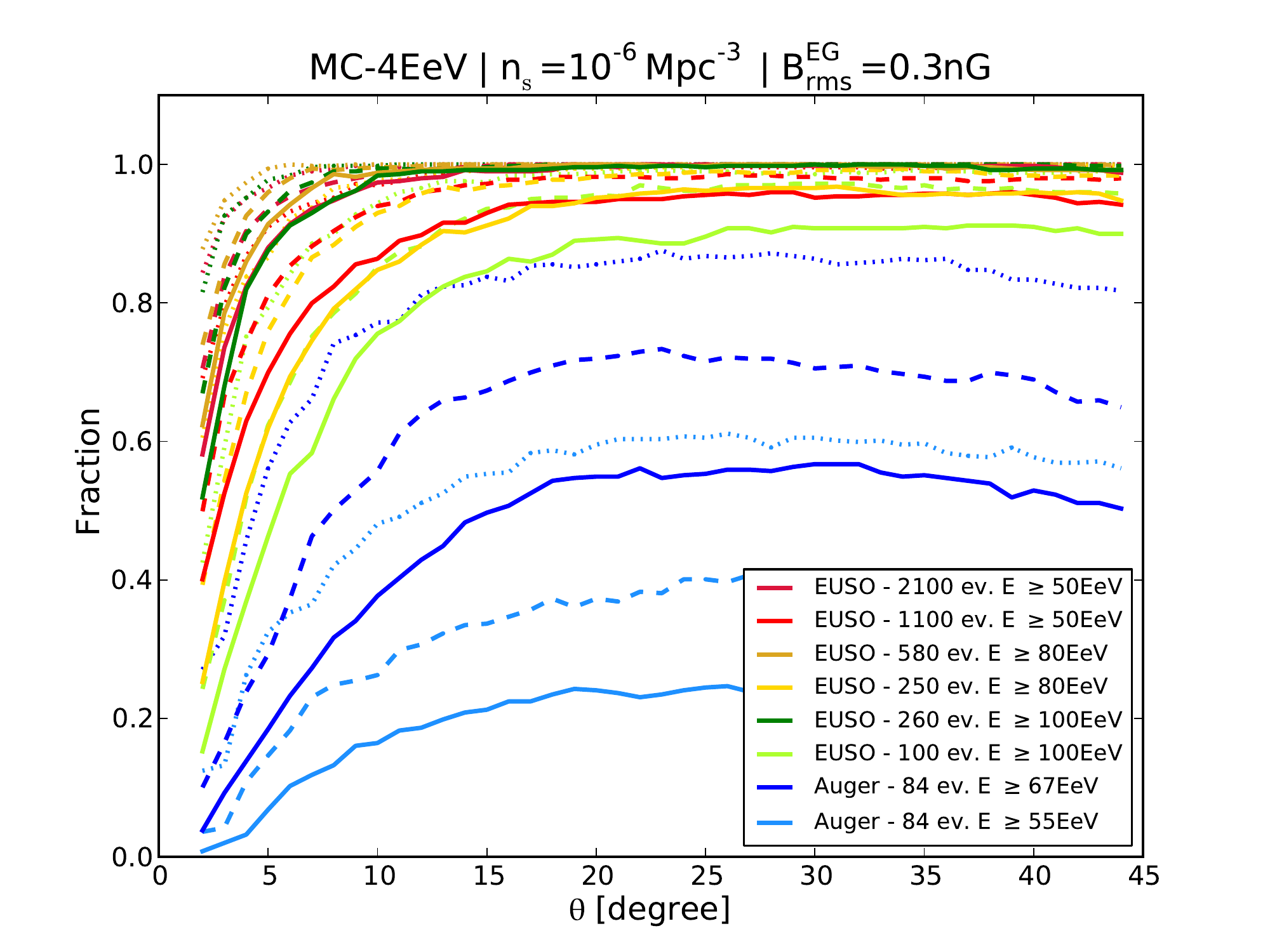}
\includegraphics[trim=0.75cm 0cm 1.5cm 0cm,clip,width=\columnwidth]{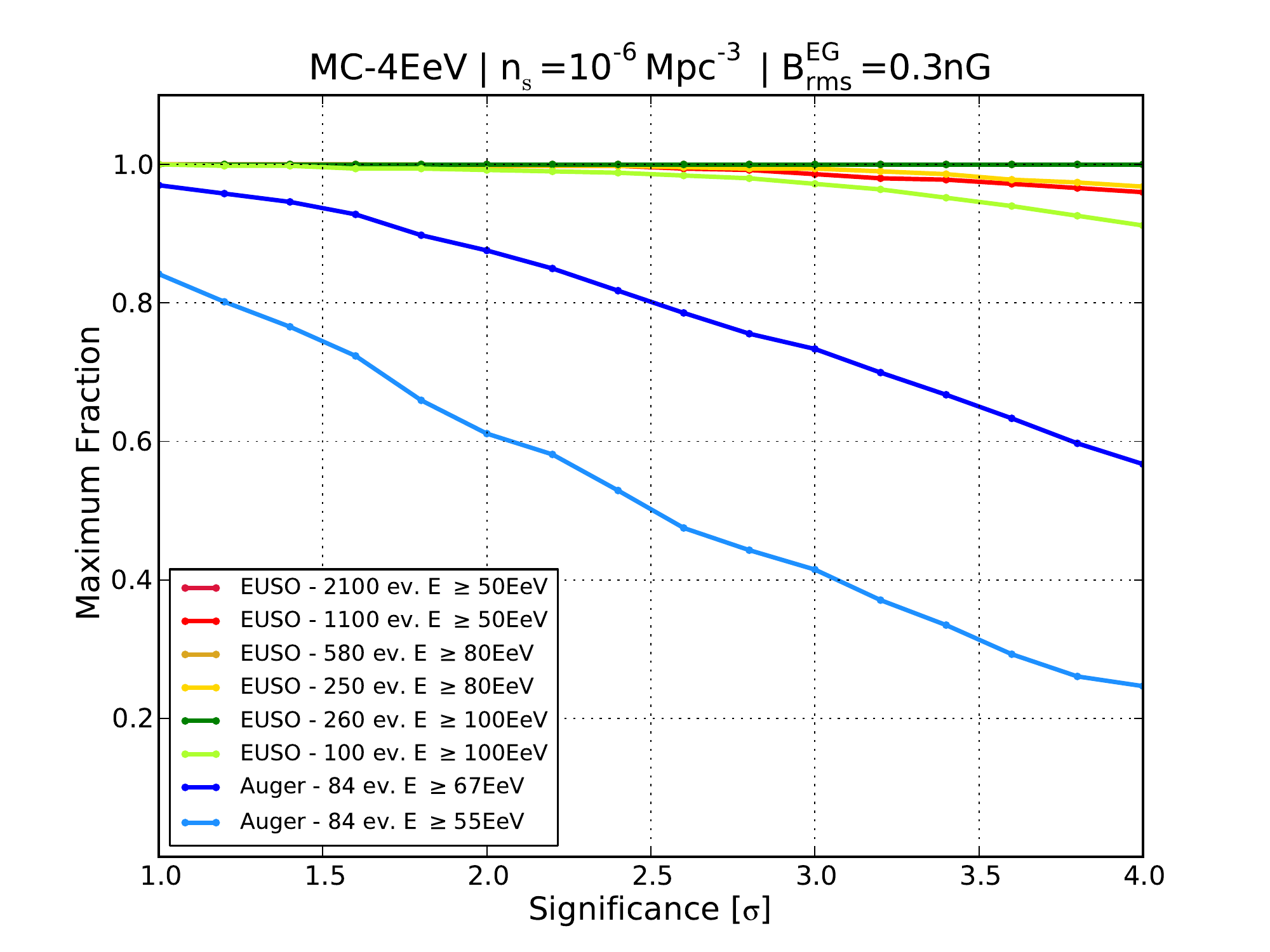}
\caption{Same as Fig.~\ref{fig:2ptFractionMC15-6}, but with a maximum proton energy of $E_{\max} = 4$~EeV.}
\label{fig:2ptFractionMC4-6}

\includegraphics[trim=0.75cm 0cm 1.5cm 0cm,clip,width=\columnwidth]{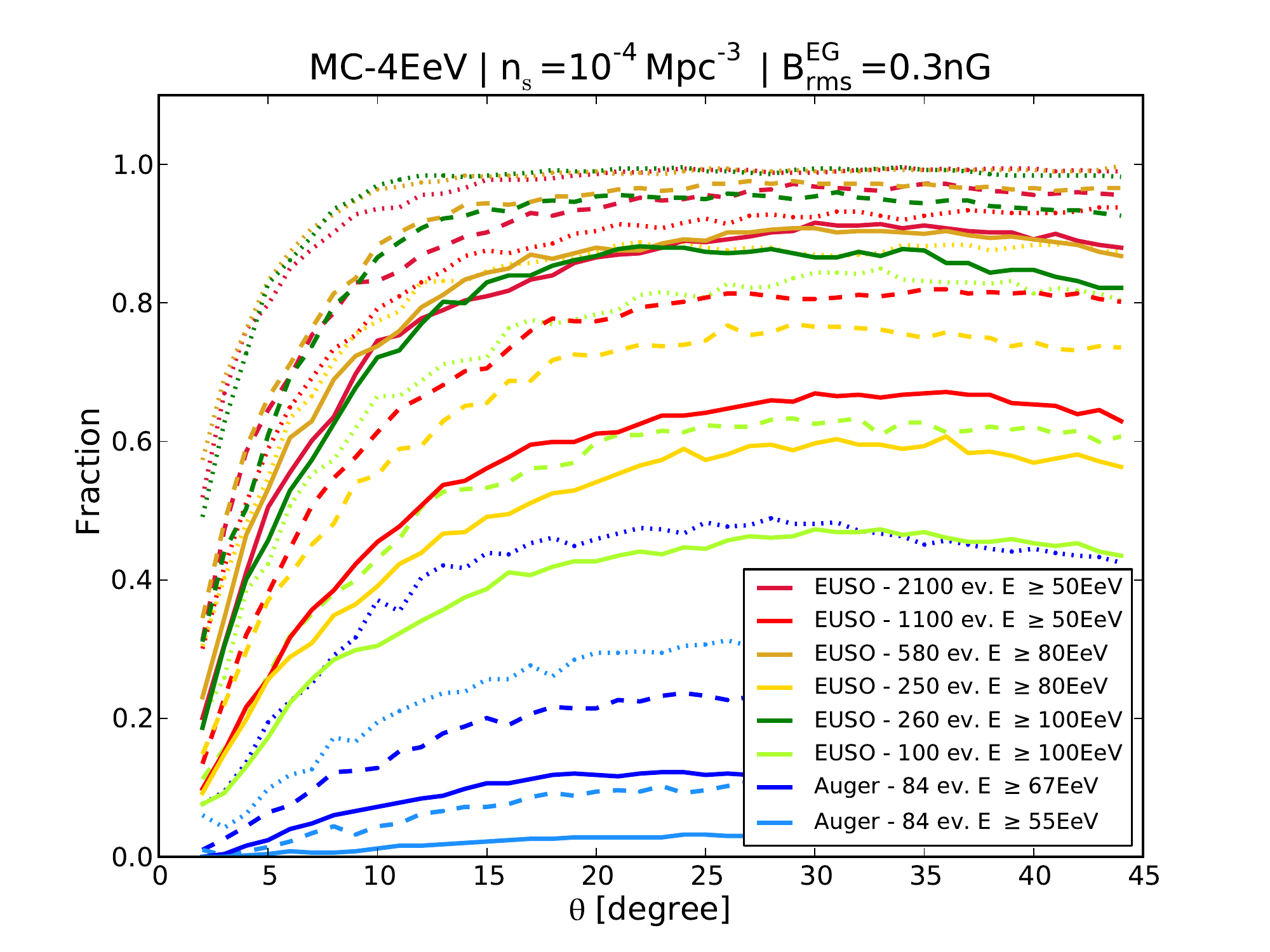}
\includegraphics[trim=0.75cm 0cm 1.5cm 0cm,clip,width=\columnwidth]{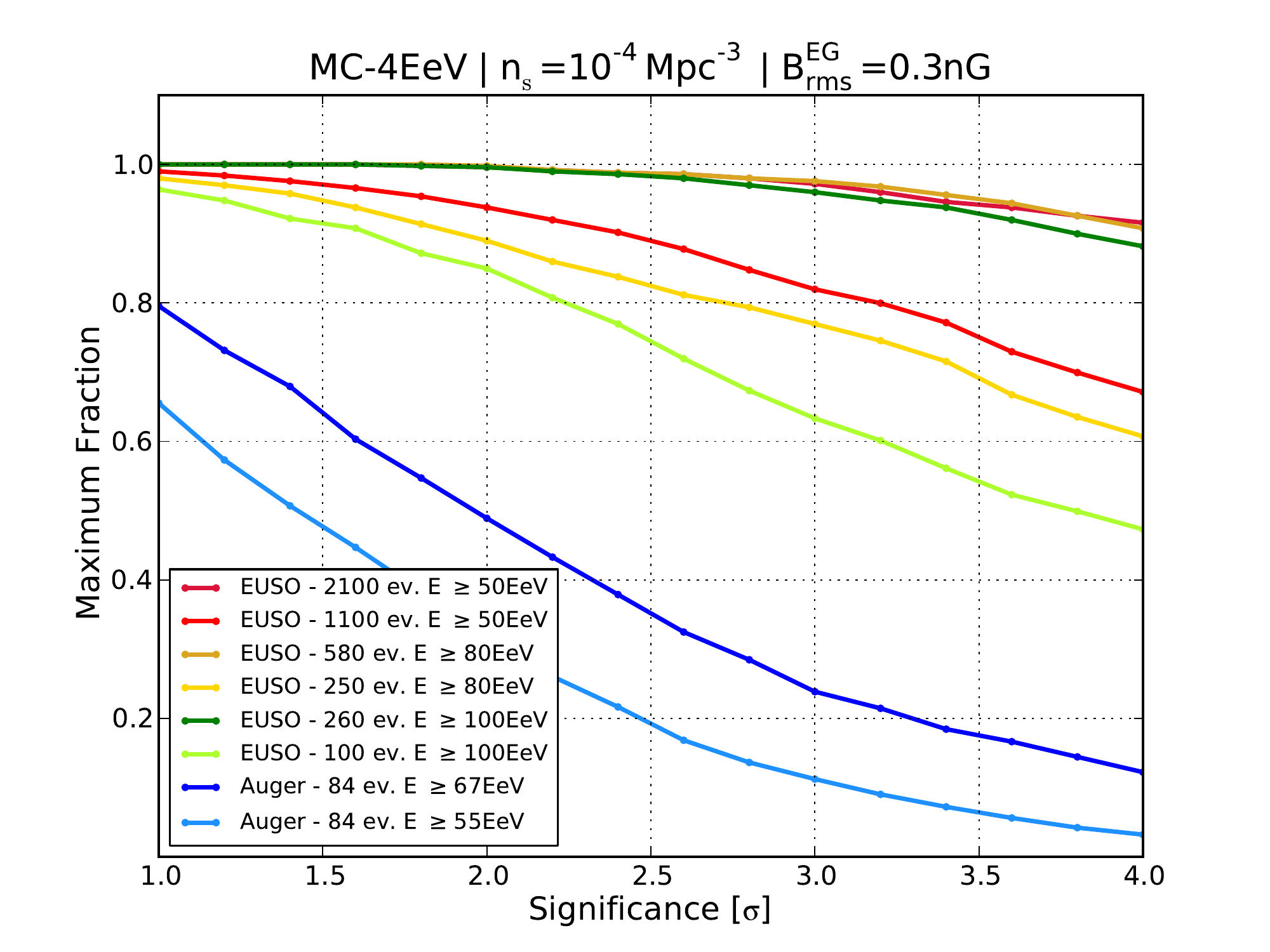}
\caption{Same as Fig.~\ref{fig:2ptFractionMC4-6}, but with the source density $n_{\mathrm{s}} = 10^{-4}\,\mathrm{Mpc}^{-3}$.}
\label{fig:2ptFractionMC4-4}
\end{figure*}

The same plots as presented in the previous subsection can be built for all the scenarios. In Fig.~\ref{fig:2ptFractionMC15-6}, we show the results corresponding to the same low proton-$E_{\max}$ model, MC-15EeV, but with a low source density: $n_{\mathrm{s}} = 10^{-6}\,\mathrm{Mpc}^{3}$. As expected, the different realizations display more significant anisotropies overall, as a result of the smaller number of sources contributing at the highest energies, and thus of the larger multiplicity of the most intense sources.

The Auger data are still compatible with such a scenario, since a majority of realizations ($\sim 60$\%) are not expected to have a two-point correlation function that departs from the isotropic expectation by more than $3\sigma$, and $\sim 40\%$ of the realizations remain within $2\sigma$ of the isotropic expectation. It is interesting to note, however, that a confirmation of the HiRes/TA energy scale would bring some tension between this astrophysical scenario and the Auger data, since only a bit more than one third of the realizations would then be compatible with a two-point correlation function less than $3\sigma$ away from the isotropic expectation.

Turning to the JEM-EUSO-like exposure, it appears clearly that such a scenario could be easily, and severely constrained with the resulting statistics, if no significant anisotropy were detected. Essentially 100\% of the realizations display anisotropies with a significance larger than $4\sigma$, at all energies considered.


\subsubsection{Comparison between the MC-15EeV and MC-4EeV models}

The results shown above correspond to the MC-15EeV model, i.e. to the intermediate value of the maximum proton energy achieved in the source, $E_{\max} = 15$~EeV. Very similar conclusions can be reached for the lower value, $E_{\max} = 4$~EeV, i.e. for the MC-4EeV model. For comparison, we show in Figs.~\ref{fig:2ptFractionMC4-6} and~\ref{fig:2ptFractionMC4-4} the fractions of realizations of the MC-4EeV model as a function of the significance of the measured anisotropy for the two extreme source densities, respectively low and high.

For the lowest density, again, all realizations display a very significant anisotropy, whatever the energy and angular scale. At the highest source density, the significance is reduced, as expected, even though a large majority of the realizations would be very significantly anisotropic, especially in the case of the HiRes/TA energy scale.

However, an interesting feature distinguishes both scenarios. The main difference between the MC-4EeV and MC-15EeV models is the presence of relatively light nuclei up to higher energies in the latter case, due to the higher energy of the proton cutoff. In particular, C, N and O nuclei, which are rather abundant in the interstellar medium, and thus in the assumed cosmic-ray source composition, are accelerated in the sources up to respectively 6, 7 and 8 times $E_{\max}$. These intermediate nuclei are thus present up to $\sim 100$~EeV in the MC-15EeV scenarios, while they disappear at $\sim 30$~EeV in the MC-4EeV scenarios.

\begin{figure*}[ht!]
\centering
\includegraphics[trim=2.75cm 0.75cm 1.25cm 2cm,clip,width=0.98\columnwidth]{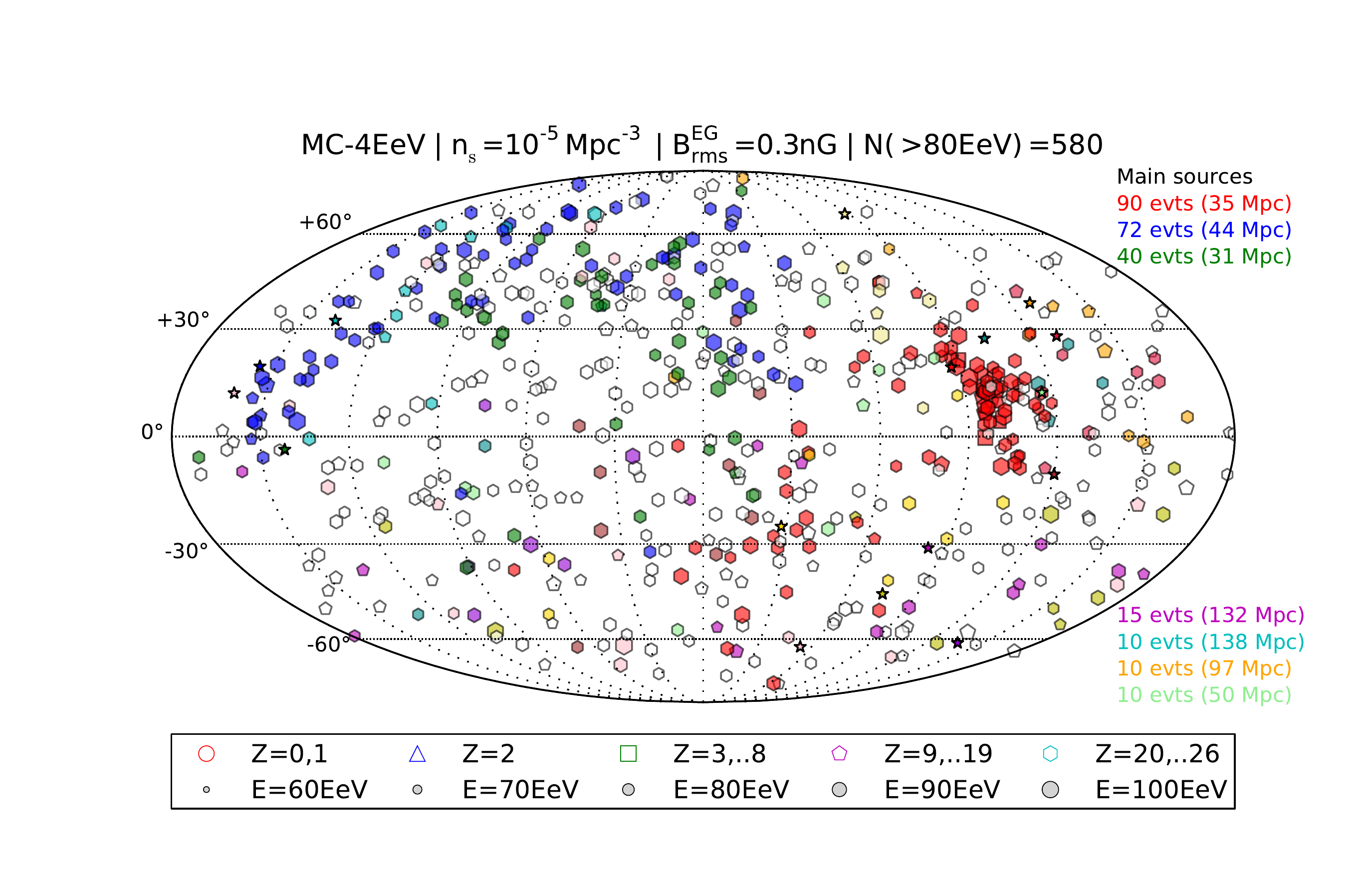}
\includegraphics[trim=2.75cm 0.75cm 1.25cm 2cm,clip,width=0.98\columnwidth]{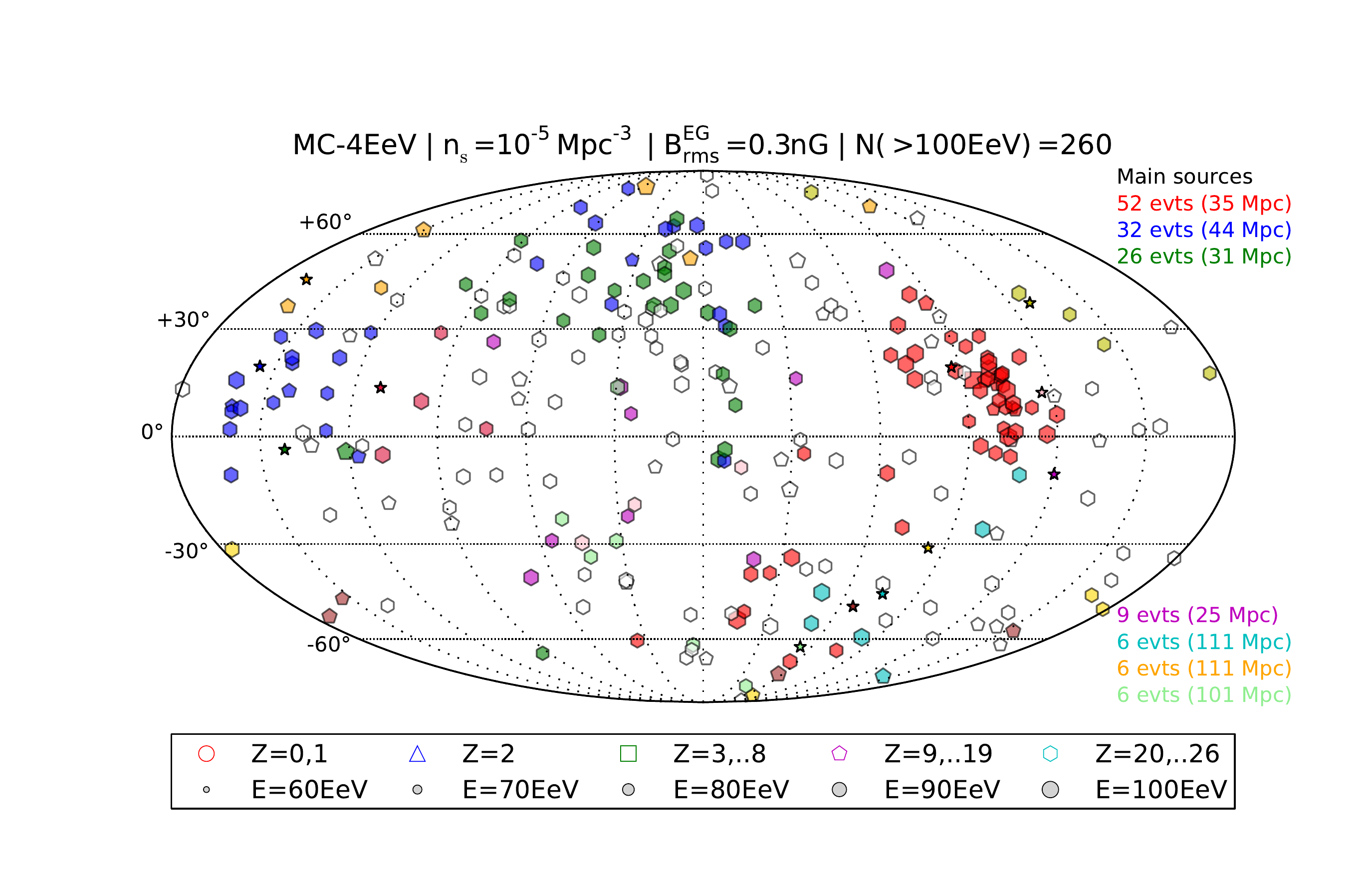}
\caption{Typical sky maps corresponding to the MC-4EeV model with a source density $n_{\mathrm{s}} = 10^{-5}\,\mathrm{Mpc}^{-3}$, for the JEM-EUSO statistics with a threshold at 80~EeV (left) and 100~EeV (right), assuming the HiRes/TA energy scale.}
\label{fig:skyMapMC4-5}

\includegraphics[trim=2.75cm 0.75cm 1.25cm 0.75cm,clip,width=0.98\columnwidth]{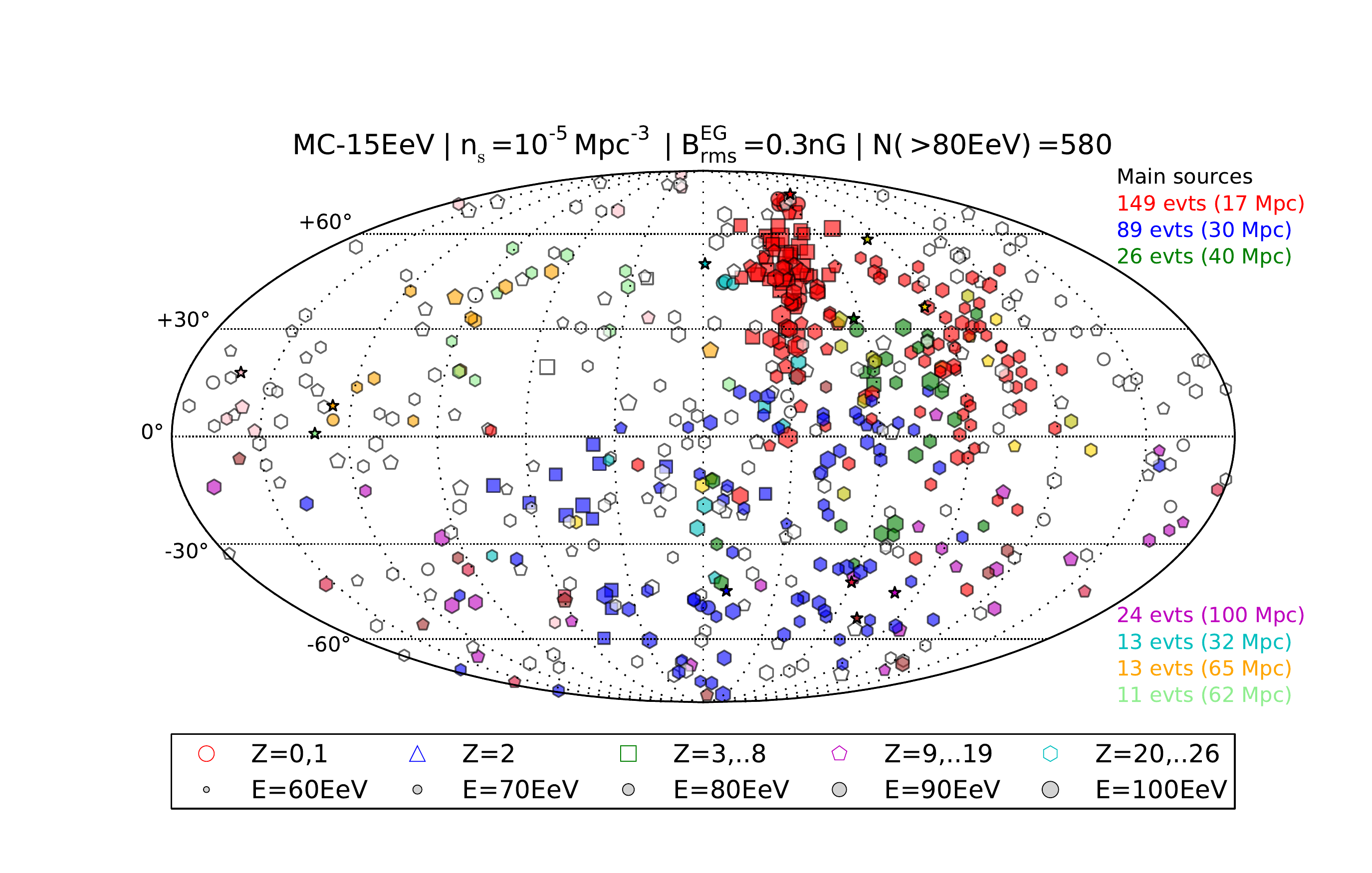}
\includegraphics[trim=2.75cm 0.75cm 1.25cm 0.75cm,clip,width=0.98\columnwidth]{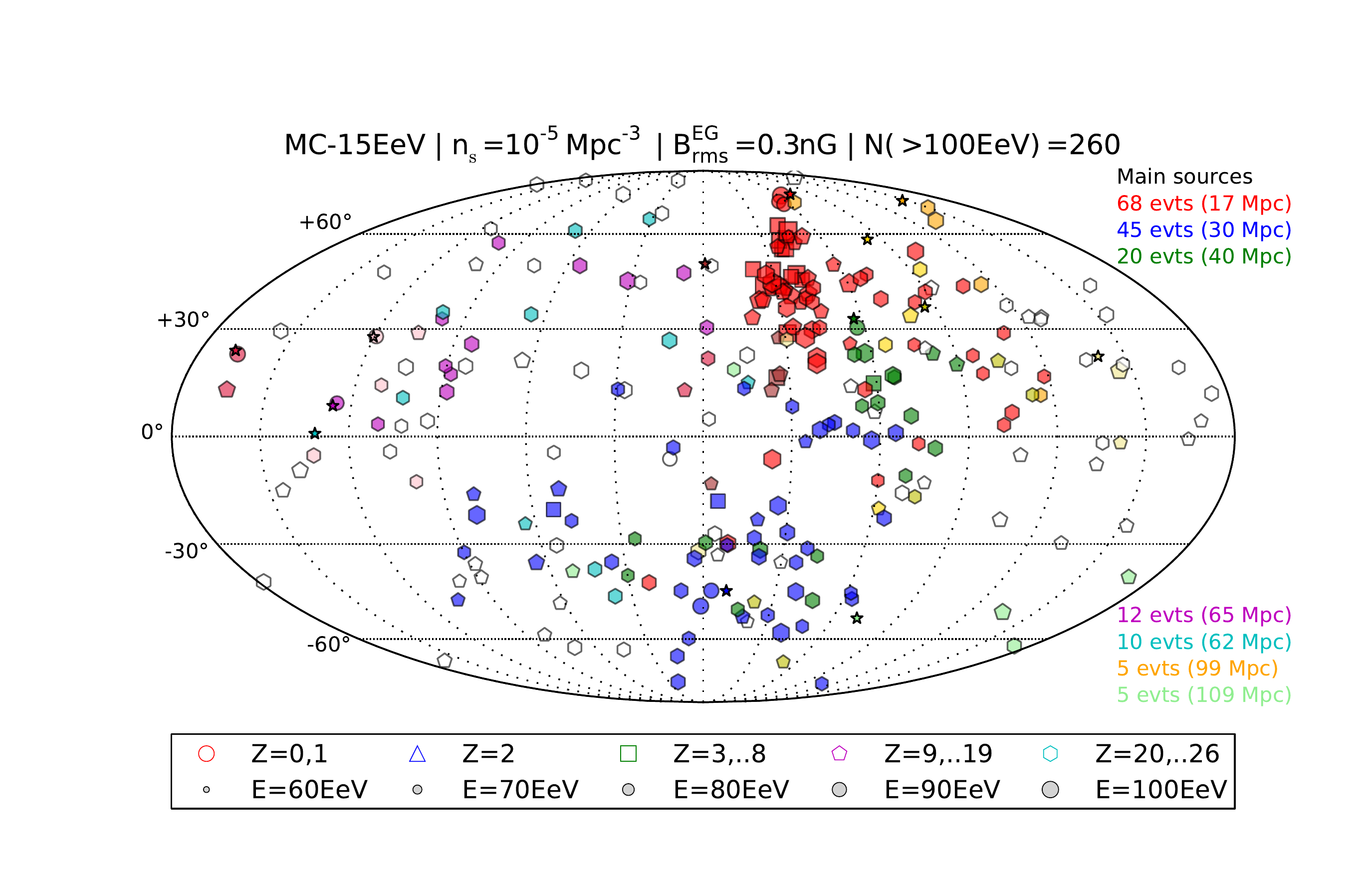}
\caption{Typical sky maps corresponding to the MC-15EeV model with a source density $n_{\mathrm{s}} = 10^{-5}\,\mathrm{Mpc}^{-3}$, for the JEM-EUSO statistics with a threshold at 80~EeV (left) and 100~EeV (right), assuming the HiRes/TA energy scale.}
\label{fig:skyMapMC15-5}

\includegraphics[trim=2.75cm 0.75cm 1.25cm 0.75cm,clip,width=0.98\columnwidth]{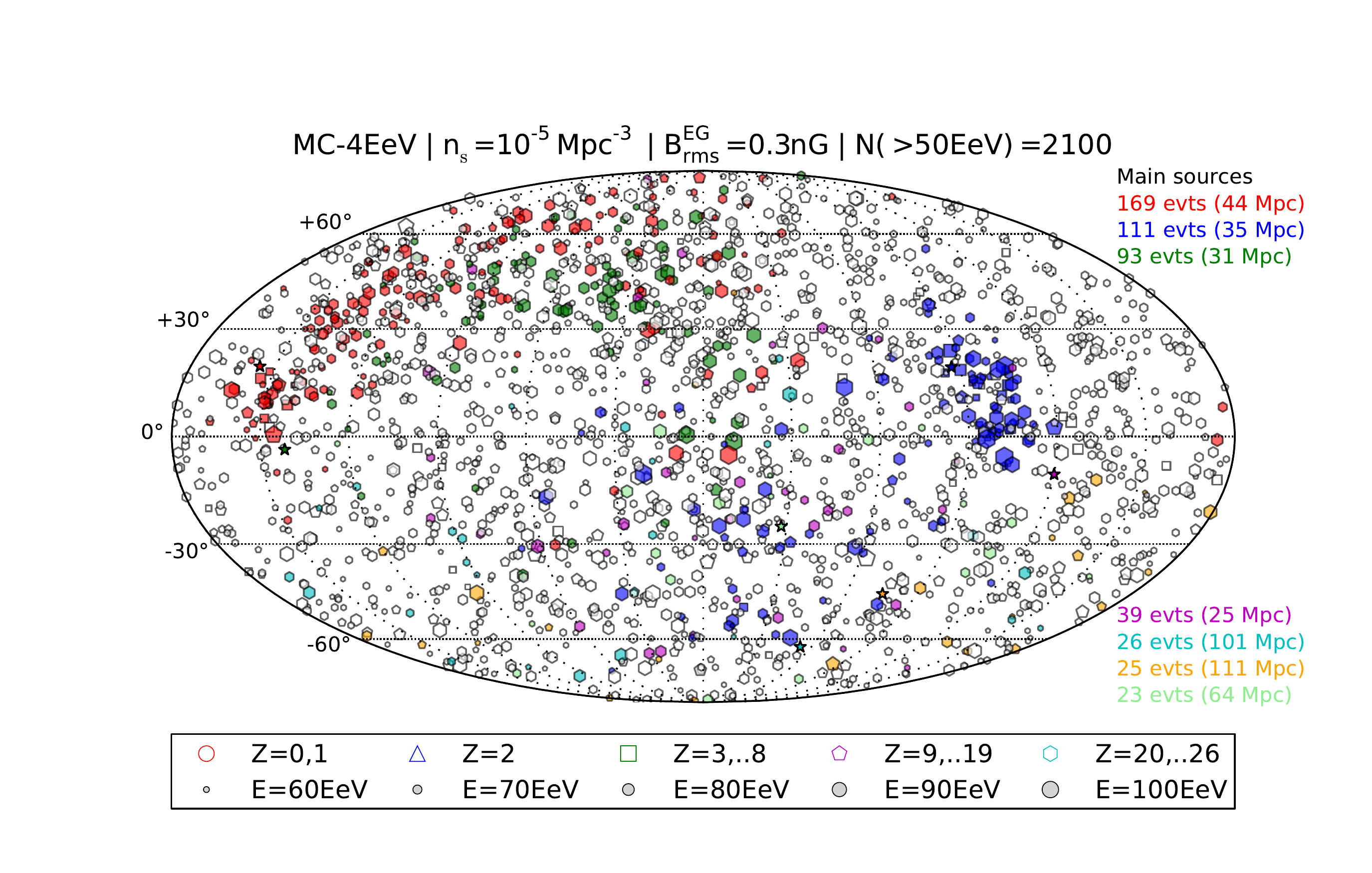}
\includegraphics[trim=2.75cm 0.75cm 1.25cm 0.75cm,clip,width=0.98\columnwidth]{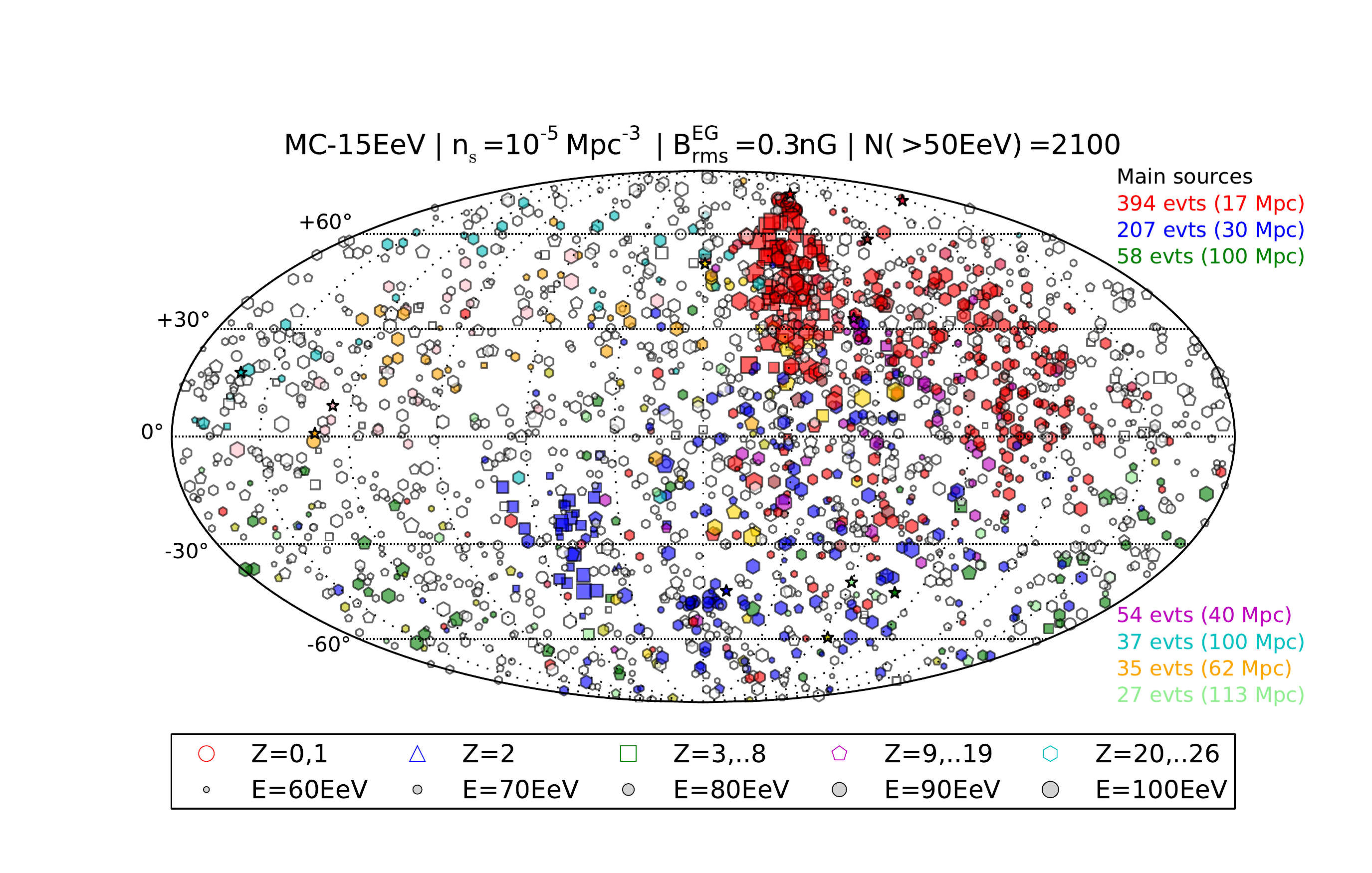}
\caption{Typical sky maps corresponding to the MC-4EeV model (left) and the MC-15EeV model (right) with a source density $n_{\mathrm{s}} = 10^{-5}\,\mathrm{Mpc}^{-3}$, for the JEM-EUSO statistics with a threshold at 50~EeV, assuming the HiRes/TA energy scale.}
\label{fig:skyMapMC415-5-50}
\end{figure*}

As an example, we show in Figs.~\ref{fig:skyMapMC4-5} and~\ref{fig:skyMapMC15-5} typical sky maps built from the MC-4EeV and MC-15EeV models at the intermediate source density, assuming the JEM-EUSO statistics with an energy threshold at 80~EeV (left figures) and 100~EeV (right figures). 
The particular realizations chosen for these sky maps are right in the middle of the distribution of the 500 realizations of the corresponding scenario relatively to the significance of their anisotropy (as measured with the two-point correlation function). In other words, half of the realizations show stronger -- i.e. more significant -- anisotropies, while the other half show weaker anisotropies. As expected, the symbols on the maps corresponding to the MC-4EeV model are mostly hexagons (Fe or sub-Fe nuclei), while many squares and pentagons (light and intermediate nuclei) contribute to the clustering on the maps corresponding to the MC-15EeV model.

\begin{figure*}[ht!]
\centering
\includegraphics[trim=0.75cm 0.25cm 1.5cm 0.5cm,clip,width=0.98\columnwidth]{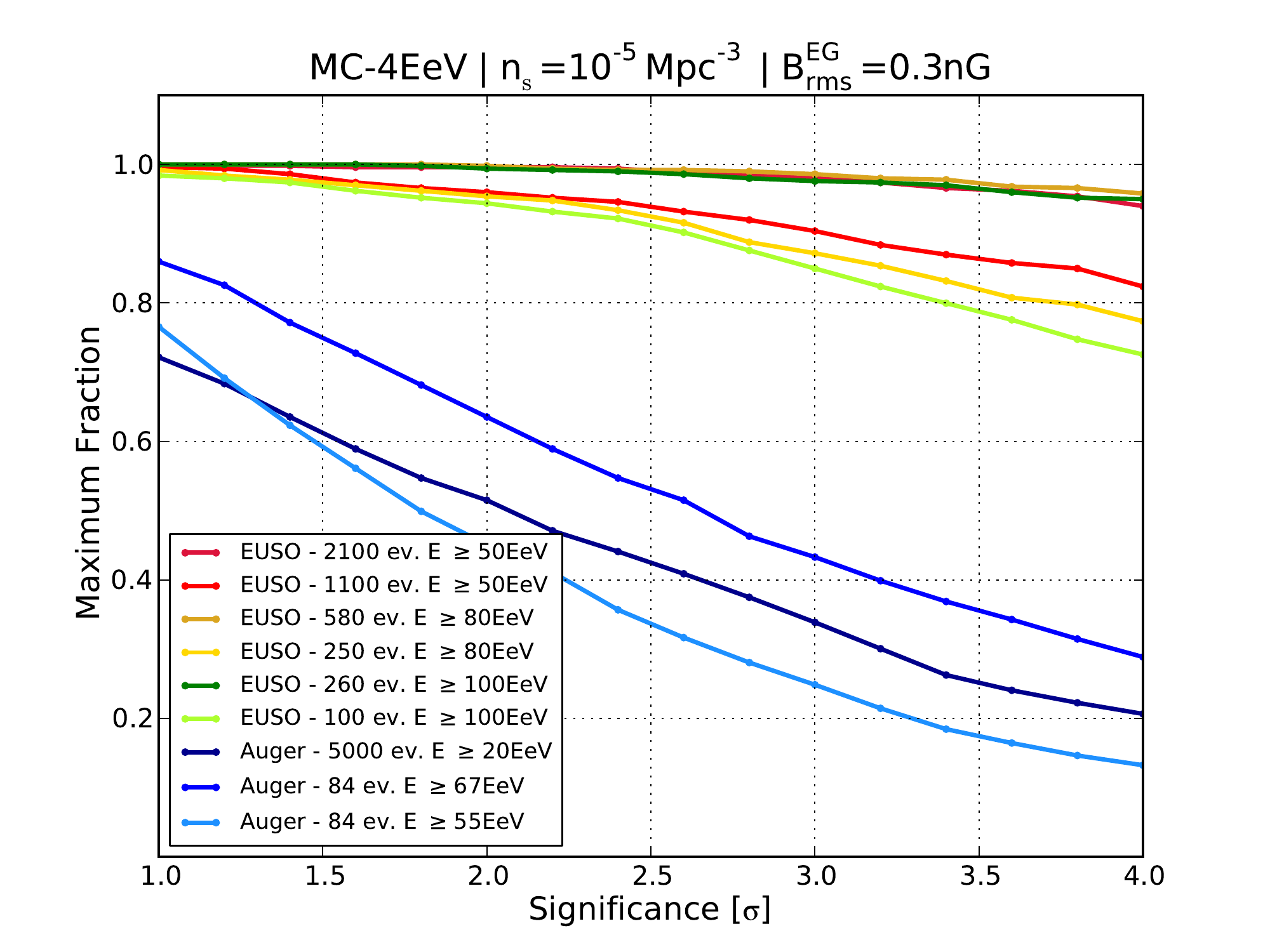}
\includegraphics[trim=0.75cm 0.25cm 1.5cm 0.5cm,clip,width=0.98\columnwidth]{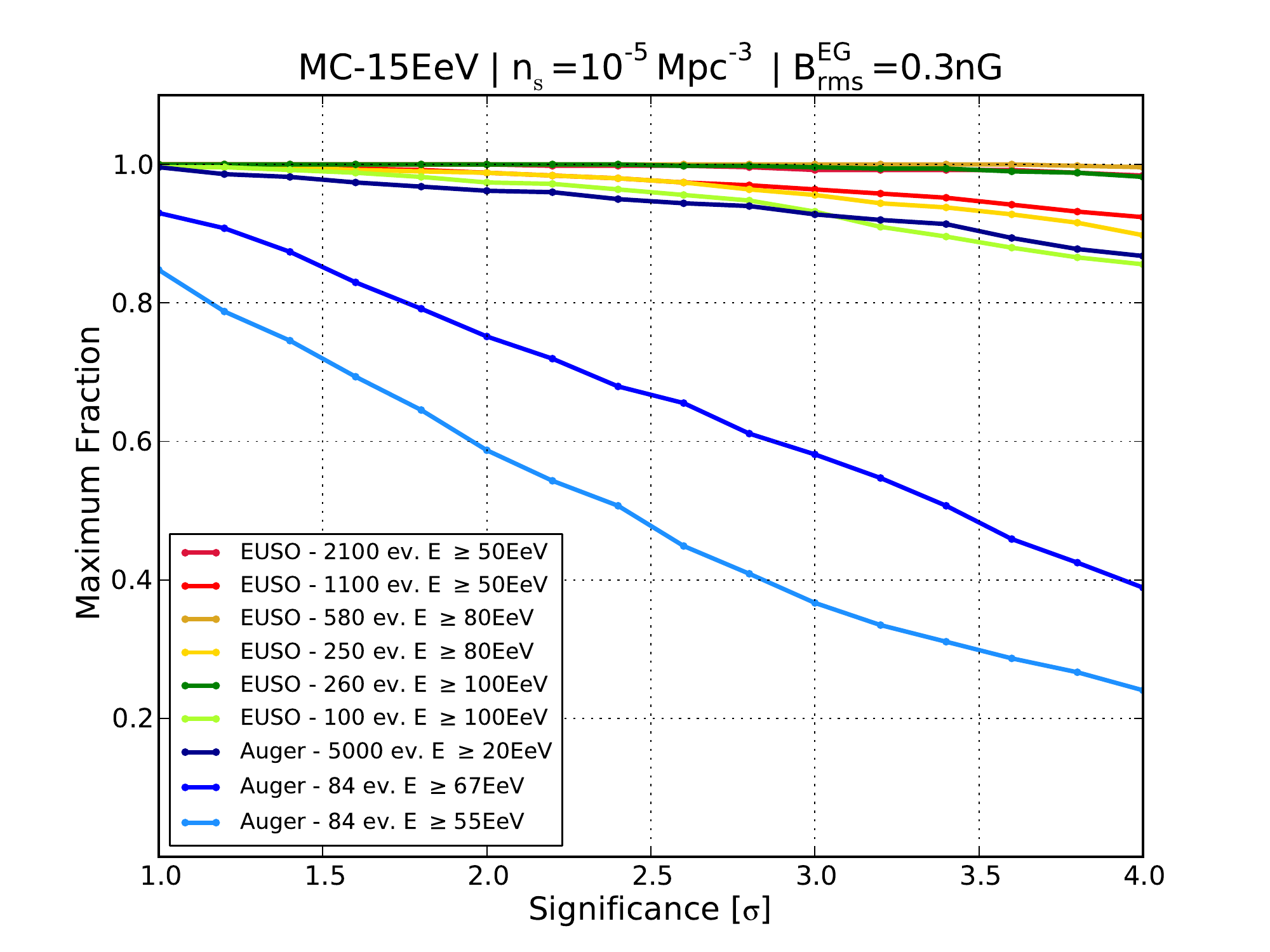}
\caption{Fractions of the 500 realizations of the MC-4EeV model (left) and of the MC-15EeV model (right) with $n_{\mathrm{s}} = 10^{-5}\,\mathrm{Mpc}^{-3}$, whose two-point correlation function shows an anisotropy with a significance larger than the significance indicated in abscissa, for different statistics (as indicated).}
\label{fig:2ptMaxMC415-5}
\end{figure*}

Going down to lower energies, the difference becomes obvious. In Fig.~\ref{fig:skyMapMC415-5-50}, we show typical sky maps for the same scenarios and the same realizations as above, with an energy threshold at 50~EeV. In the case of the MC-15EeV model (right), many protons and light nuclei are observed near their sources (in red and in blue), while the Fe events are more loosely distributed in the case of the MC-4EeV model (left). In passing, we note that the source whose events are colored in red in Fig.~\ref{fig:skyMapMC4-5} appears in blue on the left sky map of Fig.~\ref{fig:skyMapMC415-5-50}, and vice-versa. Since the color code is associated with the rank of the source in terms of apparent luminosity, this means that the sources with the highest and the second highest luminosities have been interchanged between 50~EeV and 80~EeV. Although the sources are at different distances (35~Mpc and 44~Mpc), the difference between the effect of the GZK energy losses at these distances and energies is not significant. The main reason for the change in the relative luminosity is the magnification effect due to the Galactic magnetic field, emphasized in Sect.~\ref{sec:deflectionMaps}. This effect is particularly large for Fe nuclei in this energy range, due to their intermediate rigidity. As a matter of fact, the source which appears in blue in Fig.~\ref{fig:skyMapMC4-5} is located in a region where the magnification factor is larger than that of the red source, and even more so at rigidity $R = 50/26$~EV than at rigidity $R = 80/26$~EV. The source at 44~Mpc on the far left of the sky map is thus largely ``boosted'' at 50~EeV, compared to that at 35~Mpc in the middle right of the map. This is but one of the interesting features associated with the rigidity-dependent magnification effects (see also Sect.~\ref{sec:deflectionMaps}), whose consequences and astrophysical interest will be studied in a separate paper.

Coming back to the comparison between the MC-4EeV and MC-15EeV models, it is instructive to look at the fraction of the anisotropic realizations of these models as a function of the significance of their anisotropy (as measured by the two-point correlation function). This is shown in Fig.~\ref{fig:2ptMaxMC415-5}. The curves on the right panel correspond to the MC-15EeV model and appear intermediate between those shown for the same model in Figs.~\ref{fig:2ptMaxMC15-4} and~\ref{fig:2ptFractionMC15-6} (right), as expected from the intermediate source density. The curves on the left panel correspond to the MC-4EeV model, and show that typically 5--10\% fewer realizations display a given level of anisotropy with the JEM-EUSO statistics than in the previous case, and $\sim 25$\% with the Auger statistics, due to the absence of light nuclei. Overall, the same structure can nevertheless be observed and the same conclusions can be drawn.

However, in addition to the statistics and energy thresholds shown in this type of figures for the other models, we have shown in the two plots of Fig.~\ref{fig:2ptMaxMC415-5} an additional curve corresponding to an extended Auger detector, which would collect 5,000 events above 20~EeV (corresponding roughly to a total exposure of 100,000~$\mathrm{km}^{2}\,\mathrm{sr}\,\mathrm{yr}$). The difference between the MC-4EeV and MC-15EeV is striking at this energy, with this large statistics. In the case of the MC-4EeV, 5000 events above 20 EeV do not produce a more significant anisotropy than 84 events above 55~EeV or 67~EeV -- or, more precisely, significant anisotropy is not obtained in a larger number of realizations of the MC-4EeV model (with $n_{s} = 10^{-5}\,\mathrm{Mpc}^{-3}$) for 5000 events at 20~EeV than for 84 events at 55--67~EeV. In contrast, 5000~events above 20~EeV lead to an anisotropy as significant (or lead as often to a significant anisotropy) as the JEM-EUSO statistics above 50, 80 or 100~EeV, in the case of the MC-15EeV model. This is due to the presence of protons in sufficient numbers at this low energy to produce small-scale anisotropies, despite the much more distant horizon and thus the larger number of sources contributing to the flux.

\begin{figure}[ht!]
\centering
\includegraphics[trim=0.75cm 0cm 1.5cm 0.5cm,clip,width=\columnwidth]{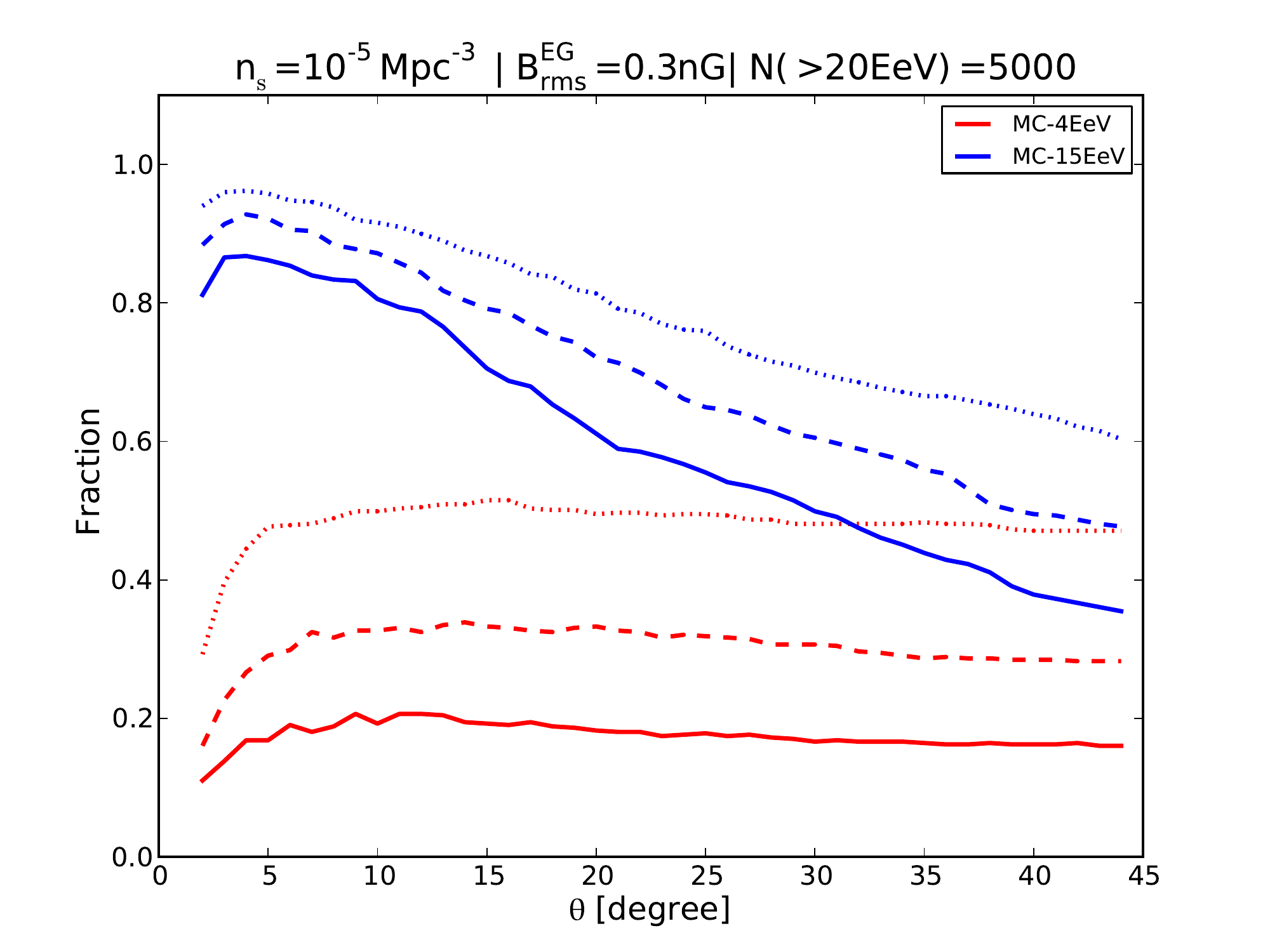}
\caption{Fraction of the 500 realizations of the MC-4EeV (red) and MC-15EeV (blue) scenarios, with $n_{\mathrm{s}} = 10^{-5}\,\mathrm{Mpc}^{-3}$, which give an anisotropic signal with a significance larger than $2 \sigma$ (dotted lines), $3 \sigma$ (dashed lines), or $4 \sigma$ (plain line), for a statistics of 5000 events above 20~EeV, with the coverage map of Auger (see text).}
\label{fig:2ptFractionMC15-5-20}
\end{figure}

This interpretation is confirmed by the curves plotted in Fig.~\ref{fig:2ptFractionMC15-5-20}, where we show the fraction of the realizations of the MC-4EeV (red) and MC-15EeV (blue) models which display anisotropies with a significance of 2$\sigma$, 3$\sigma$, 4$\sigma$, with the extended Auger statistics at 20~EeV, as a function of angular scale. Clearly the angular scale where a large fraction of realizations of the MC-15EeV model shows a large anisotropy corresponds to small deflections (thus associated with light nuclei), whereas this is not the case for the MC-4EeV model, due to the dominance of heavier nuclei with larger deflections.

In conclusion, even though both scenarios are compatible with the current data, they could in principle be distinguished through their different anisotropy patterns as a function of energy with data sets collected by experiments having a larger exposure.

\subsection{Proton-dominated models: MC-high and pure-p}

In the previous subsection, we considered models in which the UHECR sources accelerate protons only up to an energy smaller than the GZK energy scale. In these scenarios, the highest energy particles are heavy nuclei, experiencing large deflections as a result of their interactions with the magnetic field in our Galaxy. By contrast, if the sources are able to accelerate protons up to the highest energies observed, say up to $3\,10^{20}$~eV or above, then the UHECRs are likely to be dominated by protons in the GZK range. Indeed, the nuclei of hydrogen and helium are overwhelmingly dominant in the interstellar medium, so unless a strong discrimination mechanism favors heavier nuclei in the acceleration process, H and He nuclei should be dominantly injected in the intergalactic medium. Now, since the GZK horizon scale of the He nuclei (and of the light and intermediate nuclei) is much smaller than that of the protons in the energy range under consideration, the latter should dominate the UHECR composition observed at the Earth.

The scenarios in which protons can be accelerated up to 300~EeV or above are thus particularly interesting in the context of the on-going and very important quest for the UHECR sources. Figures~\ref{fig:deflectionFractionProton}, \ref{fig:deflectionMapBackwardMean} and~\ref{fig:deflectionMapForwardMean} clearly show that the deflections of protons in the GZK energy range are very small, except in a small part of the sky around the Galactic center. As a consequence, a proton-dominated scenario should easily lead to the detection of individual sources by direct pointing, once a few UHE protons are observed from the closest sources, at sufficiently high energy for the roughly isotropic background from more distant sources to be negligible.

\begin{figure}[ht!]
\centering
\includegraphics[trim=2.75cm 3.25cm 1.25cm 2cm,clip,width=\columnwidth]{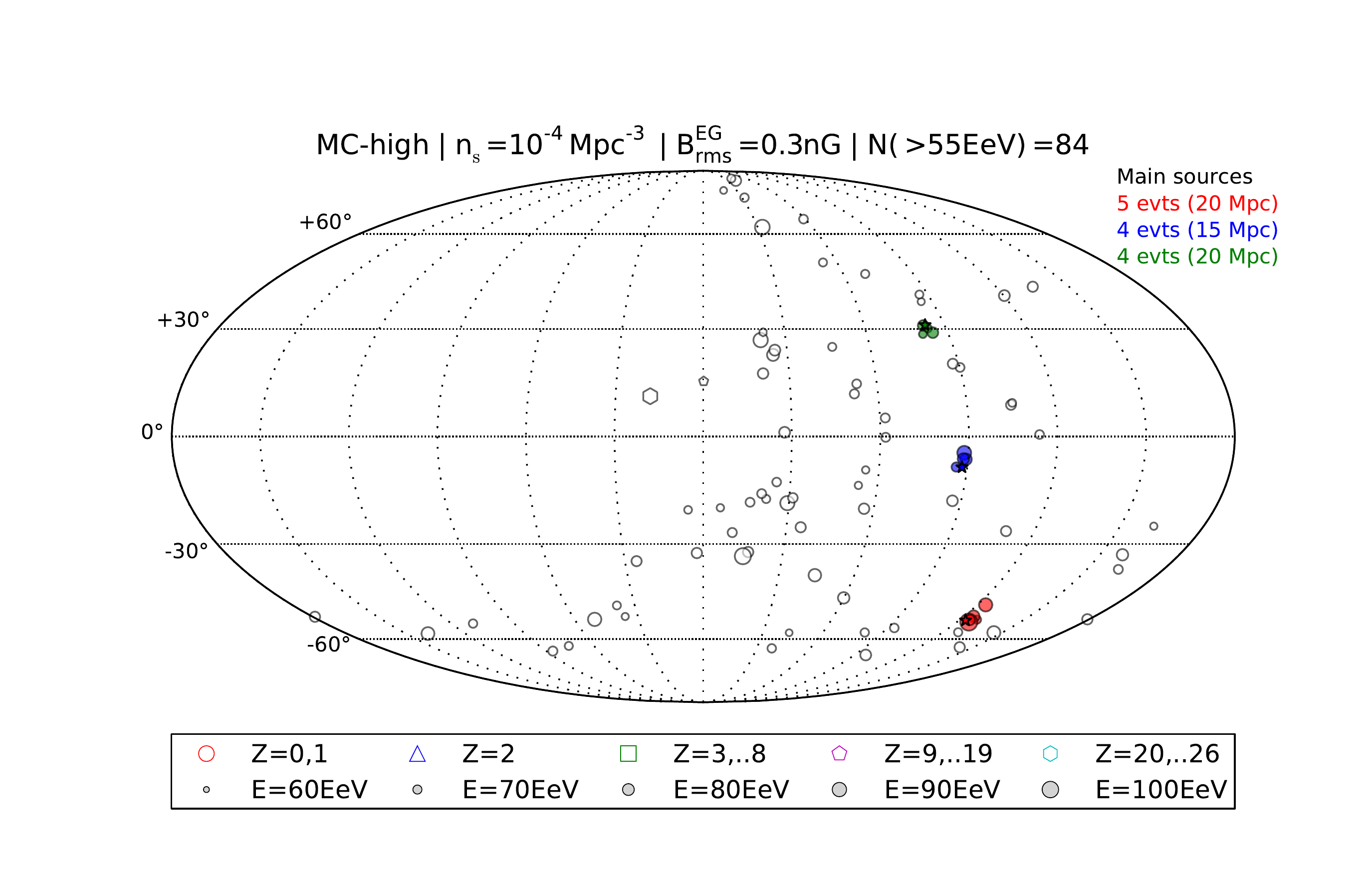}
\includegraphics[trim=2.75cm 3.25cm 1.25cm 1.75cm,clip,width=\columnwidth]{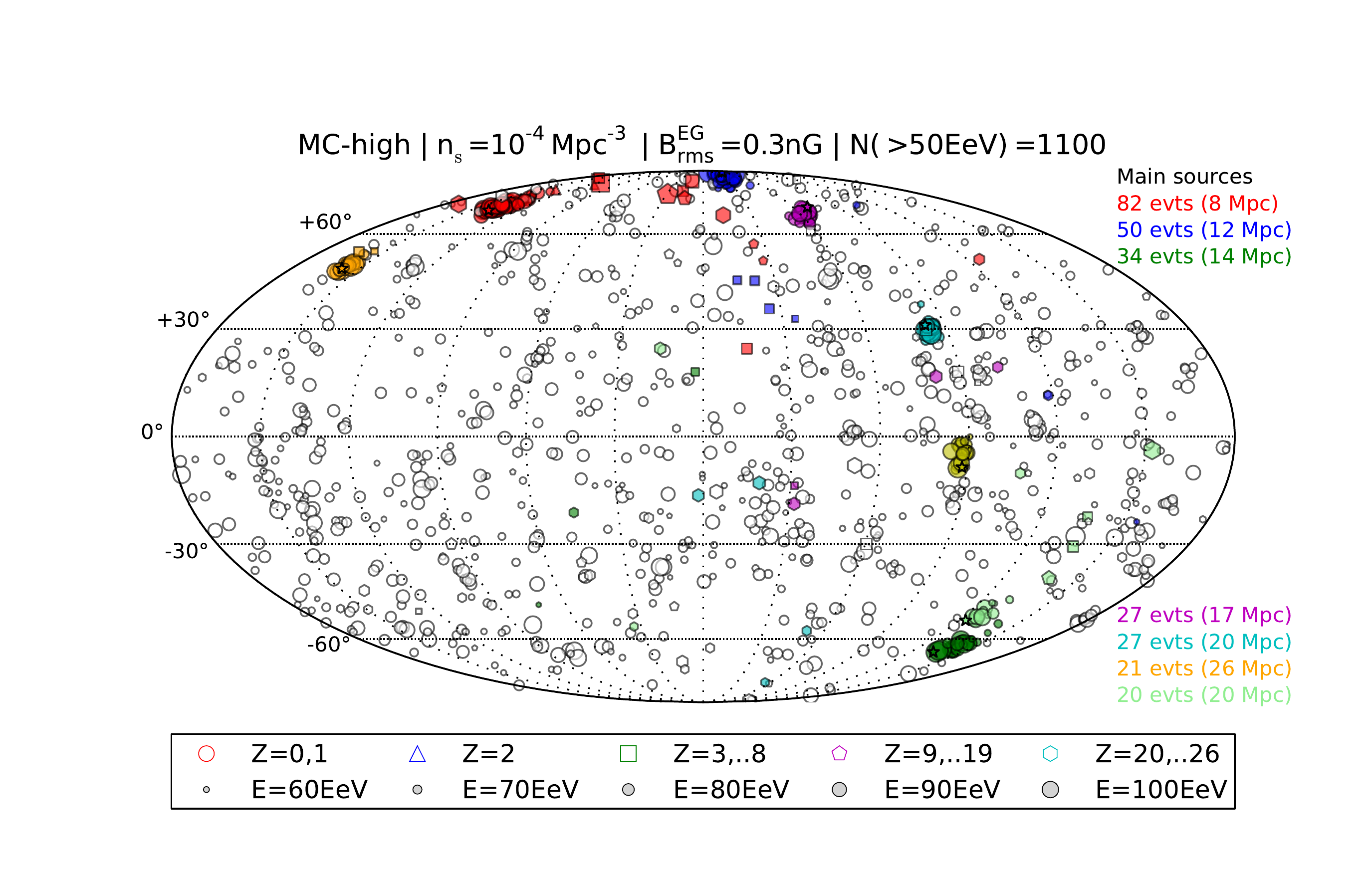}
\includegraphics[trim=2.75cm 3.25cm 1.25cm 1.75cm,clip,width=\columnwidth]{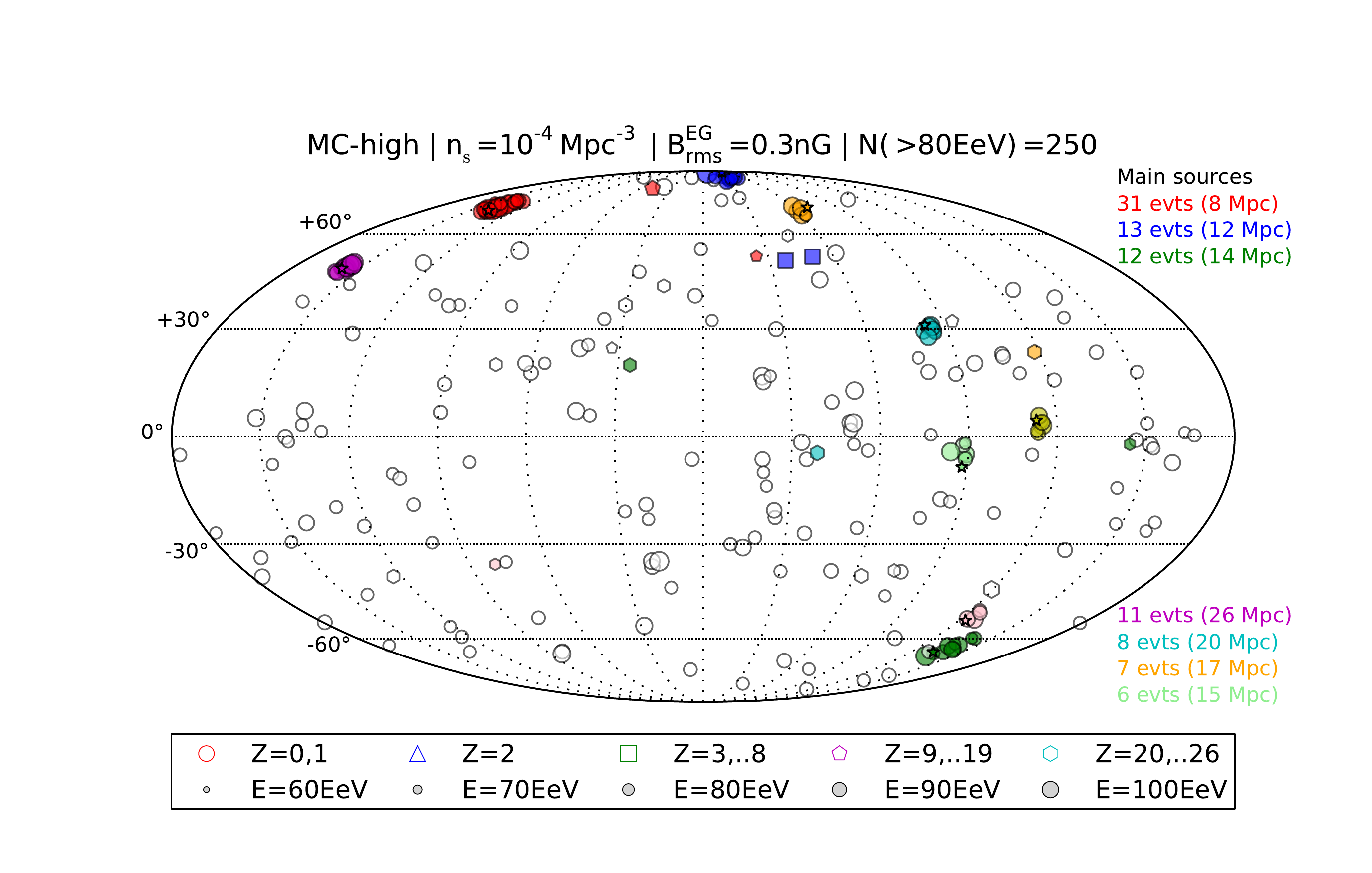}
\includegraphics[trim=2.75cm 1cm 1.25cm 1.75cm,clip=true,width=\columnwidth]{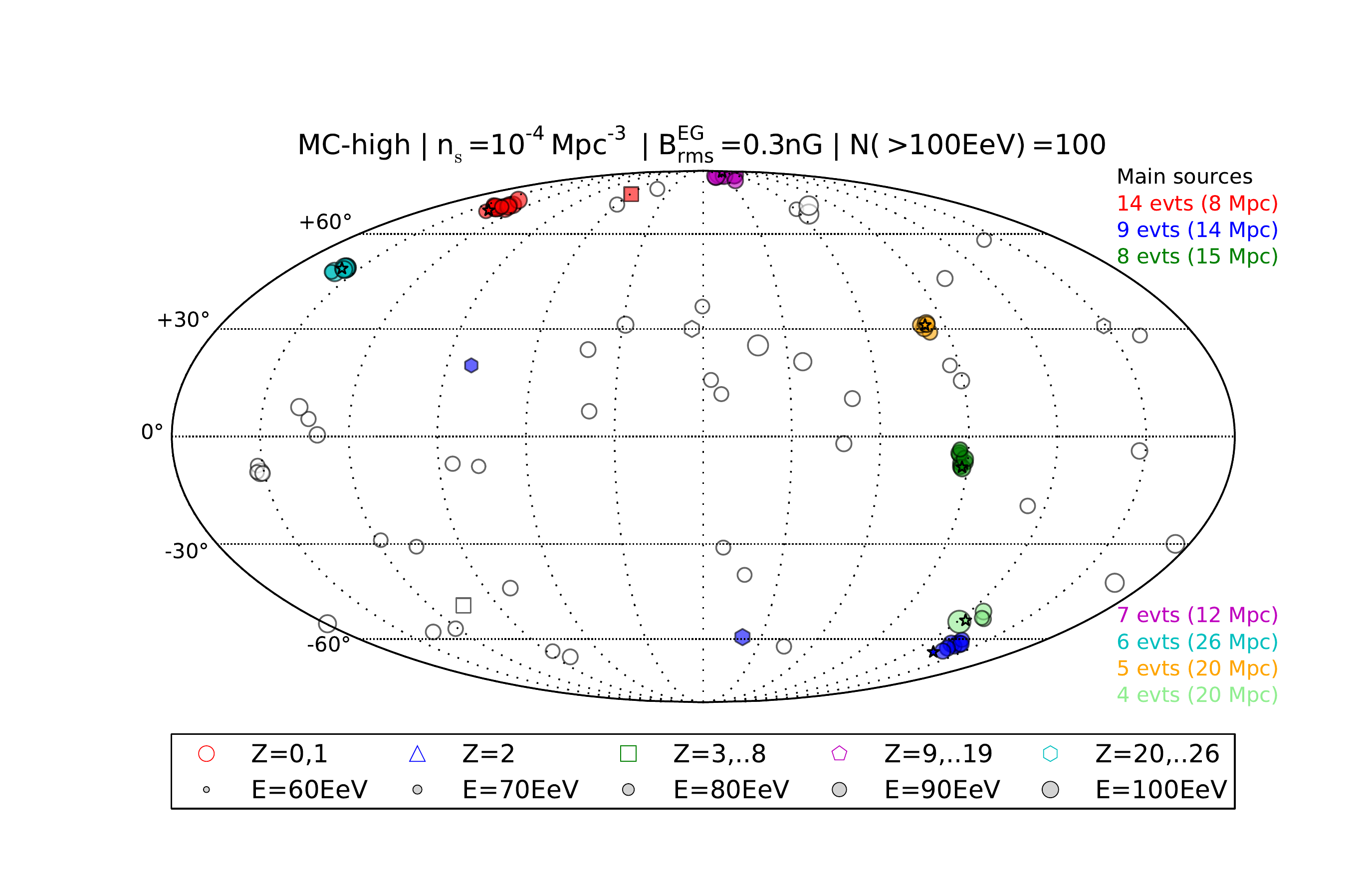}
\caption{Examples of typical sky maps corresponding to the proton-dominated MC-high model (see text), simulated for the current statistics of Auger (top panel) and for the expected statistics that JEM-EUSO would gather with a total exposure of $300,000\,\mathrm{km}^{2}\,\mathrm{sr}\,\mathrm{yr}$, assuming the flux normalization given by the Auger energy scale and a source density $n_{\mathrm{s}} = 10^{-4}\,\mathrm{Mpc}^{-3}$. The second, third and fourth maps are drawn with a (reconstructed) energy threshold of 50~EeV, 80~EeV and 100~EeV respectively.}
\label{fig:MCHighSkyMaps}
\end{figure}

In Fig.~\ref{fig:MCHighSkyMaps}, we show some typical sky maps (in the middle of the distribution, as defined above) obtained with the JEM-EUSO statistics at 50~EeV, 80~EeV and 100~EeV as well as with the current Auger statistics (top panel), for the proton-dominated model, MC-high, with a source density of $10^{-4}\,\mathrm{Mpc}^{-3}$.

As expected, individual sources are easily detected in such scenarios. Unfortunately, although these sky maps appear like a perfect situation to discover and study the UHECR sources, a simple look at the top panel of the Figure shows that such a scenario is already excluded by the existing data. Indeed, the sky map built for the same realization with the Auger coverage and current statistics shows very tight multiplets, right in the direction of the sources. Such small-angular scale clusters of events have not been observed, whereas they are expected for 100\% of the realizations of the MC-high scenario with $n_{\mathrm{s}} = 10^{-4}\,\mathrm{Mpc}^{-3}$. Note that the situation would of course be even worse if we drew the sky maps corresponding to a MC-high scenario with a lower source density, since the average multiplicities of the different sources would then be even larger, and lead to even more obvious small-angular-scale multiplets impossible to miss in the Auger-like sky maps. As already indicated, the results obtained with a pure-proton model are exactly the same in all respects.

\begin{figure*}[ht]
\centering
\includegraphics[trim=0.75cm 0cm 1.5cm 0.5cm,clip,width=\columnwidth]{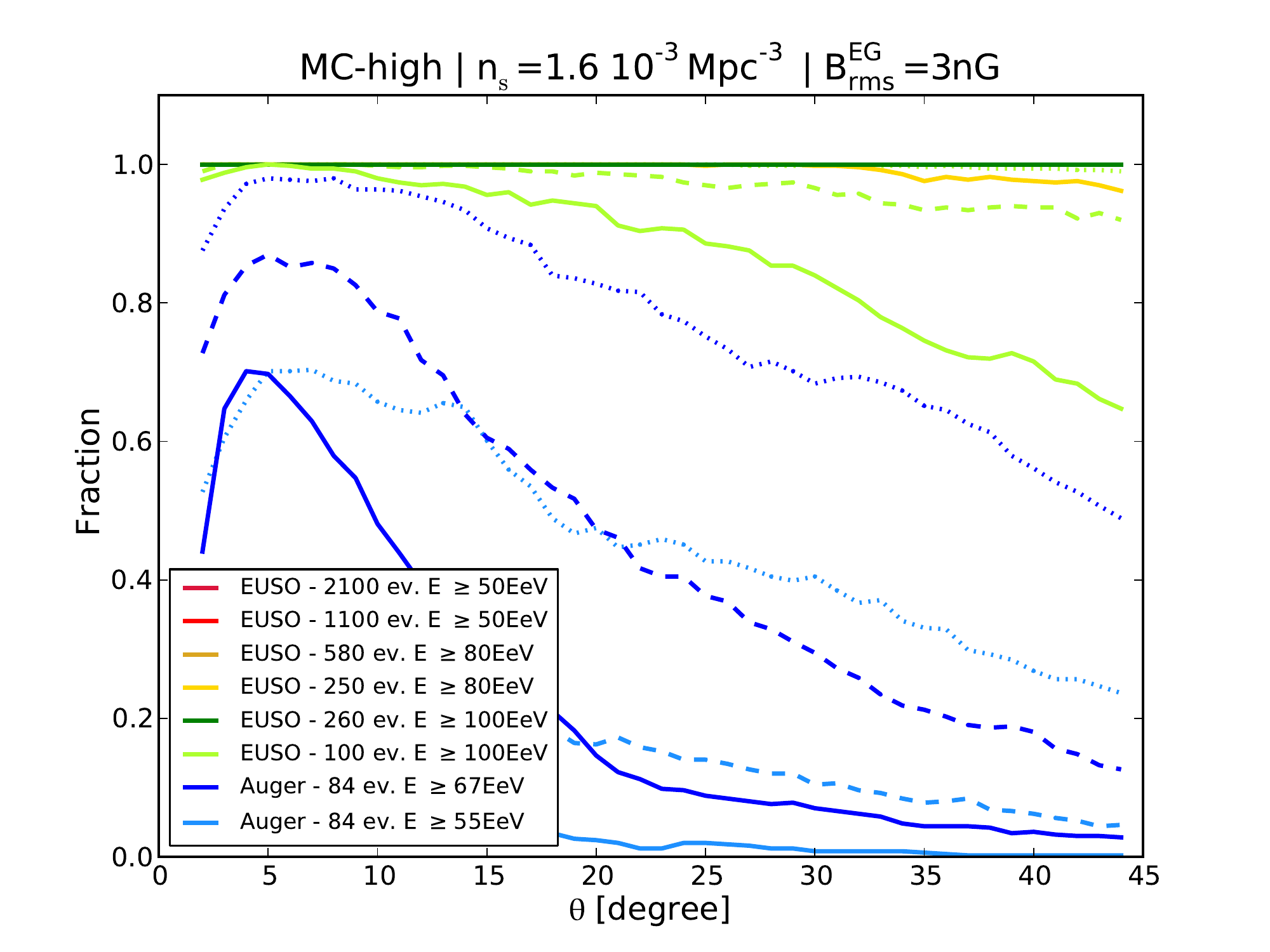}
\includegraphics[trim=0.75cm 0cm 1.5cm 0.5cm,clip,width=\columnwidth]{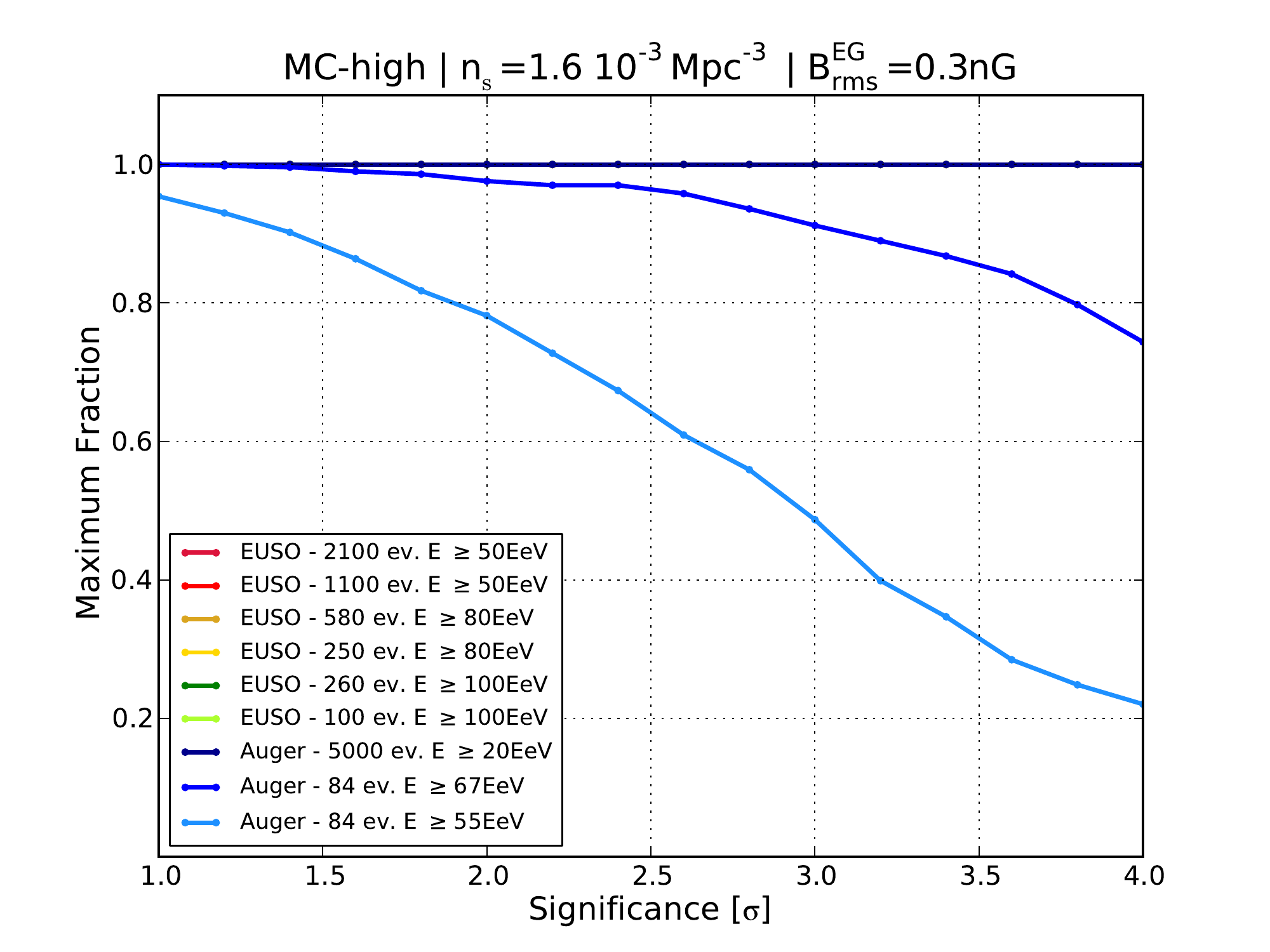}
\caption{Same as Fig.~\ref{fig:2ptFractionMC15-4} (left) and Fig.~\ref{fig:2ptMaxMC15-4} (right), but for the MC-high (proton-dominated) model, with a source density $n_{\mathrm{s}} = 1.6\,10^{-3}\,\mathrm{Mpc}^{-3}$.}
\label{fig:2ptFractionMCHigh-3}
\end{figure*}

It is nevertheless interesting to see whether some specific scenarios in which protons dominate up to highest energies can be compatible with the Auger data. For this, we allowed the source density to be extremely high, and simulated sky maps with a version of the MC-high model where all the galaxies in the catalog (see Sect.~\ref{sec:sourceDistrib}) were assumed to be sources of UHECRs. This corresponds to a density $n_{\mathrm{s}} = 1.6\,10^{-3}\,\mathrm{Mpc}^{-3}$.

\begin{figure}[ht!]
\centering
\includegraphics[trim=2.75cm 3.25cm 1.25cm 2cm,clip,width=\columnwidth]{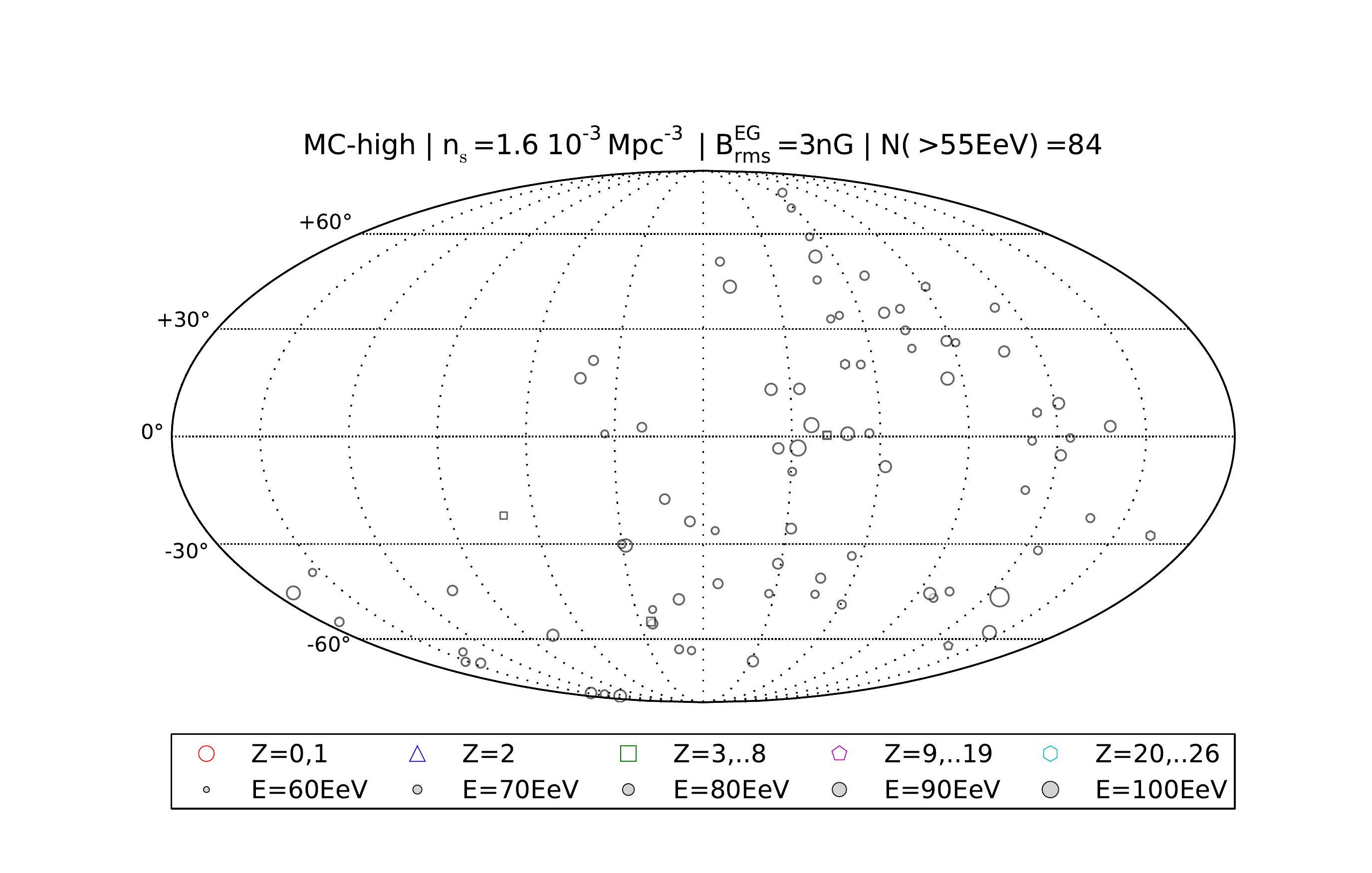}
\includegraphics[trim=2.75cm 3.25cm 1.25cm 1.75cm,clip,width=\columnwidth]{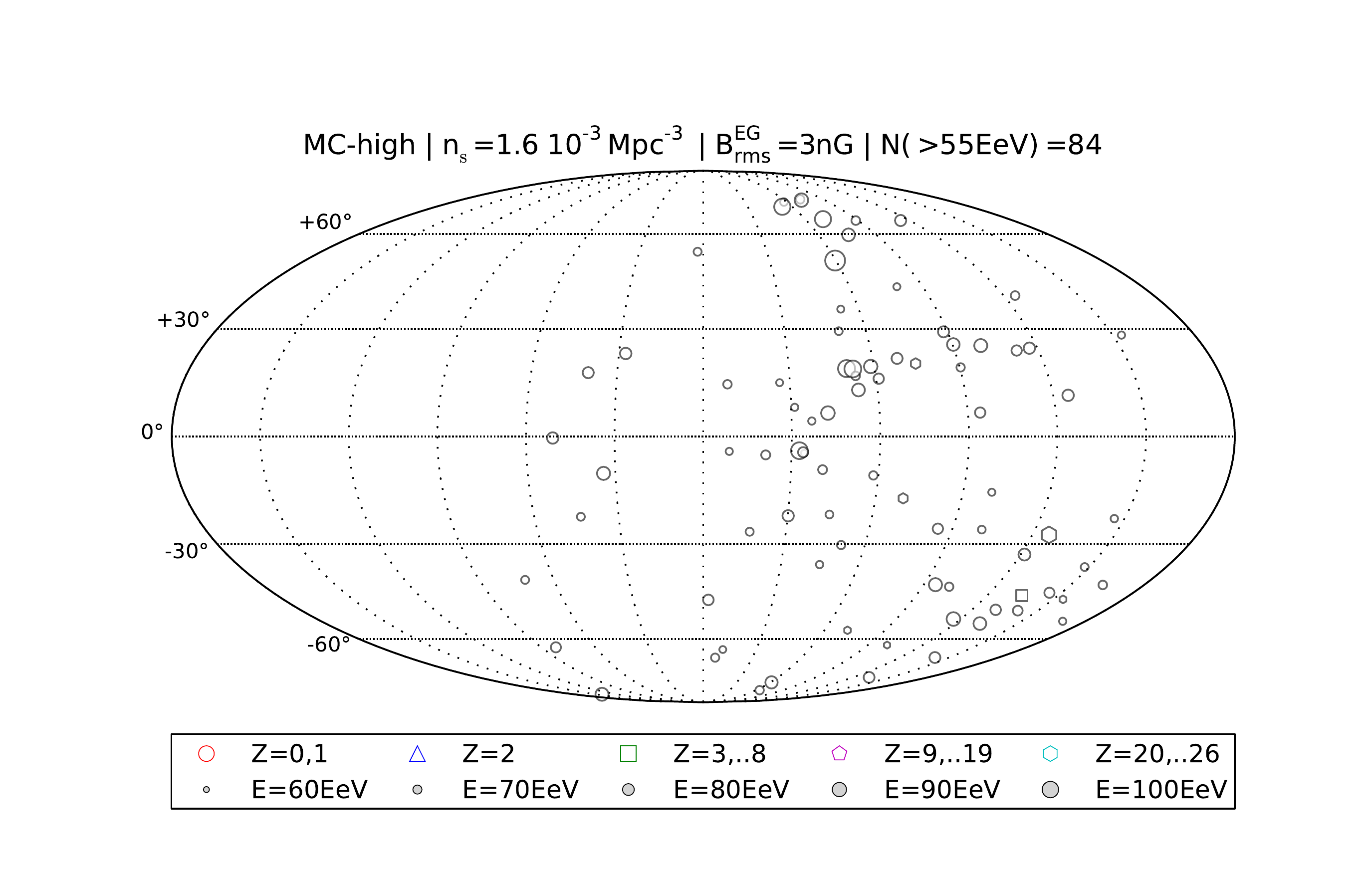}
\includegraphics[trim=2.75cm 3.25cm 1.25cm 1.75cm,clip,width=\columnwidth]{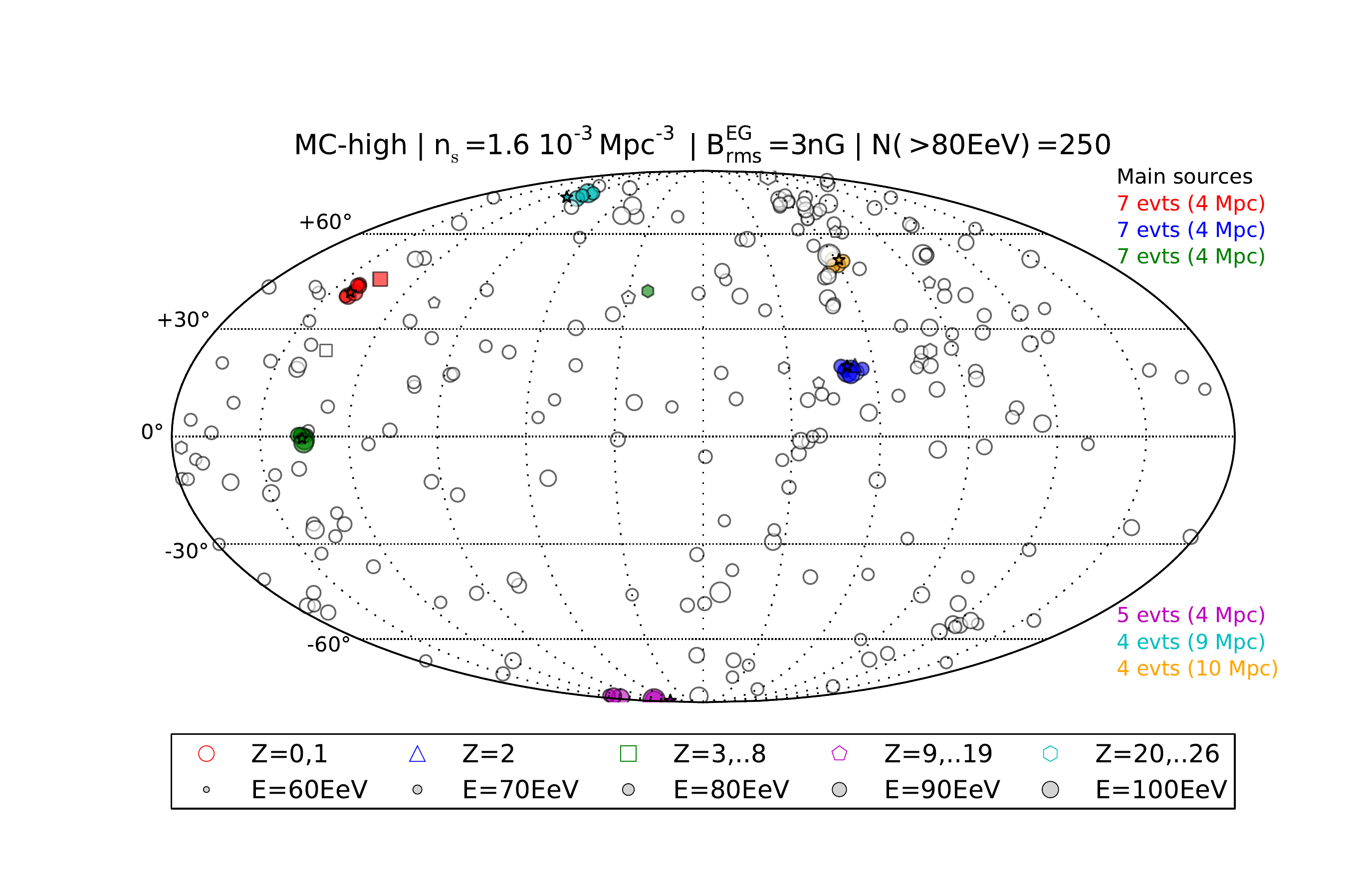}
\includegraphics[trim=2.75cm 1cm 1.25cm 1.75cm,clip=true,width=\columnwidth]{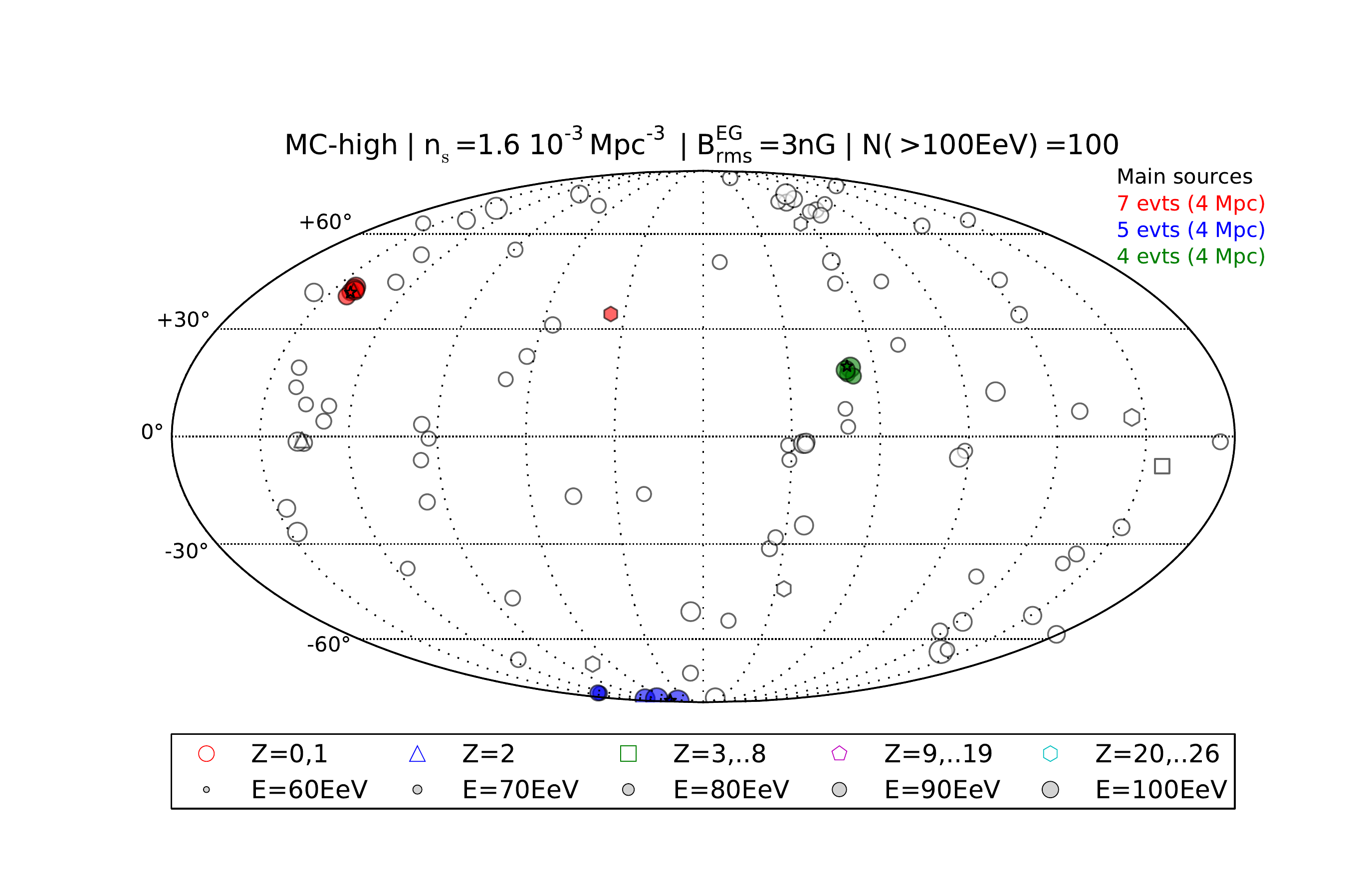}
\caption{Examples of typical sky maps corresponding to the proton-dominated MC-high model with extremely large source density,  $n_{\mathrm{s}} = 1.6\,10^{-3}\,\mathrm{Mpc}^{-3}$. The first (top) panel shows the realization which gave the smallest signal of anisotropy (out of the 500 realizations simulated) with the current Auger statistics. The second panel shows the realization lying at the middle of the distribution (see text). The third and fourth panels show the expected sky maps with the reference statistics of JEM-EUSO with a threshold at 80~EeV and 100~EeV respectively, for the same realization as in the second panel (Auger energy scale).}
\label{fig:MCHighSkyMaps-3}
\end{figure}

A minority, but substantial fraction of the realizations of this scenario were indeed found compatible with the existing data, on the basis of the two-point correlation function. The corresponding probabilities are shown in the usual way in Fig.~\ref{fig:2ptFractionMCHigh-3} (left). We also show examples of sky maps corresponding to this scenario in Fig.~\ref{fig:MCHighSkyMaps-3}. The top panel shows the sky map of the realization which gives the smallest anisotropy signal out of the 500 realizations simulated, with the reference Auger statistics and sky coverage. No obvious anisotropy can be seen on the map, which is indeed confirmed by the result of the statistical anisotropy study. From the point of view of the two-point correlation function, this sky map is similar to the actual Auger sky map at 55~EeV. The second panel of Fig.~\ref{fig:MCHighSkyMaps-3} shows the sky map obtained with the ``median realization'', i.e. the realization sitting in the middle of the distribution of the 500 realizations ordered by significance of auto-correlation anisotropy. Finally, the third and fourth panels show the expected sky maps of this latter realization with the JEM-EUSO reference statistics with a threshold in the (reconstructed) energy of the UHECR events at 80~EeV and 100~EeV, respectively. These sky maps clearly show the interest of increasing the statistics in the case of such scenarios. While the sources cannot be identified so far, they will certainly be with a ten times larger exposure.

Note also that we performed simulations with an extended Auger statistics, and found that significant multiplets should be detected by Auger in the coming years as well. These results are summarized quantitatively in Fig.~\ref{fig:2ptFractionMCHigh-3} (right). As can be seen, 100\% of the realizations give rise to extremely significant small-scale anisotropies at either the future Auger or JEM-EUSO statistics.

%


\subsection{Pure-Fe models}
\label{sec:pureFe}

We now turn to the study of an extreme case of heavy source composition, in which all the UHECRs injected in the intergalactic medium are Fe nuclei. As already noted, the GZK horizon structure for Fe nuclei is not very different from that of protons, so it is possible to obtain an equally good fit of the all-sky UHECR spectrum by using pure-Fe sources as by using a mixed-composition model, with either a high or low value of the maximum proton energy (even though the required steepness of the source spectrum is different in each case, see above).

From what we know (or think we know) of the astrophysical environments where ultra-high-energy particle acceleration might occur in the universe, one must recognize, however, that a pure-Fe model is not by itself very realistic, from the astrophysical point of view. Nevertheless, pure-Fe models, as a reference case of study, display interesting features which help understanding some important aspects of the UHECR phenomenology, and are thus worth investigating.

This is mostly due to the fact that UHE Fe nuclei, say around or above $10^{20}$~eV, are photo-dissociated into lighter nuclei as they propagate through the intergalactic medium and interact with the background microwave and infrared photons. These interactions typically eject one or a few nucleons out of the UHE nuclei, and since most of them occur near the threshold energy of the photodissociation (through giant dipolar resonance) in the rest frame of the nuclei, they roughly leave the remaining nuclei as well as the ejected nucleons with the same Lorentz factor, i.e. with the same energy per nucleon as the parent nuclei \citep[see][]{Allard:2012} for a more complete discussion of heavy nuclei propagation). As a consequence, an iron nucleus with initial energy $E$ above the GZK cutoff will produce secondary particles and eventually protons with an energy $E/A$. By this process, the secondary protons coming from primary Fe nuclei can reach energies only a factor of $56/26 \simeq 2.15$ lower than the energy which primary protons could reach in the same sources if they were accelerated there, and if the maximum energy were proportional to the charge of the nuclei, $Z$, as assumed for the mixed-composition models.

As a consequence, pure-Fe models with a conceivably high maximum energy at the sources lead to UHECR populations which contain a non negligible fraction of protons, when propagation effects are taken into account. For this reason, the pure-Fe models turn out to offer a simple way to explore more complicated situations than the typical scenarios investigated here, where all the sources have the same spectrum, composition and maximum energy.

In particular, we may use the pure-Fe models as a first hint to the phenomenology that could be relevant to scenarios in which, for example, the maximum energy of the protons is different in different sources, and can reach $10^{20}$~eV (or more) in only a small number of sources (e.g. the most powerful ones). In such a case, most of the UHECRs in the GZK energy range would be heavy nuclei and thus experience relatively large deflections, as in the cases explored in Sect.~\ref{sec:MCModels}, but protons and low-$Z$ nuclei would also be present at some level, originating from a subset of the sources. Another possibility could be that the maximum energy reached in the UHECR sources is not as well-defined as usually assumed, and does not give rise to an exponential cutoff. If the acceleration mechanism leads to a more gradual decrease of the UHECR flux injected in the extragalactic space as a function of energy, then a transition towards a heavy composition may occur, say, around $10^{19}$~eV (as possibly indicated by the Auger data) due to the progressive extinction of the protons at the sources, while a subdominant fraction of protons survives up to the highest energies, with a decreasing fraction. Finally, a distribution of maximum energies has also been shown by \cite{Blaksley:2012} to result in a modified apparent source composition, which may be important to understand the evolution of the average UHECR mass as a function of energy. In such a situation, a transition towards heavier nuclei occurs at ultra-high-energy, without implying a complete disappearance of protons or light/intermediate nuclei, as in the mixed-composition models with identical sources which are generally explored (e.g. strict MC-4EeV and MC-15 EeV models).

For all these reasons, even though pure-Fe models are not astrophysically realistic by themselves, the results of this subsection should be considered as indicative of a number of interesting situations in which a small, but non negligible fraction of protons are present among UHECRs up to the highest energies.

\begin{figure}[ht!]
\centering
\includegraphics[trim=2.75cm 3.25cm 1.25cm 2cm,clip,width=\columnwidth]{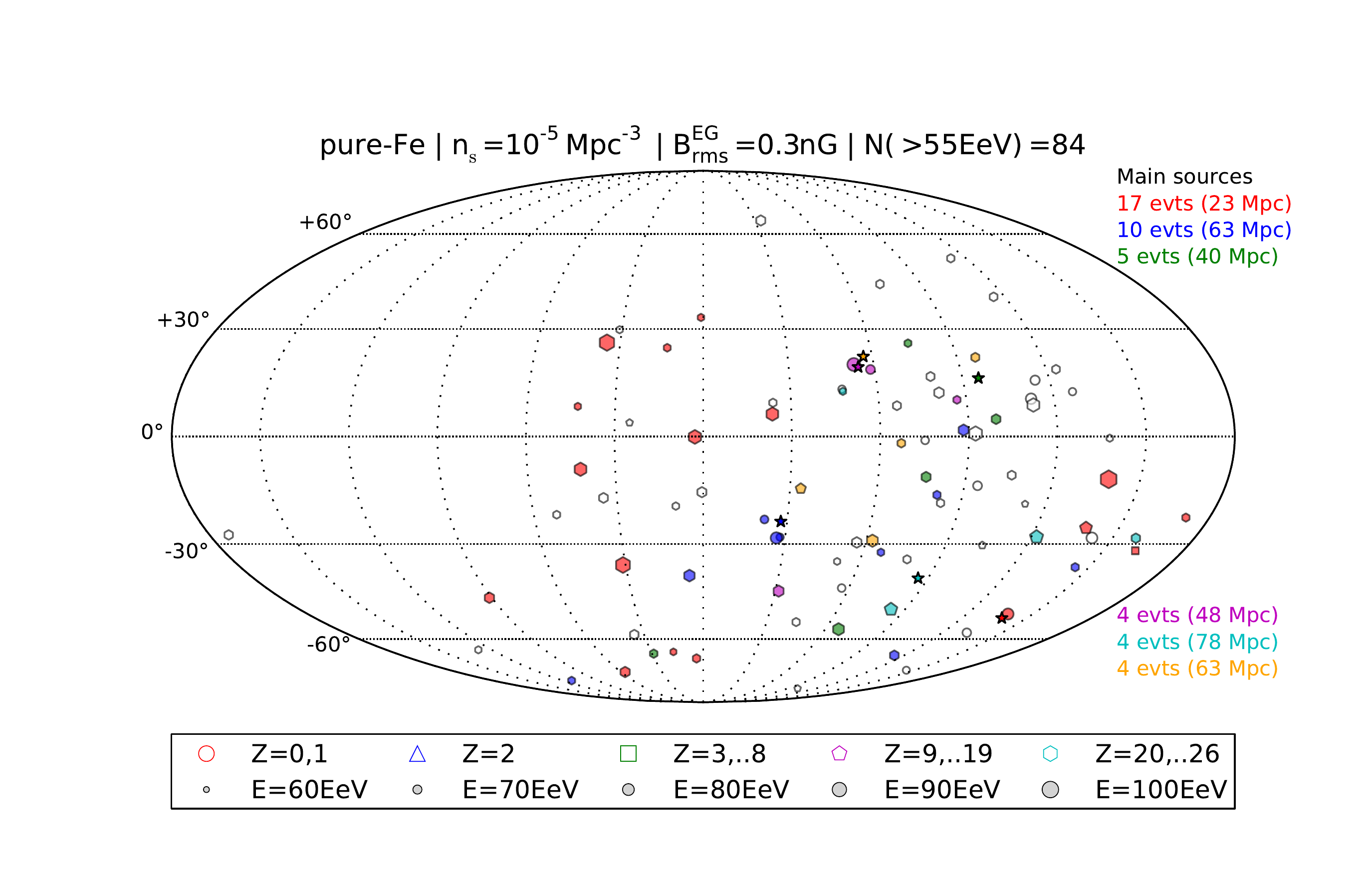}
\includegraphics[trim=2.75cm 3.25cm 1.25cm 1.75cm,clip,width=\columnwidth]{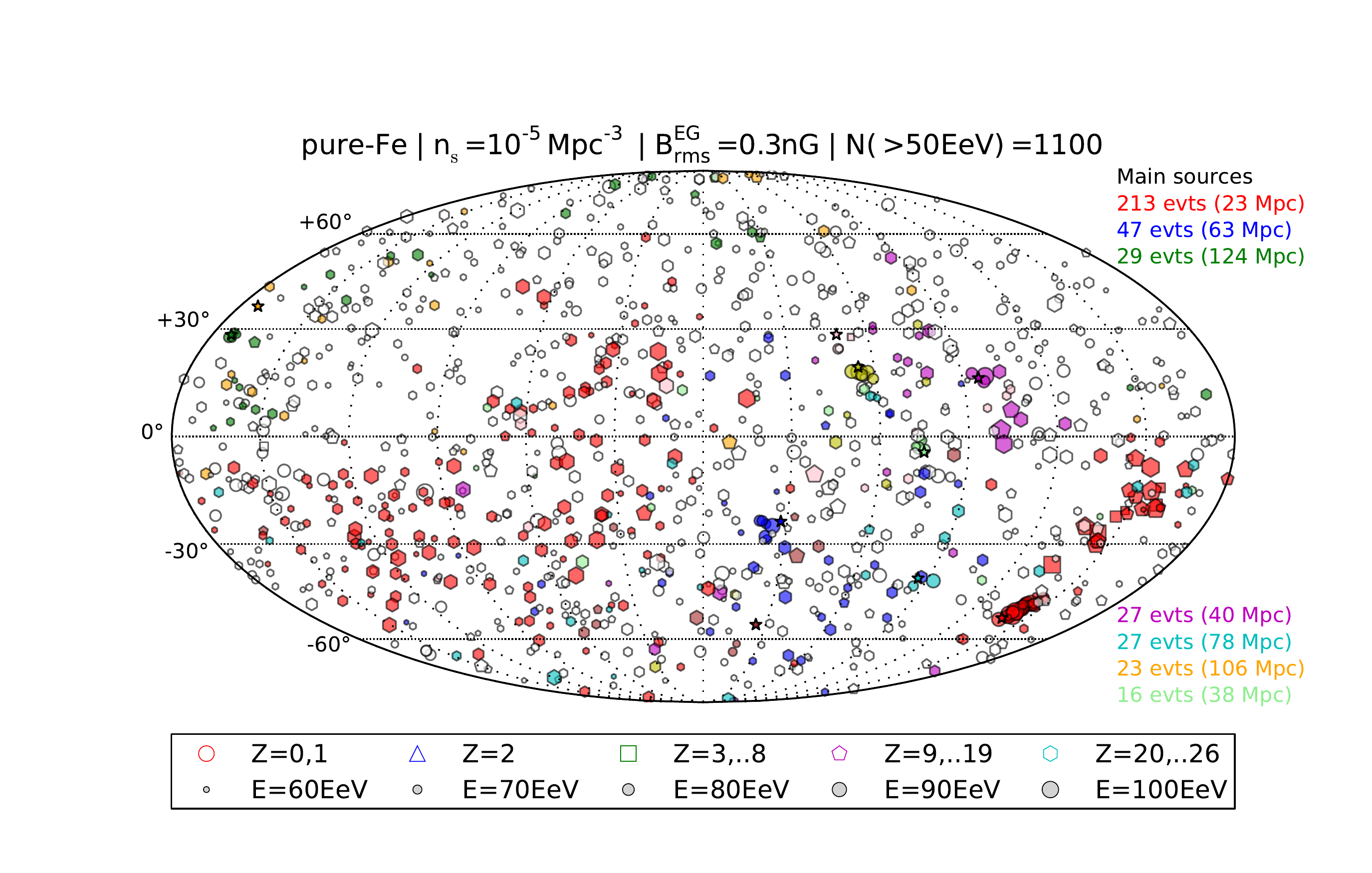}
\includegraphics[trim=2.75cm 3.25cm 1.25cm 1.75cm,clip,width=\columnwidth]{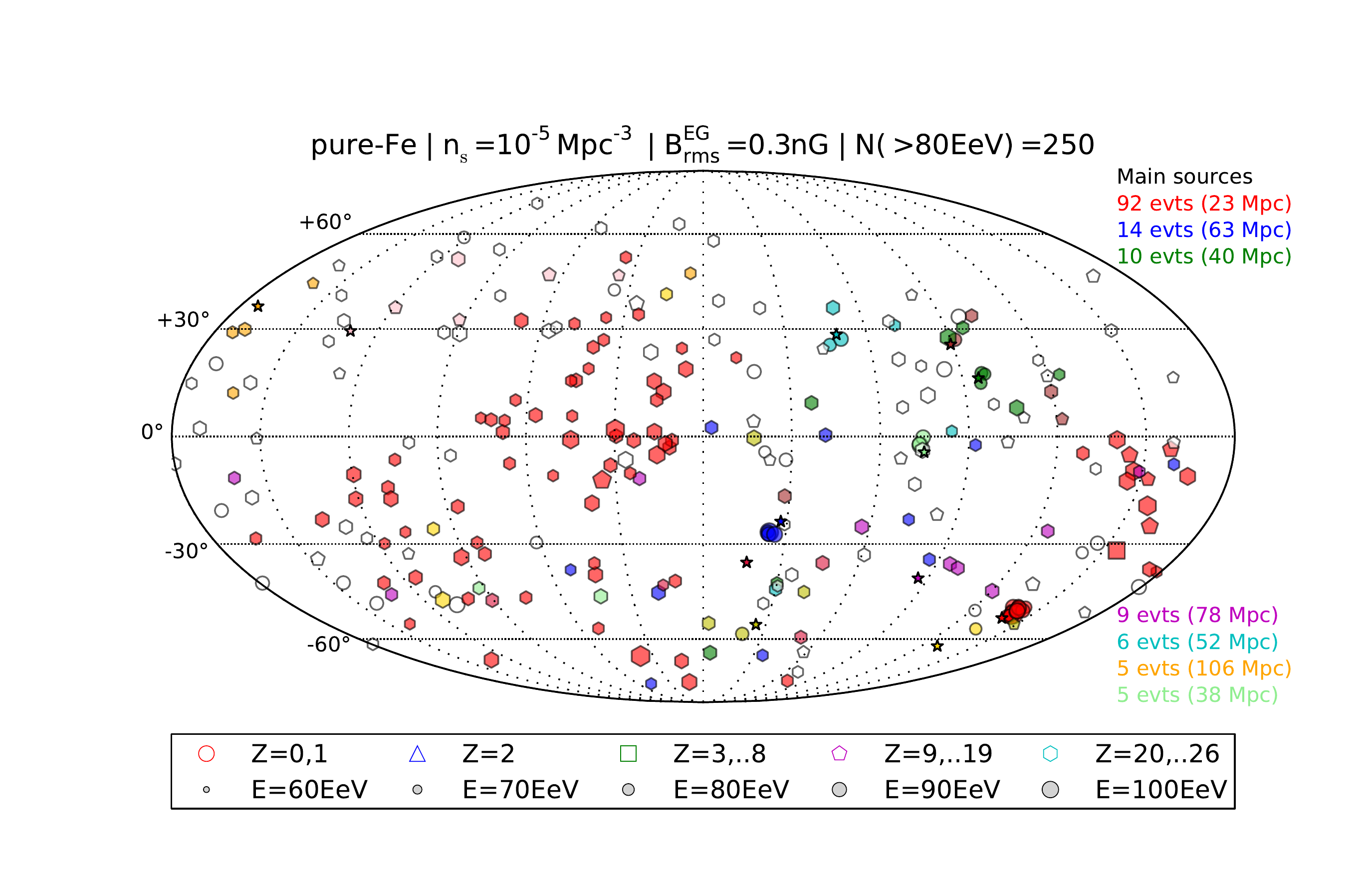}
\includegraphics[trim=2.75cm 1cm 1.25cm 1.75cm,clip=true,width=\columnwidth]{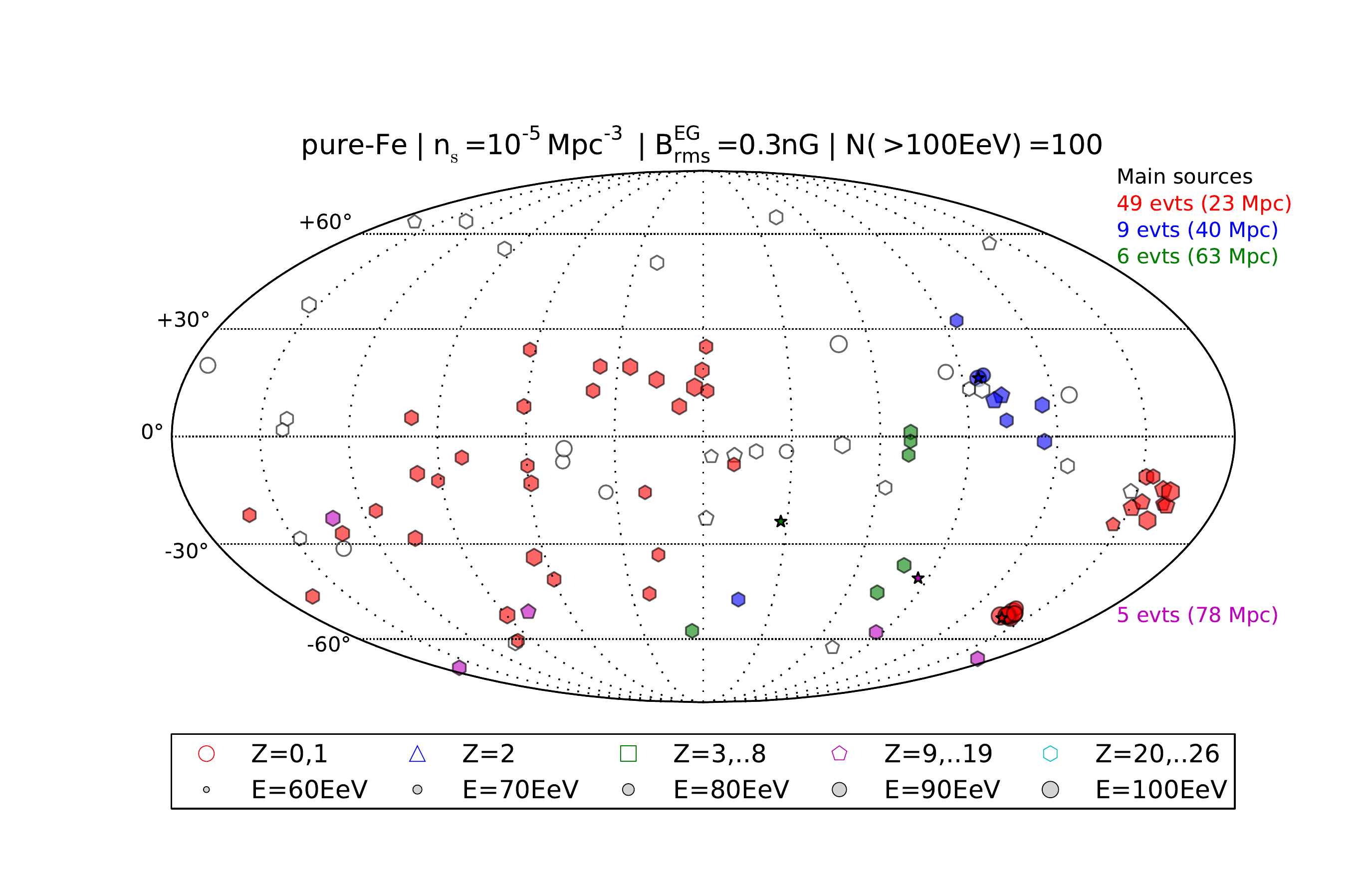}
\caption{Examples of sky maps corresponding to the pure-Fe model, with a source density $n_{\mathrm{s}} = 10^{-5}\,\mathrm{Mpc}^{-3}$, simulated with the current statistics of Auger (top panel) and with the expected statistics that JEM-EUSO would gather with a total exposure of $300,000\,\mathrm{km}^{2}\,\mathrm{sr}\,\mathrm{yr}$, assuming the flux normalization given by the Auger energy scale. The second, third and fourth maps are drawn with a (reconstructed) energy threshold of 50~EeV, 80~EeV and 100~EeV, respectively.}
\label{fig:PureFeSkyMaps}
\end{figure}

The distinctive feature of pure-Fe models (or more generally models with a subdominant fraction of protons, either primary or secondary) is evident on the sky maps shown in Fig.~\ref{fig:PureFeSkyMaps}. While no significant clustering is observed in the data set corresponding to the current Auger statistics (84 events above 55~EeV, assuming the Auger energy scale), a number of UHECR sources appear very clearly with larger statistics, thanks to the secondary protons generated during propagation.

In the Auger reference sky map,  two protons are seen in purple around their source (towards the top of the coverage map), but could be merely random coincidence. As for the most intense source in this sky map, located at 23~Mpc and marked by a red star at the bottom right of the map, it spreads many events across the sky, but only one of them is a proton. Although it is located very close to the source, it cannot be used to determine the source position, since it appears like any other events in the sky map.

It is worth noting, however, that the possibility to determine the composition of UHECRs event by event would be very important, especially for this kind of models, as a sub-sky map built by selecting only protons would give a view on the UHECR sky much easier to decipher. However, such a perspective appears quite distant at the moment, from the experimental point of view, be it only because of the so-called shower-to-shower fluctuations which can easily make a particular proton-induced atmospheric shower be very similar to an Fe-induced shower, and vice-versa. New detection techniques, e.g. based on the radio signal of the showers, may nevertheless change this situation in the future.

The identification of proton events can however be done in an indirect, but rather trivial way without any detailed study of the showers, apart from the reconstruction of their arrival direction with a precision which is usually not challenging experimentally. Indeed, looking at the bottom panel of Fig.~\ref{fig:PureFeSkyMaps}, one sees that the dominant source, which contributes 49~events above 100~EeV, spreads many UHE Fe nuclei over a large area of the map on the left hemisphere, as well as 10~events in a relatively tight cluster on the right, but leads also to the superposition of 9~events right in the direction of the source. It is statistically impossible for Fe nuclei to be observed so precisely in the same arrival direction, and since intermediate-mass nuclei (as can also be checked from the shape of the symbols on the map) are destroyed at lower energy by the GZK process, only protons can be responsible for such a tight cluster (barring a random coincidence, which may account for one or two events at most in a data set with such statistics).

\begin{figure*}[ht!]
\centering
\includegraphics[trim=0.75cm 0cm 1.5cm 0cm,clip,width=\columnwidth]{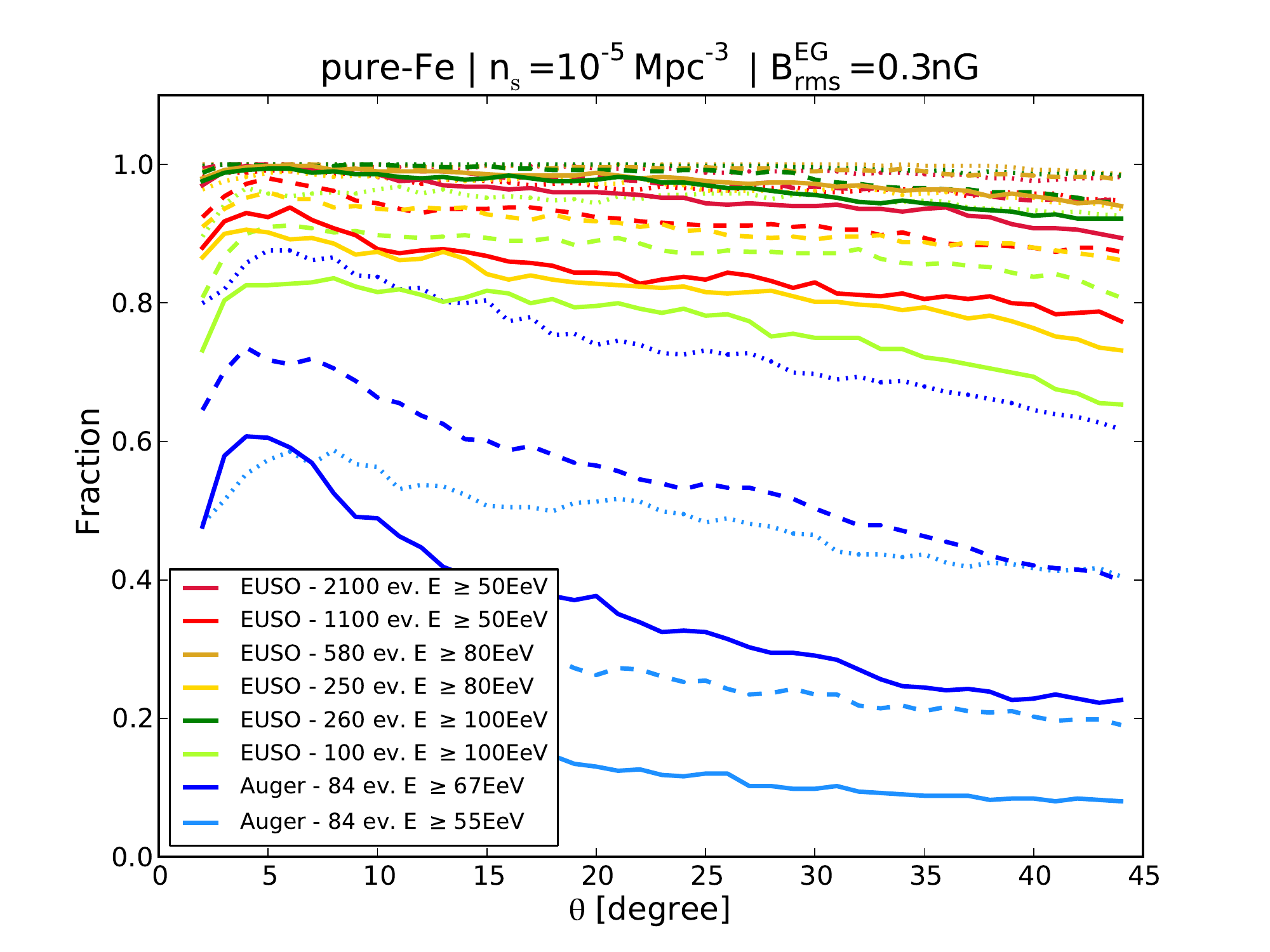}
\includegraphics[trim=0.75cm 0cm 1.5cm 0cm,clip,width=\columnwidth]{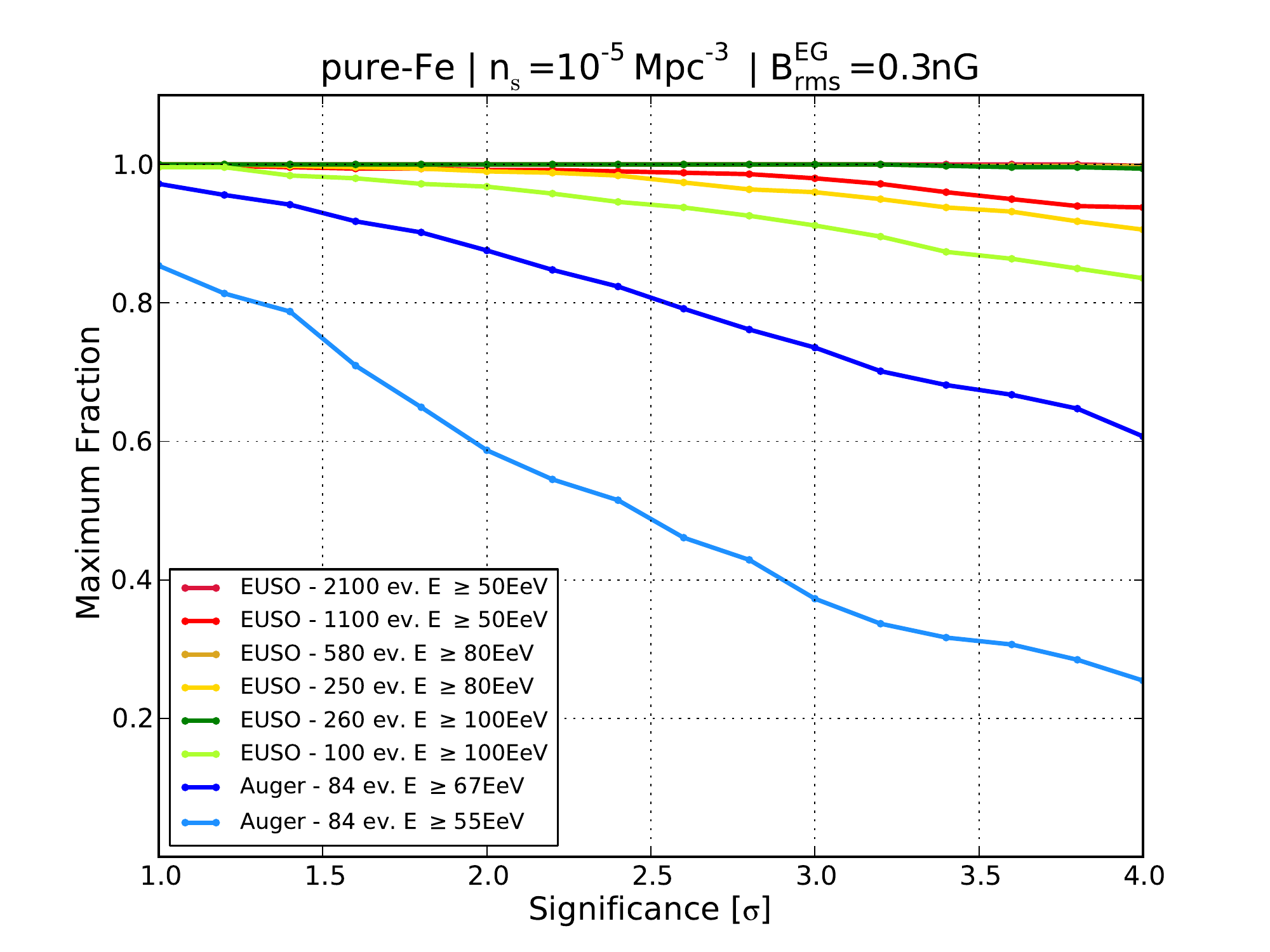}
\caption{Same as Fig.~\ref{fig:2ptFractionMC15-4} (left) and Fig.~\ref{fig:2ptMaxMC15-4} (right), but for a pure-Fe model with source density $n_{\mathrm{s}} = 10^{-4}\,\mathrm{Mpc}^{-3}$.}
\label{fig:2ptFractionPureFe-5}
\end{figure*}

The sky maps corresponding to an energy threshold of 50~EeV and 80~EeV (second and third panel of Fig.~\ref{fig:PureFeSkyMaps}) also displays some tight clusters, corresponding to the direction of individual sources. The evolution of the sky maps with energy can also give further information (and confidence) about the source location. It is worth pointing out here that each sky map is built from a different run, i.e. we draw a whole new set of events for each map. Therefore, even though the source realization of the model under study is the same for all the maps of a given figure (which means that the sources of the catalog selected as actual UHECR sources are the same), the events seen above 100~EeV in the map built with an energy threshold of 80~EeV, say, are not the same as those shown in the map built with an threshold at 100~EeV, which is in all respects a different map, from a different data set. In other words, the data sets for the different maps are all totally independent. For this reason, a direct comparison of the maps at different thresholds cannot be done here visually, but can of course be carried out by applying energy threshold on the lower-energy maps.

Note finally that we have chosen here a realization which displays less significant anisotropies than the average, in order to remain compatible with the current constraints (we could also have presented a model with a larger source density, with essentially the same features). Indeed, many pure-Fe models show significant anisotropies already with the current Auger statistics, as apparent on Fig.~\ref{fig:2ptFractionPureFe-5}, where we show the usual statistical analysis of the anisotropy properties of the model. Respectively around 60\% or 40\% of the realizations appear to display small-scale anisotropies with a 2$\sigma$ or 3$\sigma$ significance, with the Auger statistics of reference.


\section{Summary and perspective}
\label{sec:summary}

In this paper, we have studied in a systematic way some aspects of the phenomenology of a wide range of possible astrophysical scenarios for the origin of UHECRs, allowed by the current constraints and available data. We concentrated on the statistical analysis of the anisotropies that can be expected in the arrival directions of the particles at different energies, varying the source composition and spectrum, as well as the source density and the intergalactic magnetic field.

This study requires a proper treatment of the propagation of the UHECRs, whether protons or heavier nuclei, including the energy losses and photo-dissociation induced by the interaction with the various photon backgrounds, as well as the deflections caused by the interaction with the extragalactic and Galactic magnetic fields.

The main part of the angular distance between the source direction and the observed arrival direction of a given UHECR is due to the interaction with the Galactic magnetic field, which is not precisely known, unfortunately. We presented the model used for our computation and the resulting deflection properties as a function of rigidity in Sects.~\ref{sec:GMF} and~\ref{sec:particleDeflections}. Although the actual deflections are impossible to predict with confidence, we argued that the overall patterns relevant to our present study should be reliable on average, since they depend more on the typical structures present in the GMF than on the particular arrangement of the magnetic field maxima or reversals. This is also the reason why we limited our anisotropy studies to a search for intrinsic anisotropies, which only depend on the global, statistical properties of the GMF, rather than searching for features associated with particular directions in the sky. Our results are thus conservative in this respect, since other studies could in principle be carried out to derive additional constraints and information about the source models (e.g. by searching for multiplets with coherent patterns in the energy/deflection space, or correlations with particular sources or structures).

Likewise, the exact distribution of the sources in the sky, as well as the UHECR deflections recorded as a consequence of the specific location of the sources with respect to the GMF structures cannot be considered important. By exploring many realizations of each scenario, we believe that the statistical properties that we obtained are robust to reasonable variations in the GMF model or other assumptions underlying our study. As discussed in Sect.~\ref{sec:sourceDistrib}, we used the 2MRS catalog as a reference catalog of possible sources, in order to respect as much as possible the local distribution of matter, in the absence of any information about the actual UHECR sources. But the main relevant parameter is the source density, and the exact location of the sources plays no role in the results presented here.

The astrophysical models investigated have been described and commented in Sect.~\ref{sec:models}. They make use of earlier results, and explore situations in which the UHECRs are dominated by different types of particles (light, heavy, or with a transition from light to heavy), while ensuring compatibility with the existing constraints, notably as far as the UHECR spectrum and intrinsic anisotropies are concerned.

The general conclusion of our study is that the absence of a clear anisotropy signal in the current UHECR data does not imply that the UHECR sky will remain impossible to decipher and that UHECR sources cannot be found in a foreseeable future. On the contrary, we showed that a new generation of detectors, with a typical integrated exposure of 300,000 km$^{2}$~sr~yr and full sky coverage, appears in a very good position to make a major breakthrough in this part of the field of astroparticle physics, by detecting significant anisotropies in essentially all the currently favored astrophysical scenarios.

The least favorable scenarios, in which protons are totally absent at the highest energies, have been shown to lead nevertheless to anisotropies with a significance of at least $4\sigma$ in the vast majority of the realizations. We would thus fail to derive important information about the UHECR sources with an instrument like JEM-EUSO or another large exposure detector only for very unlikely (and unlucky) distributions of the nearby sources. We made this statement quantitative by exploring the so-called cosmic variance for each astrophysical model through the simulation of 500 independent realizations of the different scenarios.

In more favorable scenarios, the presence of UHE protons can be extremely helpful to detect the sources directly through the identification of tight clusters of events (within a few degrees) in given directions of the sky. Although such scenarios have been put into question by the Auger results on UHECR composition, we also showed that proton-dominated scenarios are still viable from the point of view of anisotropies, provided that the source density is particularly large, of the order of $10^{-3}\,\mathrm{Mpc}^{-3}$. For such scenarios, the value of a detector with larger statistics is particularly obvious, as the most nearby sources could be identified easily.

We have noted that the scenarios generally studied, lacking further knowledge on the sources and in order to limit the number of free parameters, have a common drawback which consists in assuming that all the sources have essentially the same parameters. As a matter of fact, we only studied standardized models, where all sources have the same composition and the same maximum energies. Although this allowed us to investigate the range of features that could be expected from a high-statistics, full-sky coverage UHECR detector in the future, the actual situation is most certainly more complicated, with some sources being able to accelerate protons up to higher energies than others, so that a small, but crucial fraction of protons may still be present at high-energy, even if the UHECRs turn out to be dominated by heavy nuclei, with smaller rigidities. A hint to how this could modify our results has been given through the study of pure-Fe models, which are characterized by a sub-dominant component of (secondary) protons. Qualitatively, it is clear that such more realistic models can only produce UHECR skies where \emph{more} significant anisotropies can be detected, and where individual sources can be identified with more precision, thanks to the possible existence of small-angular scale multiplets, which are fully compatible with the current constraints derived from low-statistics data sets. Therefore, the situations considered here are in this respect conservative. A more exhaustive study of these more realistic astrophysical models, where different sources may have different values of $E_{\max}$, is left for future work.

We also note that the tools presented in this paper can easily be applied to a wider variety of astrophysical scenarios, including those involving bursting sources, which have been left aside in the present study.

In this paper, we have used the JEM-EUSO performances as a reference for a future, large exposure detector. This allowed us to be definite regarding the detection efficiency, which will most probably depend on energy for such large detectors. However, this choice is by no means limitative. Any other detector achieving the same exposure and performances (or better), would be equally (or more) suitable to write the next page of the long history of UHECR detection and cosmic ray studies. We made sure, however, that our results would properly apply to detectors with the somewhat lower performances that can be achieved in space (at least for a first generation of instruments). Indeed, a detector with a degraded precision on the reconstructed energy will attribute a given energy to a substantial number of events which are in reality less energetic. This so-called spill-over is asymmetric, due to the steeply decreasing energy spectrum. Now, since lower energy events come from sources which are distributed over a larger volume than high-energy events, as a consequence of the GZK effect, a degraded energy resolution tends to degrade the anisotropy signal that can be detected at a given energy. To secure ourselves against a possible misinterpretation of our results as a consequence of this problem, we decided to apply to all the data sets generated in this study a (hopefully) conservative error on the energy reconstruction of 30\%. In the case of a future detector with better energy resolution, the anisotropy signals to be expected can thus only be more significant than those presented here.

In conclusion, our results give strength to the general idea that a gain of about one order of magnitude in the total exposure of UHECR detectors would make a significant difference compared to the existing, and allow considerable progress in the study of these mysterious particles, whose sources are still unknown, and which challenge both the particle acceleration scenarios in the universe and the astrophysical modeling of the sources. The very likely detection of significant anisotropies (as made quantitative by our study) means that, for the first time, a situation could be reached in which it becomes useful (because meaningful) to study separately different regions of the sky, dominated by different UHECR sources. Further astrophysical information about the UHECR sources and acceleration processes could for instance be obtained from the comparison of the fluxes and spectra obtained in different regions of the sky, and from the study of particular patterns in the UHECR energy/arrival direction space. These aspects of the problem have not been addressed here, as we focused on the first, key question of anisotropies, but should be investigated in future works.

Finally, while stressing the astrophysical value of pursuing UHECR studies with a new generation of detectors, we wish to remind the importance of linking this study to the complementary studies of various classes of high-energy sources in the universe. UHECR studies need increased data sets, notably at the highest energies where the GZK effect ensures the dominance of only a handful of sources in the whole sky, but they will also greatly benefit from the development of high-energy astrophysics in general, with the central goal of applying multi-messenger constraints to the study of what may still be regarded as the extreme universe.


\bibliographystyle{aa710} 
\bibliography{article.bib} 

\begin{thebibliography}{55}
\expandafter\ifx\csname natexlab\endcsname\relax\def\natexlab#1{#1}\fi

\bibitem[{Abu-Zayyad {et~al.}(2012)Abu-Zayyad, Aida, Allen, Anderson, Azuma,
  Barcikowski, Belz, Bergman, Blake, Cady, Cheon, Chiba, Chikawa, Cho, Cho,
  Fujii, Fujii, Fukuda, Fukushima, Gorbunov, Hanlon, Hayashi, Hayashi,
  Hayashida, Hibino, Hiyama, Honda, Iguchi, Ikeda, Ikuta, Inoue, Ishii,
  Ishimori, Ivanov, Iwamoto, Jui, Kadota, Kakimoto, Kalashev, Kanbe, Kasahara,
  Kawai, Kawakami, Kawana, Kido, Kim, Kim, Kim, Kim, Kitamoto, Kobayashi,
  Kobayashi, Kondo, Kuramoto, Kuzmin, Kwon, Lim, Machida, Martens, Martineau,
  Matsuda, Matsuura, Matsuyama, Matthews, Myers, Minamino, Miyata, Miyauchi,
  Murano, Nakamura, Nam, Nonaka, Ogio, Ohnishi, Ohoka, Oki, Oku, Okuda, Oshima,
  Ozawa, Park, Pshirkov, Rodriguez, Roh, Rubtsov, Ryu, Sagawa, Sakurai,
  Sampson, Scott, Shah, Shibata, Shibata, Shimodaira, Shin, Shin, Shirahama,
  Smith, Sokolsky, Sonley, Springer, Stokes, Stratton, Stroman, Suzuki,
  Takahashi, Takeda, Taketa, Takita, Tameda, Tanaka, Tanaka, Tanaka, Thomas,
  Thomson, Tinyakov, Tkachev, Tokuno, Tomida, Troitsky, Tsunesada, Tsutsumi,
  Tsuyuguchi, Uchihori, Udo, Ukai, Vasiloff, Wada, Wong, Wood, Yamakawa,
  Yamaoka, Yamazaki, Yang, Yoshida, Yoshii, Zollinger, \& Zundel}]{TA:2012a}
Abu-Zayyad, T., Aida, R., Allen, M., {et~al.} 2012, Nuclear Instruments and
  Methods in Physics Research A, 689, 87

\bibitem[{Abu-Zayyad {et~al.}(2013)Abu-Zayyad, Aida, Allen, Anderson, Azuma,
  Barcikowski, Belz, Bergman, Blake, Cady, Cheon, Chiba, Chikawa, Cho, Cho,
  Fujii, Fujii, Fukuda, Fukushima, Hanlon, Hayashi, Hayashi, Hayashida, Hibino,
  Hiyama, Honda, Iguchi, Ikeda, Ikuta, Inoue, Ishii, Ishimori, Ivanov, Iwamoto,
  Jui, Kadota, Kakimoto, Kalashev, Kanbe, Kasahara, Kawai, Kawakami, Kawana,
  Kido, Kim, Kim, Kim, Kim, Kitamoto, Kitamura, Kitamura, Kobayashi, Kobayashi,
  Kondo, Kuramoto, Kuzmin, Kwon, Lan, Lim, Machida, Martens, Matsuda, Matsuura,
  Matsuyama, Matthews, Minamino, Miyata, Murano, Myers, Nagasawa, Nagataki,
  Nakamura, Nam, Nonaka, Ogio, Ohnishi, Ohoka, Oki, Oku, Okuda, Oshima, Ozawa,
  Park, Pshirkov, Rodriguez, Roh, Rubtsov, Ryu, Sagawa, Sakurai, Sampson,
  Scott, Shah, Shibata, Shibata, Shimodaira, Shin, Shin, Shirahama, Smith,
  Sokolsky, Stokes, Stratton, Stroman, Suzuki, Takahashi, Takeda, Taketa,
  Takita, Tameda, Tanaka, Tanaka, Tanaka, Thomas, Thomson, Tinyakov, Tkachev,
  Tokuno, Tomida, Troitsky, Tsunesada, Tsutsumi, Tsuyuguchi, Uchihori, Udo,
  Ukai, Vasiloff, Wada, Wong, Wood, Yamakawa, Yamane, Yamaoka, Yamazaki, Yang,
  Yoneda, Yoshida, Yoshii, Zhou, Zollinger, \& Zundel}]{TA:2013a}
Abu-Zayyad, T., Aida, R., Allen, M., {et~al.} 2013, ApJL, 768, L1

\bibitem[{Adams {et~al.}(2013)Adams, Ahmad, Albert, Allard, Ambrosio,
  Anchordoqui, Anzalone, Arai, Aramo, Asano, Ave, Barrillon, Batsch, Bayer,
  Belenguer, Bellotti, Berlind, Bertaina, Biermann, Biktemerova, Blaksley,
  B{\l}e{\c c}ki, Blin-Bondil, Bl{\"u}mer, Bobik, Bogomilov, Bonamente, Briggs,
  Briz, Bruno, Cafagna, Campana, Capdevielle, Caruso, Casolino, Cassardo,
  Castellini, Catalano, Cellino, Chikawa, Christl, Connaughton, Cort{\'e}s,
  Crawford, Cremonini, Csorna, D'Olivo, Dagoret-Campagne, de~Castro, De~Donato,
  de~la Taille, Del~Peral, Dell'Oro, de~Pascale, Di~Martino, Distratis,
  Dupieux, Ebersoldt, Ebisuzaki, Engel, Falk, Fang, Fenu,
  Fern{\'a}ndez-G{\'o}mez, Ferrarese, Franceschi, Fujimoto, Galeotti, Garipov,
  Geary, Giaccari, Giraudo, Gonchar, Gonz{\'a}lez~Alvarado, Gorodetzky,
  Guarino, Guzman, Hachisu, Harlov, Haungs, Hern{\'a}ndez~Carretero, Higashide,
  Iguchi, Ikeda, Inoue, Inoue, Insolia, Isgr{\`o}, Itow, Joven, Judd, Jung,
  Kajino, Kajino, Kaneko, Karadzhov, Karczmarczyk, Katahira, Kawai, Kawasaki,
  Keilhauer, Khrenov, Kim, Kim, Kim, Kleifges, Klimov, Ko, Kolev, Kreykenbohm,
  Kudela, Kurihara, Kuznetsov, La~Rosa, Lee, Licandro, Lim, L{\'o}pez,
  Maccarone, Mannheim, Marcelli, Marini, Martin-Chassard, Mart{\'\i}nez,
  Masciantonio, Mase, Matev, Maurissen, Medina~Tanco, Mernik, Miyamoto,
  Miyazaki, Mizumoto, Modestino, Monnier~Ragaigne, Morales de~los R{\'\i}os,
  Mot, Murakami, Nagano, Nagata, Nagataki, Nakamura, Nam, Nam, Nam, Napolitano,
  Naumov, Neronov, Nomoto, Ogawa, Ohmori, Olinto, Orlea{\'n}ski, Osteria,
  Pacheco, Panasyuk, Parizot, Park, Pastircak, Patzak, Paul, Pennypacker,
  Peter, Picozza, Pollini, Prieto, Reardon, Reina, Reyes, Ricci,
  Rodr{\'\i}guez, Rodr{\'\i}guez~Fr{\'\i}as, Ronga, Rothkaehl, Roudil, Rusinov,
  Rybczy{\'n}ski, Sabau, S{\'a}ez~Cano, Saito, Sakaki, Sakata, Salazar,
  S{\'a}nchez, Santangelo, Santiago~Cr{\'u}z, Sanz~Palomino, Saprykin, Sarazin,
  Sato, Sato, Schanz, Schieler, Scotti, Scuderi, Segreto, Selmane, Semikoz,
  Serra, Sharakin, Shibata, Shimizu, Shinozaki, Shirahama,
  Siemieniec-Ozieb{\l}o, Silva~Lopez, Sledd, S{\l}omi{\'n}ska, Sobey, Sugiyama,
  Supanitsky, Suzuki, Szabelska, Szabelski, Tajima, Tajima, Tajima, Takahashi,
  Takami, Takeda, Takizawa, Tenzer, Tibolla, Tkachev, Tomida, Tone, Trillaud,
  Tsenov, Tsuno, Tymieniecka, Uchihori, Vaduvescu, Vald{\'e}s~Galicia,
  Vallania, Valore, Vankova, Vigorito, Villase{\~n}or, von Ballmoos, Wada,
  Watanabe, Watanabe, Watts, Weber, Weiler, Wibig, Wiencke, Wille, Wilms,
  W{\l}odarczyk, Yamamoto, Yamamoto, Yang, Yano, Yashin, Yonetoku, Yoshida,
  Yoshida, Young, Zamora, \& Zuccaro~Marchi}]{JEM-EUSO:2013a}
Adams, J.~H., Ahmad, S., Albert, J.~N., {et~al.} 2013, Astropart. Phys., 44, 76

\bibitem[{Allard(2012)}]{Allard:2012}
Allard, D. 2012, Astropart. Phys., 39, 33

\bibitem[{Allard {et~al.}(2008)Allard, Busca, Decerprit, Olinto, \&
  Parizot}]{Allard:2008}
Allard, D., Busca, N.~G., Decerprit, G., Olinto, A.~V., \& Parizot, E. 2008,
  JCAP, 10, 033

\bibitem[{Allard {et~al.}(2007{\natexlab{a}})Allard, Olinto, \&
  Parizot}]{Allard:2007b}
Allard, D., Olinto, A.~V., \& Parizot, E. 2007{\natexlab{a}}, A{\&}A, 473, 59

\bibitem[{Allard {et~al.}(2007{\natexlab{b}})Allard, Parizot, \&
  Olinto}]{Allard:2007a}
Allard, D., Parizot, E., \& Olinto, A.~V. 2007{\natexlab{b}}, Astropart. Phys.,
  27, 61

\bibitem[{Allard {et~al.}(2005)Allard, Parizot, Olinto, Khan, \&
  Goriely}]{Allard:2005}
Allard, D., Parizot, E., Olinto, A.~V., Khan, E., \& Goriely, S. 2005, A{\&}A,
  443, L29

\bibitem[{Beck(2009)}]{Beck:2009}
Beck, R. 2009, Ap{\&}SS, 320, 77

\bibitem[{Berezinsky {et~al.}(2006)Berezinsky, Gazizov, \&
  Grigorieva}]{Berezinsky:2006}
Berezinsky, V.~S., Gazizov, A., \& Grigorieva, S.~I. 2006, Phys. Rev. D, 74,
  43005

\bibitem[{Blaksley \& Parizot(2012)}]{Blaksley:2012}
Blaksley, C. \& Parizot, E. 2012, Astropart. Phys., 35, 342

\bibitem[{Blaksley {et~al.}(2013)Blaksley, Parizot, Decerprit, \&
  Allard}]{Blaksley:2013}
Blaksley, C., Parizot, E., Decerprit, G., \& Allard, D. 2013, A{\&}A, 552, 125

\bibitem[{Brown {et~al.}(2007)Brown, Haverkorn, Gaensler, Taylor, Bizunok,
  McClure-Griffiths, Dickey, \& Green}]{Brown:2007}
Brown, J.~C., Haverkorn, M., Gaensler, B.~M., {et~al.} 2007, ApJ, 663, 258

\bibitem[{Crook {et~al.}(2007)Crook, Huchra, Martimbeau, Masters, Jarrett, \&
  Macri}]{Crook:2007}
Crook, A.~C., Huchra, J.~P., Martimbeau, N., {et~al.} 2007, ApJ, 655, 790

\bibitem[{Das {et~al.}(2008)Das, Kang, Ryu, \& Cho}]{Das:2008}
Das, S., Kang, H., Ryu, D., \& Cho, J. 2008, ApJ, 682, 29

\bibitem[{Dawson {et~al.}(2013)Dawson, Mari{\c s}, Roth, Salamida, Abu-Zayyad,
  Ikeda, Ivanov, Tsunesada, Pravdin, \& Sabourov}]{Dawson:2013}
Dawson, B.~R., Mari{\c s}, I.~C., Roth, M., {et~al.} 2013, UHECR 2012 -
  International Symposium on Future Directions in UHECR Physics, 53, 01005

\bibitem[{Decerprit \& Allard(2011)}]{Decerprit:2012b}
Decerprit, G. \& Allard, D. 2011, A{\&}A, 535, 66

\bibitem[{Decerprit {et~al.}(2012)Decerprit, Busca, \&
  Parizot}]{Decerprit:2012a}
Decerprit, G., Busca, N.~G., \& Parizot, E. 2012, A{\&}A, 538, 16

\bibitem[{Dolag {et~al.}(2002)Dolag, Bartelmann, \& Lesch}]{Dolag:2002}
Dolag, K., Bartelmann, M., \& Lesch, H. 2002, A{\&}A, 387, 383

\bibitem[{Donnert {et~al.}(2009)Donnert, Dolag, Lesch, \&
  M{\"u}ller}]{Donnert:2009}
Donnert, J., Dolag, K., Lesch, H., \& M{\"u}ller, E. 2009, MNRAS, 392, 1008

\bibitem[{Fukushima(2013)}]{Fukushima:2013}
Fukushima, M. 2013, UHECR 2012 - International Symposium on Future Directions
  in UHECR Physics, 53, 02002

\bibitem[{Giacalone \& Jokipii(1999)}]{Giacalone:1999}
Giacalone, J. \& Jokipii, J.~R. 1999, ApJ, 520, 204

\bibitem[{Globus {et~al.}(2008)Globus, Allard, \& Parizot}]{Globus:2008}
Globus, N., Allard, D., \& Parizot, E. 2008, A{\&}A, 479, 97

\bibitem[{G{\'o}rski {et~al.}(2005)G{\'o}rski, Hivon, Banday, Wandelt, Hansen,
  Reinecke, \& Bartelmann}]{Gorski:2005}
G{\'o}rski, K.~M., Hivon, E.~F., Banday, A.~J., {et~al.} 2005, ApJ, 622, 759

\bibitem[{Greisen(1966)}]{Greisen:1966}
Greisen, K. 1966, Phys. Rev. Lett., 16, 748

\bibitem[{Harari {et~al.}(1999)Harari, Mollerach, \& Roulet}]{Harari:1999}
Harari, D.~D., Mollerach, S., \& Roulet, E. 1999, JHEP, 08, 022

\bibitem[{{High Resolution Fly's Eye Collaboration}(2008)}]{HiRes:2008a}
{High Resolution Fly's Eye Collaboration}. 2008, Phys. Rev. Lett., 100, 101101

\bibitem[{Huchra {et~al.}(2012)Huchra, Macri, Masters, Jarrett, Berlind,
  Calkins, Crook, Cutri, Erdo{\u g}du, Falco, George, Hutcheson, Lahav, Mader,
  Mink, Martimbeau, Schneider, Skrutskie, Tokarz, \& Westover}]{Huchra:2012}
Huchra, J.~P., Macri, L.~M., Masters, K.~L., {et~al.} 2012, ApJS, 199, 26

\bibitem[{Jaffe {et~al.}(2011)Jaffe, Banday, Leahy, Leach, \&
  Strong}]{Jaffe:2011}
Jaffe, T.~R., Banday, A.~J., Leahy, J.~P., Leach, S., \& Strong, A.~W. 2011,
  MNRAS, 416, 1152

\bibitem[{Jaffe {et~al.}(2010)Jaffe, Leahy, Banday, Leach, Lowe, \&
  Wilkinson}]{Jaffe:2010}
Jaffe, T.~R., Leahy, J.~P., Banday, A.~J., {et~al.} 2010, MNRAS, 401, 1013

\bibitem[{Jansson \& Farrar(2012{\natexlab{a}})}]{Jansson:2012a}
Jansson, R. \& Farrar, G.~R. 2012{\natexlab{a}}, ApJ, 757, 14

\bibitem[{Jansson \& Farrar(2012{\natexlab{b}})}]{Jansson:2012b}
Jansson, R. \& Farrar, G.~R. 2012{\natexlab{b}}, ApJL, 761, L11

\bibitem[{Kachelrie{\ss} \& Semikoz(2006)}]{Kachelriess:2006}
Kachelrie{\ss}, M. \& Semikoz, D.~V. 2006, Phys. Lett. B, 634, 143

\bibitem[{{Kampert, K.-H. for the Pierre Auger
  Collaboration}(2011)}]{Kampert:2011}
{Kampert, K.-H. for the Pierre Auger Collaboration}. 2011, in Proceedings of
  the 32$^{\rm nd}$ International Cosmic Ray Conference, Beijing, China

\bibitem[{Kotera {et~al.}(2010)Kotera, Allard, \& Olinto}]{Kotera:2010}
Kotera, K., Allard, D., \& Olinto, A.~V. 2010, JCAP, 10, 013

\bibitem[{Kotera \& Lemoine(2008)}]{Kotera:2008}
Kotera, K. \& Lemoine, M. 2008, Phys. Rev. D, 77, 123003

\bibitem[{Kotera \& Olinto(2011)}]{Kotera:2011}
Kotera, K. \& Olinto, A.~V. 2011, ARA{\&}A, 49, 119

\bibitem[{Krause(2007)}]{Krause:2007}
Krause, M. 2007, Mem. Soc. Astron. Ital., 78, 314

\bibitem[{Krause {et~al.}(2006)Krause, Wielebinski, \& Dumke}]{Krause:2006}
Krause, M., Wielebinski, R., \& Dumke, M. 2006, A{\&}A, 448, 133

\bibitem[{Letessier-Selvon \& Stanev(2011)}]{Letessier-Selvon:2011}
Letessier-Selvon, A. \& Stanev, T. 2011, Rev. Mod. Phys., 83, 907

\bibitem[{Li \& Ma(1983)}]{Li:1983}
Li, T.-P. \& Ma, Y.-Q. 1983, ApJ, 272, 317

\bibitem[{{Matthews, J. N. for the Pierre Auger and Telescope Array
  Collaborations}(2013)}]{Matthews:2013}
{Matthews, J. N. for the Pierre Auger and Telescope Array Collaborations}.
  2013, in Proceedings of the 33$^{\rm rd}$ International Cosmic Ray
  Conference, Rio de Janeiro, Brazil

\bibitem[{Nagano \& Watson(2000)}]{Nagano:2000}
Nagano, M. \& Watson, A.~A. 2000, Rev. Mod. Phys., 72, 689

\bibitem[{{Pierre Auger Collaboration}(2004)}]{Auger:2004a}
{Pierre Auger Collaboration}. 2004, Nucl. Instrum. Meth. A, 523, 50

\bibitem[{{Pierre Auger Collaboration}(2007)}]{Auger:2007a}
{Pierre Auger Collaboration}. 2007, Science, 318, 938

\bibitem[{{Pierre Auger Collaboration}(2008)}]{Auger:2008b}
{Pierre Auger Collaboration}. 2008, Astropart. Phys., 29, 188

\bibitem[{{Pierre Auger Collaboration}(2010{\natexlab{a}})}]{Auger:2010c}
{Pierre Auger Collaboration}. 2010{\natexlab{a}}, Phys. Rev. Lett., 104, 91101

\bibitem[{{Pierre Auger Collaboration}(2010{\natexlab{b}})}]{Auger:2010a}
{Pierre Auger Collaboration}. 2010{\natexlab{b}}, Phys. Lett. B, 685, 239

\bibitem[{{Pierre Auger Collaboration}(2010{\natexlab{c}})}]{Auger:2010b}
{Pierre Auger Collaboration}. 2010{\natexlab{c}}, Astropart. Phys., 34, 314

\bibitem[{Ryu {et~al.}(2010)Ryu, Das, \& Kang}]{Ryu:2010}
Ryu, D., Das, S., \& Kang, H. 2010, ApJ, 710, 1422

\bibitem[{Ryu {et~al.}(2008)Ryu, Kang, Cho, \& Das}]{Ryu:2008}
Ryu, D., Kang, H., Cho, J., \& Das, S. 2008, Science, 320, 909

\bibitem[{Sigl {et~al.}(2004)Sigl, Miniati, \& En{\ss}lin}]{Sigl:2004}
Sigl, G., Miniati, F., \& En{\ss}lin, T.~A. 2004, Phys. Rev. D, 70, 43007

\bibitem[{Tokuno {et~al.}(2012)Tokuno, Tameda, Takeda, Kadota, Ikeda, Chikawa,
  Fujii, Fukushima, Honda, Inoue, Kakimoto, Kawana, Kido, Matthews, Nonaka,
  Ogio, Okuda, Ozawa, Sagawa, Sakurai, Shibata, Taketa, Thomas, Tomida,
  Tsunesada, Udo, Abu-Zayyad, Aida, Allen, Anderson, Azuma, Barcikowski, Belz,
  Bergman, Blake, Cady, Cheon, Chiba, Cho, Cho, Fujii, Fukuda, Gorbunov,
  Hanlon, Hayashi, Hayashi, Hayashida, Hibino, Hiyama, Iguchi, Ikuta, Ishii,
  Ishimori, Ivanov, Iwamoto, Jui, Kalashev, Kanbe, Kasahara, Kawai, Kawakami,
  Kim, Kim, Kim, Kim, Kitamoto, Kobayashi, Kobayashi, Kondo, Kuramoto, Kuzmin,
  Kwon, Lim, Machida, Martens, Martineau, Matsuda, Matsuura, Matsuyama, Myers,
  Minamino, Miyata, Miyauchi, Murano, Nakamura, Nam, Ohnishi, Ohoka, Oki, Oku,
  Oshima, Park, Pshirkov, Rodriguez, Roh, Rubtsov, Ryu, Sampson, Scott, Shah,
  Shibata, Shimodaira, Shin, Shin, Shirahama, Smith, Sokolsky, Sonley,
  Springer, Stokes, Stratton, Stroman, Suzuki, Takahashi, Takita, Tanaka,
  Tanaka, Tanaka, Thomson, Tinyakov, Tkachev, Troitsky, Tsutsumi, Tsuyuguchi,
  Uchihori, Ukai, Vasiloff, Wada, Wong, Wood, Yamakawa, Yamaoka, Yamazaki,
  Yang, Yoshida, Yoshii, Zollinger, \& Zundel}]{TA:2012b}
Tokuno, H., Tameda, Y., Takeda, M., {et~al.} 2012, Nuclear Instruments and
  Methods in Physics Research A, 676, 54

\bibitem[{Tully {et~al.}(2009)Tully, Rizzi, Shaya, Courtois, Makarov, \&
  Jacobs}]{Tully:2009}
Tully, R.~B., Rizzi, L., Shaya, E.~J., {et~al.} 2009, AJ, 138, 323

\bibitem[{Zatsepin \& Kuz'min(1966)}]{Zatsepin:1966}
Zatsepin, G.~T. \& Kuz'min, V.~A. 1966, Sov. Phys. JETP Lett., 4, 78

\end{thebibliography}

\end{document}